# Computational frameworks for homogenization and multiscale stability analyses of nonlinear periodic metamaterials

Guodong Zhang*, Nan Feng* and Kapil Khandelwal§

*Graduate Students, Dept. of Civil & Env. Engg. & Earth Sci., University of Notre Dame
§Associate Professor, Dept. of Civil & Env. Engg. & Earth Sci., 156 Fitzpatrick Hall, University of Notre Dame, Notre Dame, IN 46556, United States, Email:kapil.khandelwal@nd.edu, ORCID: 0000-0002-5748-6019, (Corresponding Author).

*Preprint Submitted*

## Abstract

This paper presents a consistent computational framework for multiscale 1st order finite strain homogenization and stability analyses of rate-independent solids with periodic microstructures. Based on the principle of multiscale virtual power, the homogenization formulation is built on a priori discretized microstructure, and algorithms for computing the matrix representations of the homogenized stresses and tangent moduli are consistently derived. The homogenization results lose their validity at the onset of 1st bifurcation, which can be computed from multiscale stability analysis. The multiscale instabilities include: a) microscale structural instability which is calculated by Bloch wave analysis; and b) macroscale material instability which is calculated by rank-1 convexity checks on the homogenized tangent moduli. Details on the implementation of the Bloch wave analysis are provided, including the selection of the wave vector space and the retrieval of the real-valued buckling mode from the complex-valued Bloch wave. With the Bloch wave representation, the stability analysis of a periodic microstructure of an infinite extent can be carried out using one-unit cell that can be considered as a representative volume element (RVE). Three methods are detailed for solving the resulted constrained eigenvalue problem – two condensation methods and a null-space based projection method. Both implementations of the homogenization and stability analyses are validated using numerical examples including hyperelastic and elastoplastic metamaterials. The invariance of both homogenization and multiscale stability analysis w.r.t to the choice of RVE is demonstrated. Various microscale buckling phenomena are also demonstrated by examining several representative metamaterial examples. Aligned with theoretical results, the numerical results show that the microscopic long wavelength buckling can be equivalently detected by the loss of rank-1 convexity of the homogenized tangent moduli.

# 1 Introduction

In recent years, architected materials have received much attention and are expected to have a significant impact on how the materials are designed and discovered in future. By designing the underlying periodic microstructures, the architected materials can achieve desired engineering properties that are often not observed in nature materials, e.g. high strength-to-weight ratio [1], negative Poisson's ratio [2, 3], desirable band-gaps [4, 5], high energy absorption [6, 7], etc. For this reason, architected materials are also known as *metamaterials*. The progress in metamaterials design and application is also fueled by the advancements of additive manufacturing technologies [8], through which the fabrication of complex geometries and multiple material phases in materials microstructures is now possible. As a key tool in metamaterials analysis and design, consistent computational frameworks that are capable of describing metamaterials behavior are crucially needed to realize future metamaterials.

As a bridge between micro and macro scales, the homogenization theories can be dated back to Voigt [9] and Reuss [10] bounds, where uniform strain and stress are respectively assumed on microscale for the calculation of the homogenized material properties at macroscale. The theories were further developed and refined in a series of work [11-13] on linear composites. The extension to the nonlinear regime was initiated by the work of Hill [14] and then followed by Ogden [15], Willis and co-workers [16], among others [17-19]. In contrast to the theoretical approaches that are mostly used to estimate bounds on macroscale material properties, the computational approaches to homogenization rely on finite element modeling of a representative volume element (RVE) of the underlying microstructure and are thus capable of capturing the effect brought by detailed geometric features of the microstructure [20-22]. For random heterogeneous composites, the selection of RVE is not straightforward and usually based on statistical methods [23]. On the contrary, for periodic metamaterials, which are also the focus in this study, the definition of RVE is unambiguous and can be chosen simply as the fundamental periodic cell, i.e. the unit cell. As a result, computational homogenization methods can be used as a reliable tool for accurate and effective calculation of the overall macroscopic properties of periodic metamaterials. The transition between the macro and micro scales is governed by the Hill-Mandel condition [14, 24] that states the energy equivalence between the two scales. Extension of this scale transition to incorporate inertia and body forces has been made in [25], where the transition condition is formulated using the principle of multiscale virtual power. On the computational side, efforts have also been made to achieve consistent and efficient algorithms for calculating macroscopic properties (i.e. homogenized stress and tangent moduli), cf. Miehe's work [22] where these quantities are derived based on discretized microstructures. Also, various advancement and development on computational homogenization such as incorporation of multi-physics [26], scale effects [27], crack modeling on microscale [28], among others [29-31], have been achieved. A comprehensive review on this topic can be found in [32].

An implicit assumption for the homogenization analysis results to be valid resides on the condition that the RVE, which is usually taken to be one unit cell in periodic metamaterials, remains valid for characterizing the response during the loading process. This in turn implies that the principal loading path is stable, and no other bifurcated solution besides the principal solution exists. However, from a structural stability viewpoint, the stability of the periodic microstructures is not always guaranteed. Using homogenization analysis, Abeyaratne and Triantafyllidis [33] showed that with strong ellipticity (rank-1 convexity) preserved in the matrix material, the homogenized periodic porous solids, however, can still lose strong ellipticity at adequately large loads, which implies a possible formation of the shear band. The shear band formation due to the loss of strong ellipticity has been well studied and understood, see Ref [34]. Although it seems plausible that the deformation localization on the macroscale may be initiated by a buckling type instability on the microscale, this stability phenomenon across micro and macro scales was not immediately clear. An early effort towards understanding this issue was made by Triantafyllidis and co-workers [35], where an analytical study on the bifurcation problem of a layered hyperelastic composite was carried out. This study showed that the macroscopic stable region consistently envelops the microscopic stable region and for the considered composites the loss of ellipticity condition for the homogenized material is the same as the condition for the long wavelength buckling of the microstructure. Subsequent work by Geymonat et al. [36] presented a rigorous proof for using the Bloch wave function in the microscale stability analysis of general 3D periodic solids of infinite size, and established the connection between the instabilities at micro and macro scales. That is, the buckling mode of infinite wavelength at microscale can be equivalently detected as a loss of rank-1 convexity of the homogenized tangent moduli at macroscale. Moreover, the homogenization results are only valid before the bifurcation occurs and these results lose their validity after the onset of bifurcation. Thus, the determination of the onset of such micro/macro instabilities is critical for understanding the behavior of nonlinear metamaterials. To investigate such phenomena, the Bloch wave analysis has been used in the past studies for understanding the stability behavior of various periodic composites [37-41]. Some of the microscale buckling modes have been successfully captured in experimental studies [42, 43]. In addition to the Bloch wave method, other methods, such as stability analysis on RVEs of increasing size [44, 45], the block-diagonalization method of group-theoretic bifurcation theory [46], and others [47, 48] have also been pursued for this purpose. Although these efforts represent important contributions towards understanding the behavior of nonlinear periodic metamaterials, there is a lack of a general clear and consistent computational frameworks that can be effectively used towards this end.

The main contribution of this work is to provide a consistent computational framework for both finite strain homogenization and multiscale stability analyses of periodic microstructures of rate-independent solids with implementation details. While some of the methods were presented by the authors in the context of topology optimization and isogeometric analysis [2, 49, 50], in this study an effort is made to unify the presentation and to further develop and clarify the corresponding computational schemes. Specifically, the homogenization is formulated based on the principle of multiscale virtual power with Lagrange multipliers for periodic boundary enforcement. A consistent derivation is presented for both strain and stress driven cases based on a priori discretized microstructure. The implementation details of the Bloch wave analysis are presented, which include the selection of the wave vector space and the retrieval of the real-valued buckling mode from the complex-valued Bloch wave representation. Three different treatments for the resulted constrained eigenvalue problems – two condensation methods and a null-space projection method – are detailed. The implementation of the multiscale homogenization and stability analyses is validated through numerical examples with both hyperelastic and elastoplastic metamaterials. Different choices of RVE are discussed and it is shown that both homogenization and stability results are consistent, irrespective of the choice of the RVE. Depending on the considered metamaterial and loading conditions, different types of buckling modes are observed including short-wavelength buckling with wavelength across one or multiple unit cells and long-wavelength buckling of infinite wavelength with respect to the unit cell size. Furthermore, different types of bifurcation points – simple, double and triple – are also shown in different test cases.

The rest of this paper is organized as follows: In Section 2, a detailed formulation of the finite strain homogenization is presented. Part of the framework was presented earlier by the authors in a topology optimization context [2, 50] and is further clarified and expanded here for stress-driven homogenization. In Section 3, numerical examples are carried out that validate the homogenization framework in the current context. Important theoretical aspects and implementation details of the Bloch wave method in microscale stability analysis and the macroscale rank-1 convexity analysis are presented in Section 4. Various numerical examples that validate the stability analysis implementation and illustrate different types of microscale instability are given in Section 5. Finally, remarks and conclusions are given in Section 6.

## 2 Finite Deformation Homogenization

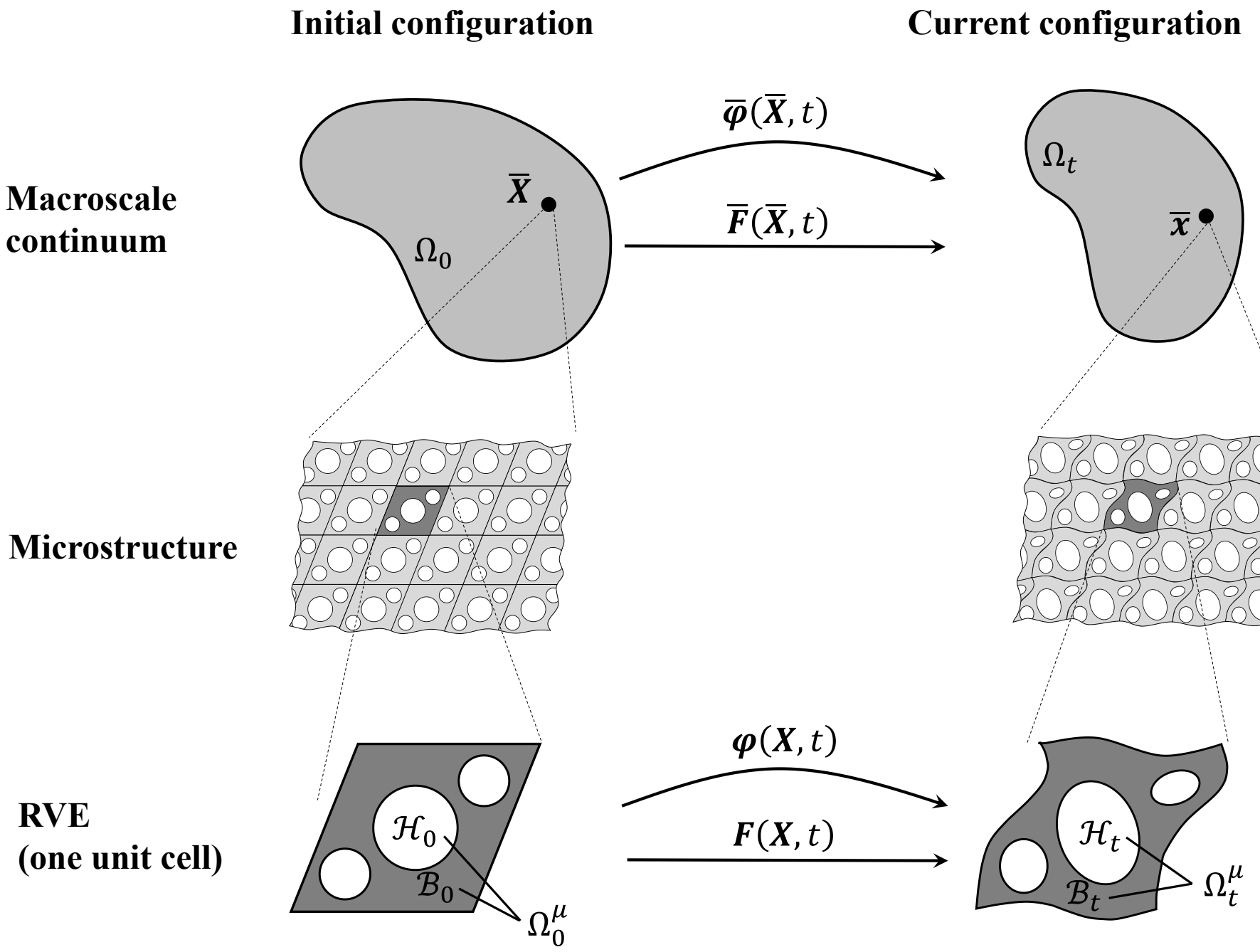

Figure 1. Illustration of the deformation of periodic solid. The motion $\boldsymbol{\varphi}$ of the RVE associated with a material point $\bar{\boldsymbol{X}}$ at the macroscale is driven by the macroscopic deformation $\bar{\boldsymbol{\varphi}}$ (or $\bar{\boldsymbol{F}}$).

Consider a metamaterial as shown in Figure 1, which is made up of periodic microstructure. The periodic microstructure can be characterized by a representative volume element (RVE). The RVE can include one or more fundamental periodic cells (i.e. unit cells). To fulfill the scale separation assumptions, the characteristic length of RVE should be much smaller than the dimensions of the macroscale continuum metamaterial [32]. In the initial configuration, the general shape of the periodic RVE in 2D is a parallelogram, see Figure 1 where $\Omega_0^\mu$ consists of a solid part $\mathcal{B}_0$ and void part $\mathcal{H}_0$, i.e. $\Omega_0^\mu = \mathcal{B}_0 \cup \mathcal{H}_0$.

Upon the application of boundary and loading conditions, the macroscale continuum undergoes a nonlinear deformation $\overline{\boldsymbol{\varphi}}$ which maps the continuum from the initial configuration $\Omega_0$ to the current configuration $\Omega_t$, i.e. $\overline{\boldsymbol{x}}(t) = \overline{\boldsymbol{\varphi}}(\overline{\boldsymbol{X}}, t)$ at $\overline{\boldsymbol{X}} \in \Omega_0$, $t \in \mathbb{R}^+$. Correspondingly, the RVE at material point $\overline{\boldsymbol{X}}$ deforms from its initial configuration $\Omega_0^\mu$ to its current configuration $\Omega_t^\mu$ through a nonlinear mapping $\boldsymbol{\varphi}$, i.e. $\boldsymbol{x}(t) = \boldsymbol{\varphi}(\boldsymbol{X}, t)$ at $\boldsymbol{X} \in \Omega_0^\mu$, $t \in \mathbb{R}^+$. Here an overbar is used to denote variables at macroscale, e.g. $\overline{\boldsymbol{X}}$ and $\overline{\boldsymbol{x}}$ are the position vectors of a material point in the initial and current configurations, respectively, at *macroscale*, while $\boldsymbol{X}$ and $\boldsymbol{x}$ are position vectors of a material point in the initial and current configurations, respectively, at *microscale*. An implicit assumption on the deformation is that the periodicity of the microstructure remains unchanged from the initial to the current configurations. Hence, the deformed RVE still serves as a periodic cell in the deformed microstructure, and can be homogenized for estimating the macroscopic metamaterial properties.

In the deformation-driven homogenization framework, deformation of the microstructure located at $\overline{\boldsymbol{X}}$ is driven by a local deformation gradient $\overline{\boldsymbol{F}}(\overline{\boldsymbol{X}}, t) = \boldsymbol{I} + \nabla_{\overline{X}} \overline{\boldsymbol{u}}$ where $\overline{\boldsymbol{u}}$ is the macroscopic displacement field satisfying $\overline{\boldsymbol{x}}(t) = \overline{\boldsymbol{X}} + \overline{\boldsymbol{u}}(t)$ and $\nabla_{\overline{X}}$ represents the gradient operator w.r.t. the macroscale coordinates $\overline{\boldsymbol{X}}$. In this study, the macroscopic deformation $\overline{\boldsymbol{F}}(\overline{\boldsymbol{X}}, t)$ is prescribed at certain fixed $\overline{\boldsymbol{X}} \in \Omega_0$, without referring to any specific macro-problem. The macroscopic material properties, i.e. homogenized/macroscopic stress and tangent moduli, are then evaluated under this deformation mode. Thus, the explicit dependence on $\overline{\boldsymbol{X}}$ is omitted in further discussions.

The microscopic displacement field $\boldsymbol{u}(\boldsymbol{X},t)$ over the RVE domain $\Omega_0^\mu$ is assumed to be driven by a prescribed macroscopic deformation $\overline{\boldsymbol{F}}(t)$, i.e.

$$\boldsymbol{u}(\boldsymbol{X},t) = (\overline{\boldsymbol{F}}(t) - \boldsymbol{I}).\boldsymbol{X} + \widetilde{\boldsymbol{u}}(\boldsymbol{X},t) \tag{1}$$

where $\widetilde{\boldsymbol{u}}(\boldsymbol{X},t)$ is the displacement fluctuation field. Correspondingly, the microscopic deformation gradient reads

$$\boldsymbol{F}(\boldsymbol{X},t) = \overline{\boldsymbol{F}}(t) + \boldsymbol{\nabla}_{\boldsymbol{X}}\widetilde{\boldsymbol{u}}(\boldsymbol{X},t) \tag{2}$$

where $\boldsymbol{\nabla}_{\boldsymbol{X}}$ denotes the gradient operator w.r.t. the microscale coordinates $\boldsymbol{X}$. Following [25], the microscale displacement field has to satisfy the kinematical admissibility constraints, which are postulated as

$$\int_{\mathcal{B}_0} \boldsymbol{u}(\boldsymbol{X},t)dV = \boldsymbol{0} \qquad \text{and} \qquad \overline{\boldsymbol{F}}(t) = \boldsymbol{I} + \frac{1}{V}\int_{\partial\Omega_0^\mu} \boldsymbol{u}(\boldsymbol{X},t) \otimes \boldsymbol{N}(\boldsymbol{X})dS \tag{3}$$

in which $V$ is the volume of the domain $\Omega_0^\mu$ and $\boldsymbol{N}$ is the unit normal vector on the boundary $\partial\Omega_0^\mu$. Applying divergence theorem to Eq. $(3)_2$ gives

$$\overline{\boldsymbol{F}}(t) = \boldsymbol{I} + \frac{1}{V}\int_{\mathcal{B}_0} \boldsymbol{\nabla}_{\boldsymbol{X}}\boldsymbol{u}(\boldsymbol{X},t)dV - \frac{1}{V}\int_{\partial\mathcal{H}_0} \boldsymbol{u}(\boldsymbol{X},t) \otimes \boldsymbol{N}(\boldsymbol{X})dS \tag{4}$$

which shows that in general the macroscopic deformation gradient $\overline{\boldsymbol{F}}(t)$ is not equal to the volume average of the microscopic deformation gradient $\boldsymbol{F}(\boldsymbol{X},t)$ due to the presence of the voids [21], where note that $\partial\mathcal{B}_0 = \partial\Omega_0^\mu \cup \partial\mathcal{H}_0$ with $\partial(\blacksquare)$ representing the boundary of $\blacksquare$. Again using the divergence theorem, it can be shown that Eqns. $(3)_1$ and $(3)_2$ are equivalent to

$$\int_{\mathcal{B}_0} \widetilde{\boldsymbol{u}}(\boldsymbol{X},t)dV = \boldsymbol{0} \qquad \text{and} \qquad \int_{\partial\Omega_0^\mu} \widetilde{\boldsymbol{u}}(\boldsymbol{X},t) \otimes \boldsymbol{N}(\boldsymbol{X})dS = \boldsymbol{0} \tag{5}$$

where it is assumed that the coordinate system on microscale is chosen such that $\int_{\mathcal{B}_0} \boldsymbol{X} dV = \boldsymbol{0}$. It can be seen that the constraint in Eq. $(3)_1$ is equivalent to removing rigid-body translation, while the constraint in Eq. $(3)_2$ implicitly removes rigid-body rotation. Thus, the kinematically admissible displacement fluctuation field $\tilde{\boldsymbol{u}}(\boldsymbol{X},t)$ is defined in a functional space $\mathcal{V}_{min}$

$$\mathcal{V}_{min} = \left\{ \tilde{\boldsymbol{u}}(\boldsymbol{X},t) \middle| \tilde{\boldsymbol{u}}(\boldsymbol{X},t) \in H^1(\mathcal{B}_0), t \in \mathbb{R}^+, \int_{\mathcal{B}_0} \tilde{\boldsymbol{u}}(\boldsymbol{X},t) dV = \boldsymbol{0},\ \int_{\partial\Omega_0^\mu} \tilde{\boldsymbol{u}}(\boldsymbol{X},t) \otimes \boldsymbol{N}(\boldsymbol{X}) dS = \boldsymbol{0} \right\} \tag{6}$$

where $H^1(\mathcal{B}_0) = \{ \boldsymbol{v} | v_i \in L^2(\mathcal{B}_0), \partial v_i / \partial X_j \in L^2(\mathcal{B}_0), i,j = 1,2,\ldots,d \}$ and $L^2(\mathcal{B}_0)$ represents the space of square-integrable functions defined on $\mathcal{B}_0$ and $d$ is the number of space dimensions. The subscript *min* means that this set of constraints is the minimal set required for kinematical admissibility. As shown in Appendix A, this set of constraints corresponds to the *constant traction boundary conditions*, i.e.

$$\boldsymbol{t}_0(\boldsymbol{X},t) = \overline{\boldsymbol{P}}(t).\boldsymbol{N}(\boldsymbol{X}) \quad \text{on } \partial\Omega_0^\mu \tag{7}$$

where $\boldsymbol{t}_0(\boldsymbol{X},t) \stackrel{\text{def}}{=} \boldsymbol{P}(\boldsymbol{X},t).\boldsymbol{N}(\boldsymbol{X})$ represents the 1st Piola-Kirchhoff (nominal) traction acting on the reference surface with normal $\boldsymbol{N}(\boldsymbol{X})$; $\boldsymbol{P}(\boldsymbol{X},t)$ denotes the 1st Piola-Kirchhoff (PK) stress on the microscale at position $\boldsymbol{X}$ at time instant $t$, while $\overline{\boldsymbol{P}}(t)$ represents the macroscopic/homogenized 1st PK stress at time $t$ (see Eq. (13)). Additional constraints can be introduced in a consistent way that may lead to periodic boundary conditions or linear displacement boundary conditions.

*Periodic boundary condition*

For RVE, the boundary can be divided into a pair of negative and positive sides, denoting as $\partial\Omega_0^{\mu-}$ and $\partial\Omega_0^{\mu+}$, respectively, see Figure 2 where points on the positive side can be reached by translating the corresponding points on the negative side using a periodic lattice vector $\boldsymbol{a}_1$ or $\boldsymbol{a}_2$ or $\pm(\boldsymbol{a}_1 \pm \boldsymbol{a}_2)$. For the periodic boundary conditions, the displacement fluctuations on the negative side equal the corresponding ones on the positive side, i.e.,

$$\widetilde{\boldsymbol{u}}^{+} = \widetilde{\boldsymbol{u}}^{-} \quad \text{on } \partial\Omega_0^{\mu} \tag{8}$$

which can be proved to automatically satisfy the constraints in Eqns. $(3)_2$ or $(5)_2$. As a result, the kinematically admissible displacement fluctuation field considering periodic boundary condition is defined in a functional space $\mathcal{V}_p$

$$\mathcal{V}_p = \left\{ \widetilde{\boldsymbol{u}}(\boldsymbol{X},t) \middle| \widetilde{\boldsymbol{u}}(\boldsymbol{X},t) \in H^1(\mathcal{B}_0), t \in \mathbb{R}^+, \int_{\mathcal{B}_0} \widetilde{\boldsymbol{u}}(\boldsymbol{X},t) dV = \boldsymbol{0},\ \widetilde{\boldsymbol{u}}^{+}(t) = \widetilde{\boldsymbol{u}}^{-}(t) \text{ on } \partial\Omega_0^{\mu} \right\} \tag{9}$$

*Linear displacement boundary condition*

The linear displacement boundary condition requires zero displacement fluctuations on the boundaries, i.e.

$$\widetilde{\boldsymbol{u}} = \boldsymbol{0} \quad \text{on } \partial\Omega_0^{\mu} \tag{10}$$

which satisfies the constraints in Eq. (8) and also constraints in Eqns. (3) or (5). The corresponding functional space $\mathcal{V}_l$ for the kinematically admissible displacement fluctuation field reads

$$\mathcal{V}_l = \left\{ \widetilde{\boldsymbol{u}}(\boldsymbol{X},t) \middle| \widetilde{\boldsymbol{u}}(\boldsymbol{X},t) \in H^1(\mathcal{B}_0), t \in \mathbb{R}^+,\ \widetilde{\boldsymbol{u}}(t) = \boldsymbol{0} \text{ on } \partial\Omega_0^{\mu} \right\} \tag{11}$$

The transition between the micro and macro scales is governed by the principle of multiscale virtual power [25], which is expressed as, $\forall\, t \in \mathbb{R}^+$

$$\overline{\boldsymbol{P}} : \delta\overline{\boldsymbol{F}} = \frac{1}{V} \int_{\mathcal{B}_0} \boldsymbol{P} : \delta\boldsymbol{F}\, dV \qquad \forall\, \delta\overline{\boldsymbol{F}} \in \text{Lin}, \qquad \delta\widetilde{\boldsymbol{u}} \in \mathcal{V} \tag{12}$$

where $\overline{\boldsymbol{P}}$ and $\boldsymbol{P}$ are the 1st PK stress tensors on macro and micro scales, respectively, and the space of the virtual fluctuation field $\delta\widetilde{\boldsymbol{u}}$ is identical to that of the fluctuation field $\widetilde{\boldsymbol{u}}$, $\mathcal{V}$, which can be $\mathcal{V}_{min}$, $\mathcal{V}_p$ or $\mathcal{V}_l$. Eq. (12) can be seen as the variational form of the Hill-Mandel condition [14, 24] that states the equivalence of the incremental virtual work between the micro and macro scales.

The stress homogenization relation

$$\overline{\boldsymbol{P}} = \frac{1}{V}\int_{\mathcal{B}_0} \boldsymbol{P}\, dV = \frac{1}{V}\int_{\partial\mathcal{B}_0} \boldsymbol{t}_0 \otimes \boldsymbol{X}\, dS \equiv \frac{1}{V}\int_{\partial\Omega_0^{\mu}} \boldsymbol{t}_0 \otimes \boldsymbol{X}\, dS \qquad \text{with } \boldsymbol{t}_0 = \boldsymbol{P}.\boldsymbol{N} \tag{13}$$

and the microscale equilibrium equation

$$\int_{\mathcal{B}_0} \boldsymbol{P}: \boldsymbol{\nabla}_{\boldsymbol{X}} \delta\widetilde{\boldsymbol{u}}\, dV = 0 \qquad \forall\, \delta\widetilde{\boldsymbol{u}} \in \mathcal{V} \tag{14}$$

can be obtained from Eq. (12) by choosing $\delta\widetilde{\boldsymbol{u}} = \boldsymbol{0}$ and $\delta\overline{\boldsymbol{F}} = \boldsymbol{0}$, respectively. Here, the second equality in Eq. (13) can be proved using divergence theorem and the fact that $\boldsymbol{\nabla}_{\boldsymbol{X}}.\boldsymbol{P} = \boldsymbol{0}$, while the third equality is due to the traction-free void boundaries, i.e. $\boldsymbol{t}_0 = \boldsymbol{0}$ on $\partial\mathcal{H}_0$.

### *2.1 Deformation driven homogenization*

In this section, a deformation-driven homogenization formulation for computing the homogenized stresses and tangent moduli of a priori discretized microstructure is presented. For ease of the derivation of homogenized quantities, the Lagrange multiplier is adopted to enforce the boundary condition. Other methods such as the condensation method can also be used [32]. As illustrated in Figure 1 that the geometry of RVE must satisfy certain constraints to be compatible with periodicity. For instance, the most general RVE shape for 2D problems is the parallelogram (Figure 1 or Figure 2a), and square or rectangular shapes are special cases of a parallelogram. The hexagonal unit cell (Figure 2b) can also be equivalently recast into a parallelogram. In this study, 2D problems (plane strain) are considered in the numerical implementations, however, all the presented methods can be canonically extended to 3D cases. Since the underlying microstructure of the metamaterials is always assumed to be periodic with repeating unit cells (Figure 1), the periodic boundary conditions are chosen [36], i.e. $\mathcal{V} = \mathcal{V}_p$ in Eq. (12).

Consider now for a given discretized RVE (Figure 2), the constraints in Eq. (8) are discretized as

$$\widetilde{\boldsymbol{u}}_q^{+} = \widetilde{\boldsymbol{u}}_q^{-}\,, \quad q = 1,2,\ldots,m \tag{15}$$

where $m$ pairs of nodes lying on the negative and positive boundary sides are identified. For example, $m$ = 17 for the parallelogram RVE in Figure 2a including 14 pairs on the inner positive/negative sides and 3 pairs at the corners (see Remarks in Section 2.1.1). The rigid-body translation constraint (Eq. (5)$_1$) can be equivalently replaced by fixing one arbitrary point, e.g. $\widetilde{\boldsymbol{u}}_o = \boldsymbol{0}$ in $\mathcal{B}_0$. Thus, the discretized functional space $\mathcal{V}^h$ is defined by

$$\mathcal{V}^h = \left\{\widetilde{\boldsymbol{u}}(\boldsymbol{X},t) \middle| \widetilde{\boldsymbol{u}}(\boldsymbol{X},t) \in H^1(\mathcal{B}_0), t \in \mathbb{R}^+,\ \widetilde{\boldsymbol{u}}_o = \boldsymbol{0},\ \widetilde{\boldsymbol{u}}_q^+ = \widetilde{\boldsymbol{u}}_q^-\ (q = 1,2,\dots,m)\right\} \tag{16}$$

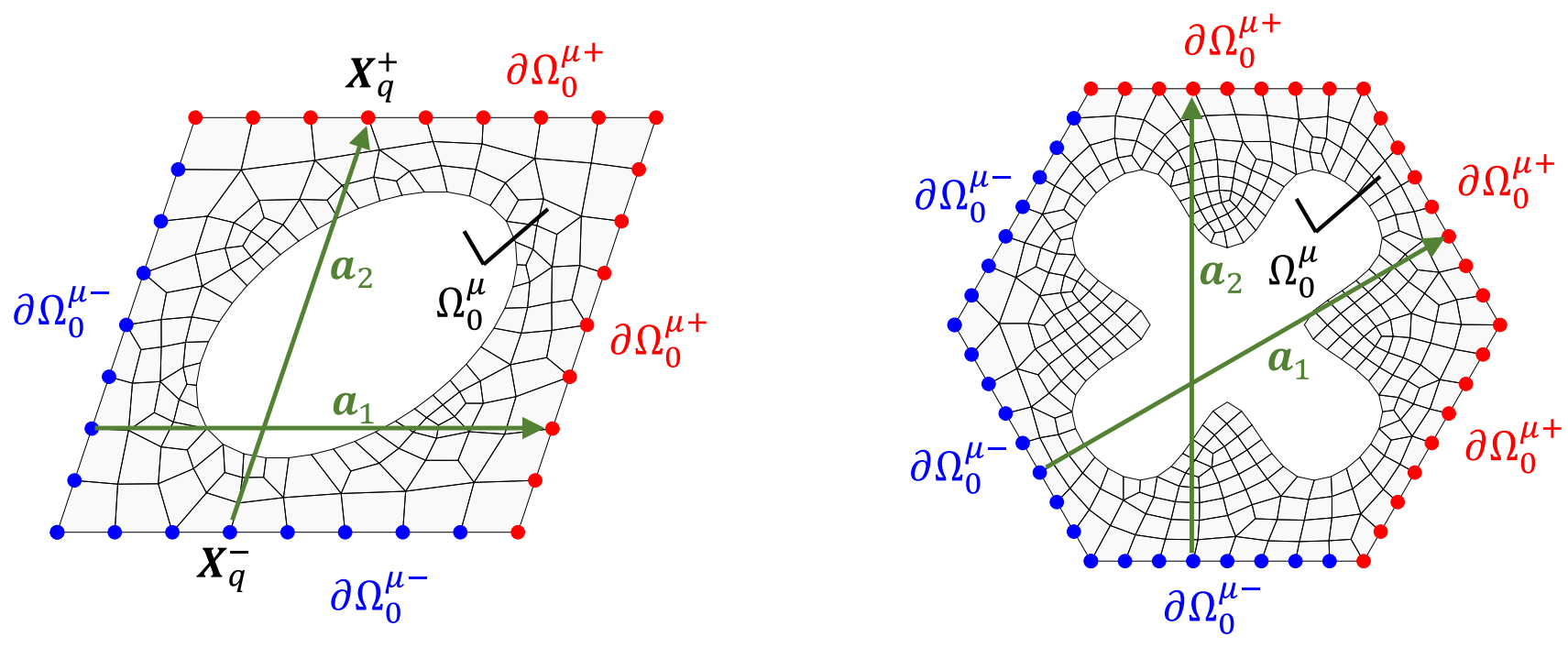

(a) Parallelogram RVE (b) Hexagonal RVE

Figure 2. Geometries and partitioning of boundary nodes of discretized microstructures of RVE (blue color denotes the negative nodes and red color denotes the positive nodes).

### *2.1.1 Principle of multiscale virtual power with Lagrange multiplier – discrete form*

Using the Lagrange multipliers to enforce the constraints in Eq. (16), the discretized version of the principle of multiscale virtual power is expressed as

$$\begin{aligned} &-V(\overline{\boldsymbol{P}}:\delta\overline{\boldsymbol{F}}) + \int_{\mathcal{B}_0} \boldsymbol{P}:\delta\boldsymbol{F}dV - \delta\boldsymbol{\lambda}^T\boldsymbol{u}_o - \boldsymbol{\lambda}^T\delta\boldsymbol{u}_o - \sum_{q=1}^{m}\delta\boldsymbol{\mu}_q^T\left[\boldsymbol{u}_q^+ - \boldsymbol{u}_q^- - (\overline{\boldsymbol{F}} - \boldsymbol{I}).\boldsymbol{L}_q\right] \\ &\qquad - \sum_{q=1}^{m}\boldsymbol{\mu}_q^T\left[\delta\boldsymbol{u}_q^+ - \delta\boldsymbol{u}_q^- - \delta\overline{\boldsymbol{F}}.\boldsymbol{L}_q\right] = 0 \end{aligned} \tag{17}$$

$$\forall\ \delta\overline{\boldsymbol{F}} \in \mathrm{Lin},\ \ \delta\boldsymbol{u} \in H^1(\mathcal{B}_0),\ \ \delta\boldsymbol{\lambda},\ \ \delta\boldsymbol{\mu}$$

where $\boldsymbol{\lambda}$ and $\boldsymbol{\mu} = [\boldsymbol{\mu}_1, \ldots, \boldsymbol{\mu}_m]^T$ are the Lagrange multipliers, and the constraints are restated in terms of the displacement field $\boldsymbol{u}(\boldsymbol{X}, t)$ instead of fluctuation field $\widetilde{\boldsymbol{u}}(\boldsymbol{X}, t)$. Note that $\boldsymbol{u}_o(t) = \boldsymbol{0}$ is equivalent to $\widetilde{\boldsymbol{u}}_o(t) = \boldsymbol{0}$ in the sense of removing rigid-body translations. The vector $\boldsymbol{L}_q$ in Eq. (17) represents the translation vector that satisfies $\boldsymbol{X}_q^+ = \boldsymbol{X}_q^- + \boldsymbol{L}_q$ where $\boldsymbol{X}_q^+$ and $\boldsymbol{X}_q^-$ are the coordinates of the nodes on a pair of positive and negative sides, see Figure 2.

***Remark:*** For the $m$ pairs of periodic boundary condition constraints, the inner nodes on each positive and negative sides can be easily identified and related through the periodic translation vectors. For the end nodes on the sides, i.e. corner nodes, special care is needed for identifying sufficient and necessary constraints that represent the periodic boundary conditions. For example, Figure 3 shows the parallelogram and hexagon cases where only the periodic constraints that are related to the corner nodes are demonstrated. As can be seen, after the removal of the repeated constraints, three constraints are needed for parallelogram RVE that identify three pairs of negative and positive corner nodes, while four constraints are needed for hexagon RVE that identify four pairs of negative and positive corner nodes. It should be noted that the constraints listed in Figure 3 can be restated in multiple ways as long as the constraints are linearly independent, e.g. the second equation in Figure 3(a) can be restated as $\boldsymbol{u}_3 = \boldsymbol{u}_1 + \boldsymbol{\nabla}_{\bar{X}}\overline{\boldsymbol{u}}.(\boldsymbol{a}_1 + \boldsymbol{a}_2)$. Also noted is that corner nodes are not necessarily present, see Figure 15c, Figure 24a, Figure 24c and Figure 34b.

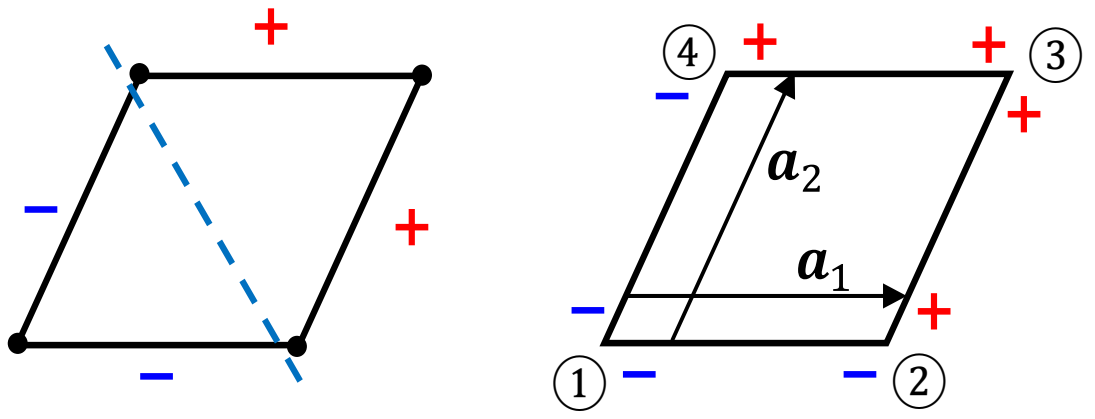

Periodic boundary conditions for corner nodes

$$\begin{cases} \boldsymbol{u}_2 = \boldsymbol{u}_1 + \nabla_{\overline{X}}\overline{\boldsymbol{u}}.\boldsymbol{a}_1 \\ \boldsymbol{u}_3 = \boldsymbol{u}_4 + \nabla_{\overline{X}}\overline{\boldsymbol{u}}.\boldsymbol{a}_1 \\ \boldsymbol{u}_4 = \boldsymbol{u}_1 + \nabla_{\overline{X}}\overline{\boldsymbol{u}}.\boldsymbol{a}_2 \\ \boldsymbol{u}_3 = \boldsymbol{u}_2 + \nabla_{\overline{X}}\overline{\boldsymbol{u}}.\boldsymbol{a}_2 \longleftarrow \text{(repeated, remove!)} \end{cases}$$

(a) Parallelogram unit cell

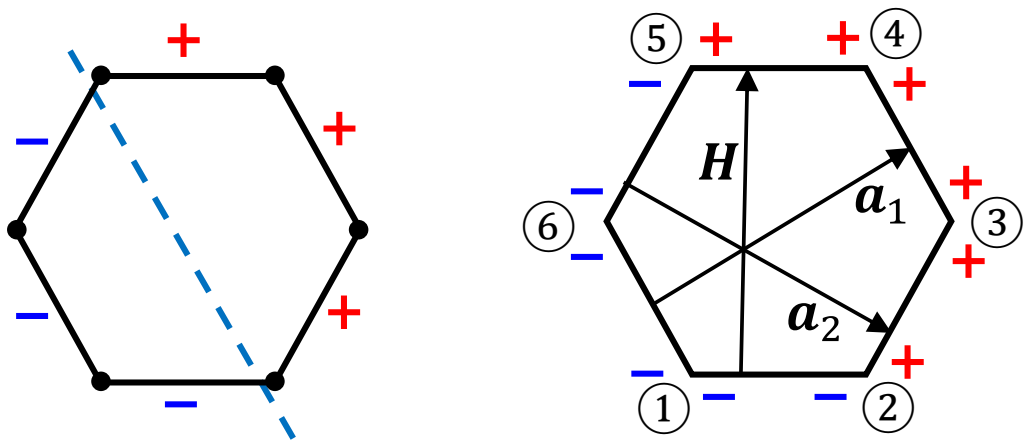

Periodic boundary conditions for corner nodes

$$\begin{cases} \boldsymbol{u}_5 = \boldsymbol{u}_1 + \nabla_{\overline{X}}\overline{\boldsymbol{u}}.\boldsymbol{H} \\ \boldsymbol{u}_4 = \boldsymbol{u}_2 + \nabla_{\overline{X}}\overline{\boldsymbol{u}}.\boldsymbol{H} \longleftarrow \text{(repeated, remove!)} \\ \boldsymbol{u}_3 = \boldsymbol{u}_1 + \nabla_{\overline{X}}\overline{\boldsymbol{u}}.\boldsymbol{a}_1 \\ \boldsymbol{u}_4 = \boldsymbol{u}_6 + \nabla_{\overline{X}}\overline{\boldsymbol{u}}.\boldsymbol{a}_1 \\ \boldsymbol{u}_2 = \boldsymbol{u}_6 + \nabla_{\overline{X}}\overline{\boldsymbol{u}}.\boldsymbol{a}_2 \\ \boldsymbol{u}_3 = \boldsymbol{u}_5 + \nabla_{\overline{X}}\overline{\boldsymbol{u}}.\boldsymbol{a}_2 \longleftarrow \text{(repeated, remove!)} \end{cases}$$

(b) Hexagon unit cell

Figure 3. Periodic boundary conditions for corner nodes ($\boldsymbol{H} = \boldsymbol{a}_1 - \boldsymbol{a}_2$ in (b)).

### *2.1.2 Interpretation of Lagrange multipliers*

The Lagrange multipliers $\boldsymbol{\lambda}$ and $\boldsymbol{\mu}$ can be interpreted as discrete nodal forces on the boundary. For instance, assuming $\delta\overline{\boldsymbol{F}} = \boldsymbol{0}$, $\delta\boldsymbol{\lambda} = \boldsymbol{0}$ and $\delta\boldsymbol{\mu} = \boldsymbol{0}$ and $\delta\boldsymbol{u} = \boldsymbol{c}_0$ (with $\boldsymbol{c}_0$ constant in $\mathcal{B}_0$) in Eq. (17) gives $\boldsymbol{\lambda}^T\boldsymbol{c}_0 = 0$. Therefore,

$$\boldsymbol{\lambda} = \boldsymbol{0} \tag{18}$$

has to be satisfied, which means that for a self-equilibrated system, fixing one arbitrary point for removing rigid-body translation does not create any reaction forces. Next, taking $\delta\overline{\boldsymbol{F}} = \boldsymbol{0}$, $\delta\boldsymbol{\lambda} = \boldsymbol{0}$ and $\delta\boldsymbol{\mu} = \boldsymbol{0}$ with $\delta\boldsymbol{u}(\boldsymbol{X}) = \boldsymbol{A}_0.\boldsymbol{X}$ in $\mathcal{B}_0$ where $\boldsymbol{A}_0 \in \text{Lin}$ (a constant 2nd-order tensor) in Eq. (17), it can be shown that

$$\left(\int_{\mathcal{B}_0} \boldsymbol{P} dV - \sum_{q=1}^{m} \boldsymbol{\mu}_q \otimes \boldsymbol{L}_q\right) : \boldsymbol{A}_0 = 0 \qquad \forall\, \boldsymbol{A}_0 \in \text{Lin} \tag{19}$$

where $\boldsymbol{\mu}_q$ and $\boldsymbol{L}_q$ $(q = 1, \ldots, m)$ are both vectors (or 1st-order tensors) while $\boldsymbol{P}$ and $\boldsymbol{A}_0$ are 2nd-order tensors. Since $\boldsymbol{A}_0$ can be chosen arbitrarily, it follows from Eqns. (13) and (19) that

$$\overline{\boldsymbol{P}} = \frac{1}{V}\int_{\mathcal{B}_0} \boldsymbol{P} dV = \frac{1}{V}\sum_{q=1}^{m} \boldsymbol{\mu}_q \otimes \boldsymbol{L}_q \tag{20}$$

which when combined with Eq. (13)₃ shows that $\boldsymbol{\mu}_q$ represents the traction force at node $q$. Therefore, it can be seen that the homogenized stress can be computed from the Lagrange multipliers $\boldsymbol{\mu}$.

### *2.1.3 Finite element formulation*

Considering the unknown variables to be solved as $\boldsymbol{u}$, $\boldsymbol{\lambda}$ and $\boldsymbol{\mu}$, the resulting set of nonlinear constrained equilibrium equations, from Eq. (17), can be written as

$$\boldsymbol{R}(\boldsymbol{u},\boldsymbol{\lambda},\boldsymbol{\mu}) = \begin{bmatrix} \boldsymbol{R}_1(\boldsymbol{u},\boldsymbol{\lambda},\boldsymbol{\mu}) \\ \boldsymbol{R}_2(\boldsymbol{u}) \\ \boldsymbol{R}_3(\boldsymbol{u}) \end{bmatrix} = \begin{bmatrix} \boldsymbol{F}_{int}(\boldsymbol{u}) - \boldsymbol{A}_1^T\boldsymbol{\lambda} - \boldsymbol{A}_2^T\boldsymbol{\mu} \\ -\boldsymbol{A}_1\boldsymbol{u} \\ -\boldsymbol{A}_2\boldsymbol{u} \end{bmatrix} + \begin{bmatrix} \boldsymbol{0} \\ \boldsymbol{0} \\ \boldsymbol{h} \end{bmatrix} = \boldsymbol{0} \tag{21}$$

where $\boldsymbol{F}_{int}$ represents the global internal force vector defined by

$$\boldsymbol{F}_{int}(\boldsymbol{u}) = \mathop{\mathbf{A}}_{e=1}^{n_{ele}} \boldsymbol{F}_{int}^{e} \quad \text{with} \quad \boldsymbol{F}_{int}^{e} = \int_{\Omega^e} \boldsymbol{B}^T \boldsymbol{P}\, dV \tag{22}$$

where $\boldsymbol{B}$ is the shape function derivative matrix, $\Omega^e$ represents the $e^{th}$ element integration domain satisfying $\mathcal{B}_0 = \bigcup_{e=1}^{n_{ele}} \Omega^e$ and $n_{ele}$ are the total number of elements in the RVE.

The matrices $\boldsymbol{A}_1$ and $\boldsymbol{A}_2$, and vector $\boldsymbol{h}$ are constructed such that

$$\boldsymbol{u}_o = \boldsymbol{A}_1\boldsymbol{u}$$

$$\boldsymbol{u}^+ - \boldsymbol{u}^- = \boldsymbol{A}_2\boldsymbol{u}$$

$$\boldsymbol{h} = \begin{bmatrix} (\overline{\boldsymbol{F}} - \boldsymbol{I}).\boldsymbol{L}_1 \\ \vdots \\ (\overline{\boldsymbol{F}} - \boldsymbol{I}).\boldsymbol{L}_m \end{bmatrix} = [\boldsymbol{L}_M]([\overline{\boldsymbol{F}}] - [\boldsymbol{I}]) = \begin{bmatrix} \tilde{X}_1 & 0 & \tilde{Y}_1 & 0 \\ 0 & \tilde{X}_1 & 0 & \tilde{Y}_1 \\ \vdots & \vdots & \vdots & \vdots \\ \tilde{X}_m & 0 & \tilde{Y}_m & 0 \\ 0 & \tilde{X}_m & 0 & \tilde{Y}_m \end{bmatrix}_{2m\times 4} \left( \begin{bmatrix} \bar{F}_{11} \\ \bar{F}_{21} \\ \bar{F}_{12} \\ \bar{F}_{22} \end{bmatrix} - \begin{bmatrix} 1 \\ 0 \\ 0 \\ 1 \end{bmatrix} \right) \tag{23}$$

where $\boldsymbol{u}$ is the global nodal displacement vector, $\boldsymbol{u}^{+}=[\boldsymbol{u}_1^{+},\dots,\boldsymbol{u}_m^{+}]^T$ and $\boldsymbol{u}^{-}=[\boldsymbol{u}_1^{-},\dots,\boldsymbol{u}_m^{-}]^T$ includes $m$ nodal displacements defined on the positive and negative boundary sides, respectively. $\boldsymbol{L}_q=[\tilde{X}_q,\tilde{Y}_q]^T$ is the translational vector from the $q^{th}$ node on the negative side to the $q^{th}$ node on the positive side. The expression of $\boldsymbol{h}$ vector is written for 2D case in Eq. (23).

The nonlinear system in Eq. (21) is solved using the Newton-Raphson (NR) method and the Jacobian matrix, which is needed for NR solver, can be calculated as

$$[\mathbf{J}_T]=\begin{bmatrix}\partial\boldsymbol{R}_1/\partial\boldsymbol{u} & \partial\boldsymbol{R}_1/\partial\boldsymbol{\lambda} & \partial\boldsymbol{R}_1/\partial\boldsymbol{\mu}\\ \partial\boldsymbol{R}_2/\partial\boldsymbol{u} & \partial\boldsymbol{R}_2/\partial\boldsymbol{\lambda} & \partial\boldsymbol{R}_2/\partial\boldsymbol{\mu}\\ \partial\boldsymbol{R}_3/\partial\boldsymbol{u} & \partial\boldsymbol{R}_3/\partial\boldsymbol{\lambda} & \partial\boldsymbol{R}_3/\partial\boldsymbol{\mu}\end{bmatrix}=\begin{bmatrix}\boldsymbol{K}_T & -\boldsymbol{A}_1^T & -\boldsymbol{A}_2^T\\ -\boldsymbol{A}_1 & \mathbf{0} & \mathbf{0}\\ -\boldsymbol{A}_2 & \mathbf{0} & \mathbf{0}\end{bmatrix} \tag{24}$$

where the term $\boldsymbol{K}_T$ is the tangent structural stiffness matrix calculated by

$$\boldsymbol{K}_T=\frac{\partial\boldsymbol{F}_{int}}{\partial\boldsymbol{u}}=\mathop{\mathbf{A}}_{e=1}^{n_{ele}}\boldsymbol{k}_T^e \quad\text{with}\quad \boldsymbol{k}_T^e=\int_{\Omega^e}\boldsymbol{B}^T[\mathbb{A}]\boldsymbol{B}\,dV \quad\text{and}\quad \mathbb{A}=\frac{\partial\boldsymbol{P}}{\partial\boldsymbol{F}} \tag{25}$$

in which the tangent moduli $\mathbb{A}$ is obtained from material subroutine. Here the bracket outside $\mathbb{A}$ means that it is arranged in a matrix form, e.g. 9×9 matrix in 3D case and 4×4 matrix for 2D plane strain case.

#### *2.1.4 Homogenized stress and tangent moduli*

Using Eq. $(20)_2$ and the definition of matrix $[\boldsymbol{L}_M]$ given in Eq. (23), the homogenized stress $\overline{\boldsymbol{P}}$ is computed as

$$[\overline{\boldsymbol{P}}]=\frac{1}{V}[\boldsymbol{L}_M]^T\boldsymbol{\mu} \tag{26}$$

where the bracket outside $\overline{\boldsymbol{P}}$ means that it is arranged in a 4×1 vector form (2D), similarly as $[\overline{\boldsymbol{F}}]$ used in Eq. (23).

The 4$^{th}$-order tensor homogenized tangent moduli $\overline{\mathbb{A}}$ is defined by

$$\bar{\mathbb{A}} = \frac{\partial \bar{\boldsymbol{P}}}{\partial \bar{\boldsymbol{F}}} \tag{27}$$

and can be rephrased in a matrix form as $[\bar{\mathbb{A}}] = \partial[\bar{\boldsymbol{P}}]/\partial[\bar{\boldsymbol{F}}]$, which is of size 4×4 for 2D case. From Eq. (26), it is clear that $[\bar{\mathbb{A}}]$ is determined by the derivative of Lagrange multiplier $\boldsymbol{\mu}$ with respect to $\bar{\boldsymbol{F}}$. To this end, the set of global equilibrium equation (Eq. (21)) is perturbed at the equilibrium state by a perturbation $\Delta\bar{\boldsymbol{F}}$, i.e.,

$$\begin{bmatrix} \boldsymbol{K}_T & -\boldsymbol{A}_1^T & -\boldsymbol{A}_2^T \\ -\boldsymbol{A}_1 & \boldsymbol{0} & \boldsymbol{0} \\ -\boldsymbol{A}_2 & \boldsymbol{0} & \boldsymbol{0} \end{bmatrix} \begin{bmatrix} \Delta\boldsymbol{u} \\ \Delta\boldsymbol{\lambda} \\ \Delta\boldsymbol{\mu} \end{bmatrix} + \begin{bmatrix} \boldsymbol{0} \\ \boldsymbol{0} \\ \boldsymbol{L}_M \end{bmatrix} [\Delta\bar{\boldsymbol{F}}] = \boldsymbol{0} \tag{28}$$

which results in

$$\begin{bmatrix} \Delta\boldsymbol{u} \\ \Delta\boldsymbol{\lambda} \\ \Delta\boldsymbol{\mu} \end{bmatrix} = -[\mathbf{J}_T]^{-1} \begin{bmatrix} \boldsymbol{0} \\ \boldsymbol{0} \\ \boldsymbol{L}_M \end{bmatrix} [\Delta\bar{\boldsymbol{F}}] \tag{29}$$

Combining Eq. (27) with Eqns. (26) and (29), it can be shown that

$$[\bar{\mathbb{A}}] = -\frac{1}{V}\left[\hat{\boldsymbol{L}}_M\right]^T [\mathbf{J}_T]^{-1} \left[\hat{\boldsymbol{L}}_M\right] \tag{30}$$

where the matrix $\left[\hat{\boldsymbol{L}}_M\right]$ is of size $(N + 2 + 2m) \times 4$ for a 2D case and is defined by

$$\left[\hat{\boldsymbol{L}}_M\right] = \begin{bmatrix} \boldsymbol{0}_{N\times 4} \\ \boldsymbol{0}_{2\times 4} \\ [\boldsymbol{L}_M]_{2m\times 4} \end{bmatrix} \tag{31}$$

where $N$ is the number of total DOFs in the displacement field, i.e. the size of $\boldsymbol{u}$ vector.

### *2.2 Stress driven homogenization*

In the stress driven homogenization, the macro Kirchhoff stress tensor ($\bar{\boldsymbol{\tau}} = \bar{\boldsymbol{P}}.\bar{\boldsymbol{F}}^T$) is prescribed and drives the homogenization analysis. Due to the principle of conservation of angular momentum, the macro Kirchhoff stress tensor is symmetric, i.e. $\bar{\boldsymbol{\tau}} = \bar{\boldsymbol{\tau}}^T$. In order to use the deformation driven framework presented in Section 2.1, the deformation gradient $\bar{\boldsymbol{F}}$ has to be solved for the prescribed $\bar{\boldsymbol{\tau}}$. Without loss of generality, the macroscopic rigid-body rotation is ignored, i.e. $\overline{\boldsymbol{R}} = \boldsymbol{I}$ or $\overline{\boldsymbol{F}} = \overline{\boldsymbol{U}}$, and thus $\overline{\boldsymbol{F}}$ is symmetric. Hence, the problem is now rephrased as determining $\bar{F}_{11}$, $\bar{F}_{22}$ and $\bar{F}_{12}$ $(= \bar{F}_{21})$ for a given, $\bar{\boldsymbol{\tau}}$, i.e. $\bar{\tau}_{11}$, $\bar{\tau}_{22}$ and $\bar{\tau}_{12}$ $(= \bar{\tau}_{21})$. It is noted that the macro Kirchhoff stress tensor $\bar{\boldsymbol{\tau}}$ is used as the applied macroscopic stress instead of the macro Cauchy stress as they differ only by a scaler $\bar{\mathrm{J}}$ $(= \det \overline{\boldsymbol{F}})$, which does not change the principal stress directions, and the focus in this study is on the multiscale stabilities under different loading paths in the principal stress space. To this end, suppose the principal macro stresses $\bar{\tau}_a$ ($a$ = 1, 2 for 2D case) are at a fixed angle $\theta$ with respect to the standard Euclidean bases $\{\boldsymbol{e}_a\}$, i.e.,

$$\bar{\boldsymbol{\tau}} = \boldsymbol{Q}\bar{\boldsymbol{\tau}}'\boldsymbol{Q}^T \quad \text{with} \quad \boldsymbol{Q}(\theta) = \begin{bmatrix} \cos\theta & -\sin\theta \\ \sin\theta & \cos\theta \end{bmatrix} \quad \text{and} \quad \bar{\boldsymbol{\tau}}' = \begin{bmatrix} \bar{\tau}'_{11} & \bar{\tau}'_{12} \\ \bar{\tau}'_{21} & \bar{\tau}'_{22} \end{bmatrix} = \begin{bmatrix} \bar{\tau}_1 & 0 \\ 0 & \bar{\tau}_2 \end{bmatrix} \tag{32}$$

with $\boldsymbol{Q}$ the bases transformation matrix, or written equivalently in matrix-vector form as

$$[\bar{\boldsymbol{\tau}}] = [\mathbb{Q}][\bar{\boldsymbol{\tau}}'] \quad \text{or} \quad [\bar{\boldsymbol{\tau}}'] = [\mathbb{Q}]^T[\bar{\boldsymbol{\tau}}] \tag{33}$$

where $[\bar{\boldsymbol{\tau}}] = [\bar{\tau}_{11} \quad \bar{\tau}_{21} \quad \bar{\tau}_{12} \quad \bar{\tau}_{22}]^T$ and $[\bar{\boldsymbol{\tau}}'] = [\bar{\tau}'_{11} \quad \bar{\tau}'_{21} \quad \bar{\tau}'_{12} \quad \bar{\tau}'_{22}]^T$ $( = [\bar{\tau}_1 \quad 0 \quad 0 \quad \bar{\tau}_2]^T )$ are the vector forms of $\bar{\boldsymbol{\tau}}$ and $\bar{\boldsymbol{\tau}}'$, respectively, and

$$[\mathbb{Q}] = \begin{bmatrix} \cos^2\theta & -\sin\theta\cos\theta & -\sin\theta\cos\theta & \sin^2\theta \\ \sin\theta\cos\theta & \cos^2\theta & -\sin^2\theta & -\sin\theta\cos\theta \\ \sin\theta\cos\theta & -\sin^2\theta & \cos^2\theta & -\sin\theta\cos\theta \\ \sin^2\theta & \sin\theta\cos\theta & \sin\theta\cos\theta & \cos^2\theta \end{bmatrix} \tag{34}$$

Without loss of generality, the principal stresses $\bar{\tau}_a$ are parameterized as

$$\bar{\tau}_1 = -\lambda\cos\phi \quad \text{and} \quad \bar{\tau}_2 = -\lambda\sin\phi \tag{35}$$

where $\lambda$ is the amplitude of load (positive in compression) and $\phi$ controls the ratio of the macro principal stresses. As a result, the stress state ($\bar{\boldsymbol{\tau}}$) is fully described by three parameters $\lambda$, $\phi$ and $\theta$. Thus, the pure stress driven homogenization can be formulated as two nested loops, with inner loop formulated as deformation driven homogenization and outer loop formulated as a system of nonlinear equations that is solved for deformation gradient ($\overline{\boldsymbol{F}}$) for a prescribed stress tensor ($\bar{\boldsymbol{\tau}}$). The nonlinear equations for the outer loop can be expressed as

$$\boldsymbol{R}_{\tau} = \begin{bmatrix} \bar{\tau}'_{11}(\bar{F}_{11}, \bar{F}_{12}, \bar{F}_{22}) \\ \bar{\tau}'_{12}(\bar{F}_{11}, \bar{F}_{12}, \bar{F}_{22}) \\ \bar{\tau}'_{22}(\bar{F}_{11}, \bar{F}_{12}, \bar{F}_{22}) \end{bmatrix} - \begin{bmatrix} \lambda \cos\phi \\ 0 \\ \lambda \sin\phi \end{bmatrix} = \boldsymbol{0} \tag{36}$$

where $\bar{\tau}'_{ij}$ are computed from Eq. $(33)_2$. The Newton-Raphson method is used to solve above nonlinear equations for $\bar{F}_{11}$, $\bar{F}_{22}$ and $\bar{F}_{12}$ ($= \bar{F}_{21}$). To calculate the Jacobian matrix, it is easy to denote

$$\begin{bmatrix} \bar{\tau}'_{11} \\ \bar{\tau}'_{12} \\ \bar{\tau}'_{22} \end{bmatrix} = \begin{bmatrix} \cos^2\theta & -\sin\theta\cos\theta & -\sin\theta\cos\theta & \sin^2\theta \\ \sin\theta\cos\theta & -\sin^2\theta & \cos^2\theta & -\sin\theta\cos\theta \\ \sin^2\theta & \sin\theta\cos\theta & \sin\theta\cos\theta & \cos^2\theta \end{bmatrix} [\bar{\boldsymbol{\tau}}] = [\boldsymbol{T}][\bar{\boldsymbol{\tau}}] \tag{37}$$

where matrix $[\boldsymbol{T}]$ is $[\mathbb{Q}]^T$ with the 2nd row being removed. Next, due to the symmetry assumption, the deformation gradient is expressed as

$$[\bar{\boldsymbol{F}}] = \begin{bmatrix} \bar{F}_{11} \\ \bar{F}_{21} \\ \bar{F}_{12} \\ \bar{F}_{22} \end{bmatrix} = \begin{bmatrix} 1 & 0 & 0 \\ 0 & 0 & 1 \\ 0 & 0 & 1 \\ 0 & 1 & 0 \end{bmatrix} \begin{bmatrix} \bar{F}_{11} \\ \bar{F}_{22} \\ \bar{F}_{12} \end{bmatrix} = [\boldsymbol{I}_{43}]\left[\hat{\bar{\boldsymbol{F}}}\right] \tag{38}$$

where $\left[\hat{\bar{\boldsymbol{F}}}\right]$ are the unknown variables that are to be solved. Moreover, from $\bar{\boldsymbol{\tau}} = \bar{\boldsymbol{P}}.\bar{\boldsymbol{F}}^T$, it can be shown that

$$[\bar{\boldsymbol{\tau}}] = [\bar{\mathbb{P}}][\bar{\boldsymbol{F}}] = [\bar{\mathbb{F}}][\bar{\boldsymbol{P}}] \tag{39}$$

with

$$[\bar{\mathbb{P}}] = \begin{bmatrix} \bar{P}_{11} & 0 & \bar{P}_{12} & 0 \\ 0 & \bar{P}_{11} & 0 & \bar{P}_{12} \\ \bar{P}_{21} & 0 & \bar{P}_{22} & 0 \\ 0 & \bar{P}_{21} & 0 & \bar{P}_{22} \end{bmatrix} \quad \text{and} \quad [\bar{\mathbb{F}}] = \begin{bmatrix} \bar{F}_{11} & 0 & \bar{F}_{12} & 0 \\ \bar{F}_{21} & 0 & \bar{F}_{22} & 0 \\ 0 & \bar{F}_{11} & 0 & \bar{F}_{12} \\ 0 & \bar{F}_{21} & 0 & \bar{F}_{22} \end{bmatrix} \tag{40}$$

With the above notations, the Jacobian matrix $[\mathbf{J}_{\tau}] = \partial \boldsymbol{R}_{\tau} / \partial \hat{\bar{\boldsymbol{F}}}$ can be obtained as

$$[\mathbf{J}_{\tau}] = [\boldsymbol{T}] \frac{\partial [\bar{\boldsymbol{\tau}}]}{\partial [\bar{\boldsymbol{F}}]} [\boldsymbol{I}_{43}] \tag{41}$$

with

$$\frac{\partial[\bar{\boldsymbol{\tau}}]}{\partial[\bar{\boldsymbol{F}}]} = [\bar{\mathbb{P}}] + [\bar{\mathbb{F}}][\bar{\mathbb{A}}] \tag{42}$$

which is derived from Eq. (39).

## 3 Numerical Validation – Homogenization Analysis

This section numerically investigates the sufficiency of using one unit cell as RVE for homogenization analysis as well as the equivalency of different choices of unit cells from the same periodic metamaterial. Both hyperelastic and elastoplastic constituent phases are considered. Specifically, the hyperelastic material phase is modeled by the regularized neo-Hookean hyperelastic model for which the free energy is expressed as

$$\psi(\boldsymbol{C}) = \frac{1}{2}\kappa(J-1)^2 + \frac{\mu}{2}(\bar{I}_1 - 3) \tag{43}$$

where $\boldsymbol{C}$ is the right Cauchy-Green tensor and $\bar{I}_1$ is the first invariant of $\bar{\boldsymbol{C}}$, i.e. $\bar{I}_1 = \mathrm{tr}\bar{\boldsymbol{C}}$ with $\bar{\boldsymbol{C}} = J^{-2/3}\boldsymbol{C}$ and $J$ the determinant of the deformation gradient. $\kappa$ and $\mu$ are bulk and shear modulus of the material. The elastoplastic material phase is modeled by a finite strain $J_2$ plasticity and the model details are provided in Appendix B. All the computations in this study are carried out using an in-house Matlab based finite element library *CPSSL-FEA* developed at the University of Notre Dame.

### *3.1 Finite elastic deformations of a metamaterial*

The first example examines a hyperelastic metamaterial that represents path-independent material cases. The periodic metamaterial (Figure 4) consists of soft matrix and hard circular inclusions with diameter $d$ = 0.4, both of which are modeled by a regularized neo-Hookean hyperelastic model (Eq. (43)). The model parameters are $\kappa_m$ = 17.5 and $\mu_m$ = 8.0 for matrix and $\kappa_I = 100\kappa_m$ and $\mu_I = 100\mu_m$ for inclusion. The unit cell is of unit side length and unit thickness with inclusion located at either center or position (-0.2, 0.2). It is clear that the two unit cells with circular inclusions located at different positions represent the same metamaterial. With four-node (Q4) plane strain FE discretization, the FE meshes of the two unit cells are shown in Figure 4.

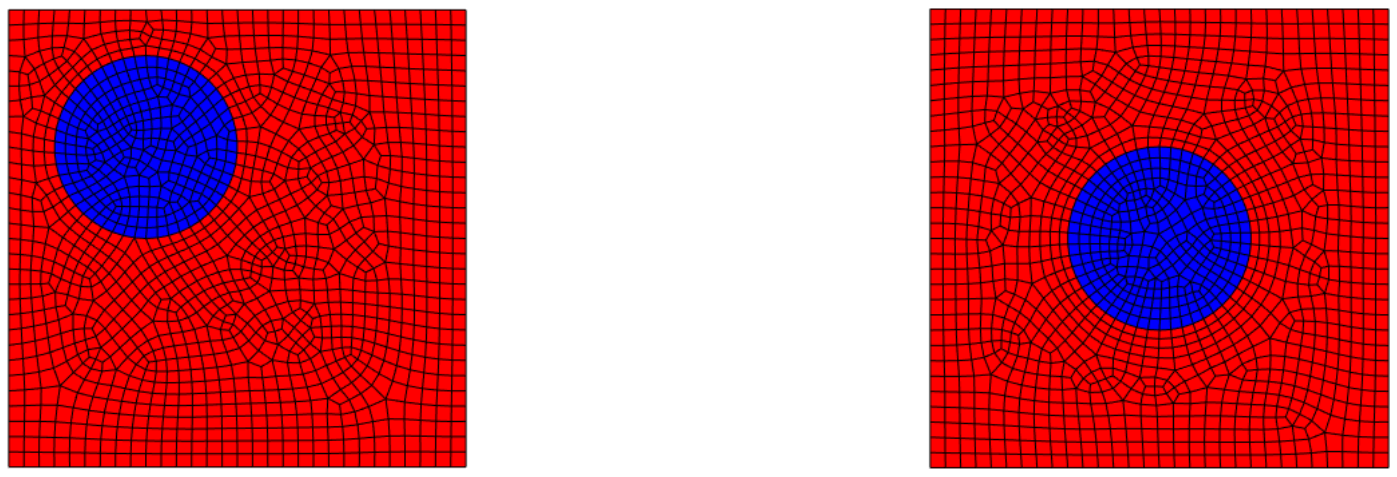

(a) Unit cell-1 ($n_{ele}$ = 1388) (b) Unit cell-2 ($n_{ele}$ = 1352)

Figure 4. FE meshes of two equivalent unit cells of the hyperelastic metamaterial.

The first two tests are carried out on unit cell-1 (Figure 4a) to demonstrate that it is sufficient to choose the smallest repeating cell (unit cell) as RVE for homogenization analysis. To this end, both simple tension $[\bar{\boldsymbol{F}}] = [\bar{F}_{11} \quad \bar{F}_{21} \quad \bar{F}_{12} \quad \bar{F}_{22}]^T = [1.4 \quad 0 \quad 0 \quad 1]^T$ and simple shear $[\bar{\boldsymbol{F}}] = [\bar{F}_{11} \quad \bar{F}_{21} \quad \bar{F}_{12} \quad \bar{F}_{22}]^T = [1 \quad 0 \quad 0.4 \quad 1]^T$ are considered for homogenization analysis on RVE that consists of a different number of unit cell-1. The results are shown in Table 1 and Figure 5 for simple tension and Table 2 and Figure 6 for simple shear. As shown in Table 1 and Table 2, the homogenized stress, tangent moduli, and strain energy for RVEs with a different number of unit cell match exactly with each other. Besides, the periodicity with respect to one unit cell of the deformed shape of the metamaterial can be seen from Figure 5 and Figure 6.

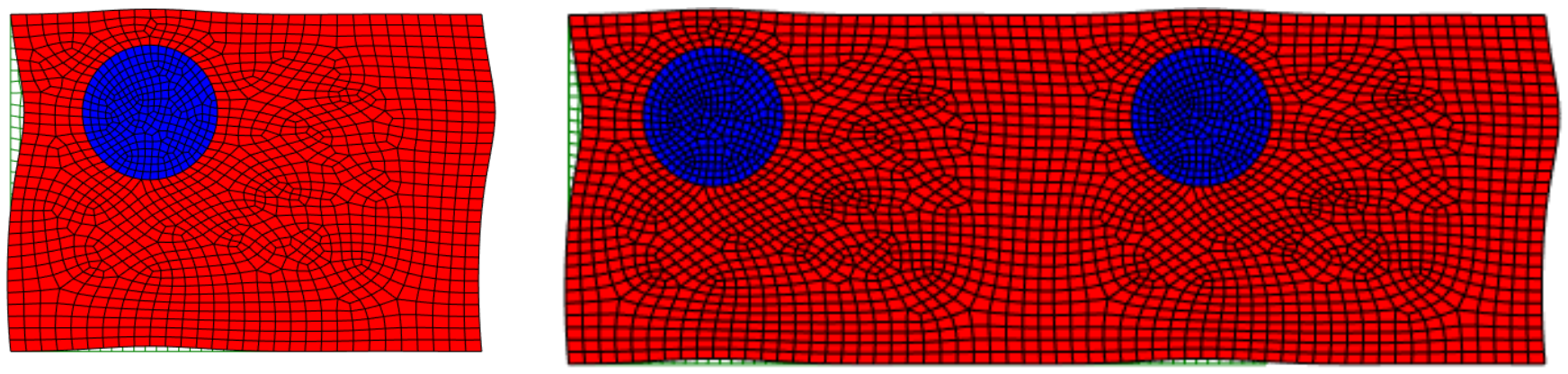

(a) 1×1 RVE (b) 2×1 RVE

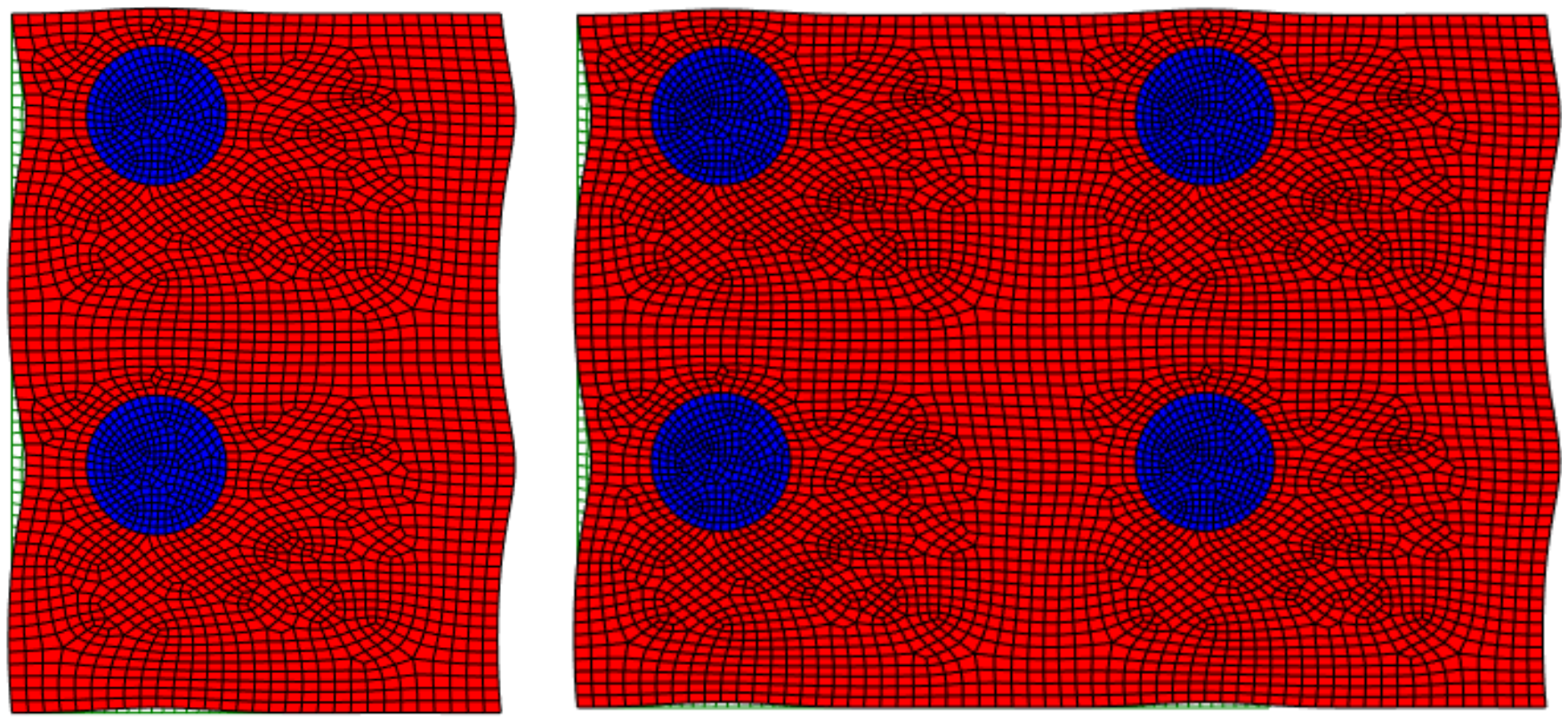

(c) 1×2 RVE (d) 2×2 RVE

Figure 5. Deformed shapes of the RVEs consisting of different number of unit cells (unit cell-1) under simple tension.

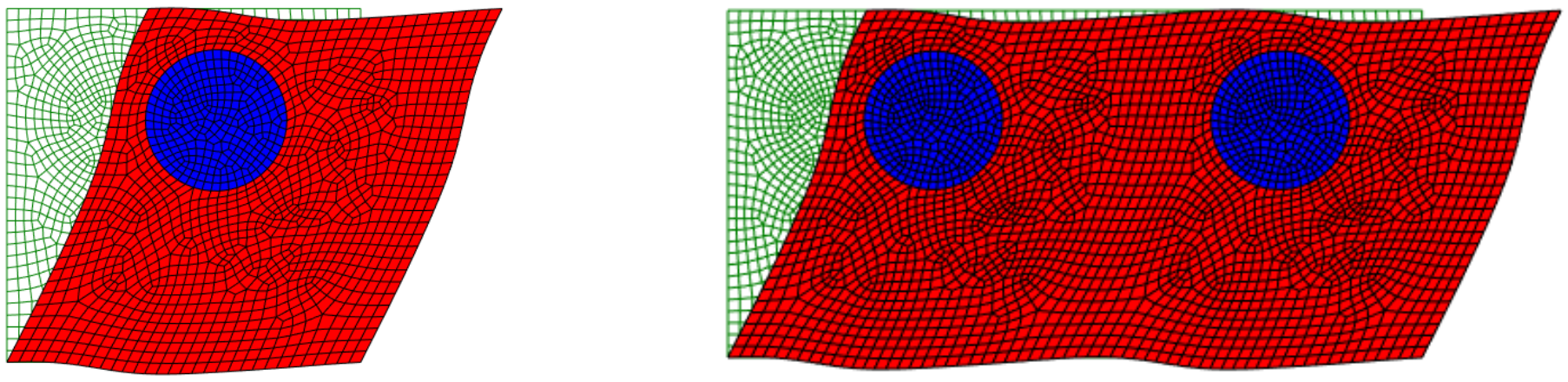

(a) 1×1 RVE (b) 2×1 RVE

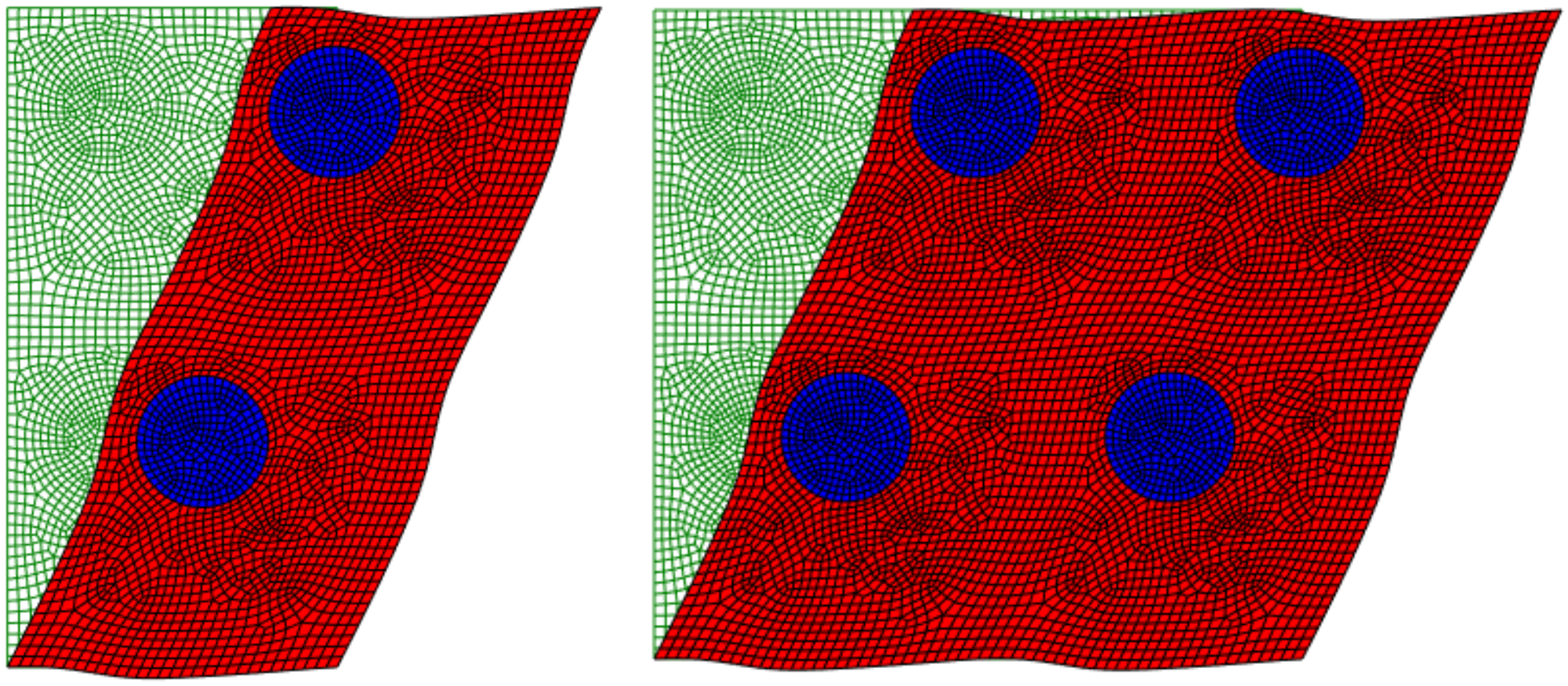

(c) 1×2 RVE (d) 2×2 RVE

Figure 6. Deformed shapes of the RVEs consisting of different number of unit cells (unit cell-1) under simple shear.

Table 1. Homogenized quantities from the RVEs consisting of different number of unit cells (unit cell-1) under simple tension.

| RVE | Homogenized 1st PK stress $[\bar{\boldsymbol{P}}]$ | Homogenized tangent moduli $[\bar{\mathbb{A}}]$ | | | | Homogenized stored energy $\bar{\psi}$ |
|---|---|---|---|---|---|---|
| 1×1 | 11.6573 | 26.0259 | -0.0002 | -0.0003 | 30.2150 | 2.4347 |
| 1×2 | 0 | -0.0002 | 7.5260 | -1.1210 | 0 | |
| 2×1 | 0 | -0.0003 | -1.1210 | 7.3172 | 0 | |
| 2×2 | 8.8865 | 30.2150 | 0 | 0 | 54.4927 | |

Table 2. Homogenized quantities from the RVEs consisting of different number of unit cells (unit cell-1) under simple shear.

| RVE | Homogenized 1st PK stress $[\bar{\boldsymbol{P}}]$ | Homogenized tangent moduli $[\bar{\mathbb{A}}]$ | | | | Homogenized stored energy $\bar{\psi}$ |
|---|---|---|---|---|---|---|
| 1×1 | -0.5929 | 35.1466 | -12.9755 | -2.9184 | 14.8978 | 0.7710 |
| 1×2 | 4.0980 | -12.9755 | 16.2037 | 11.5362 | -13.1511 | |
| 2×1 | 3.8687 | -2.9184 | 11.5362 | 9.7712 | -2.9799 | |
| 2×2 | -0.5731 | 14.8978 | -13.1511 | -2.9799 | 35.0999 | |

For periodic metamaterial, the size of the smallest repeating cell (unit cell) can be uniquely determined, whereas the choice of unit cell is not unique. For example, the two different unit cells shown in Figure 4 (a) and (b) both represent the same metamaterial. Theoretically, the homogenization result should be independent of the choice of the unit cell. To demonstrate this feature, the two different unit cells are analyzed under both simple tension and simple shear. The results are given in Table 3 and Figure 7 for tension while Table 4 and Figure 8 for shear. In this case, the deformed shapes are different due to the different unit cell geometric features as shown in Figure 7 and Figure 8. Moreover, there are also small differences in the calculated homogenized quantities, i.e. stress, tangent moduli, and strain energy, as shown in Table 3 and Table 4, which can be attributed to different FE meshes and geometric approximations.

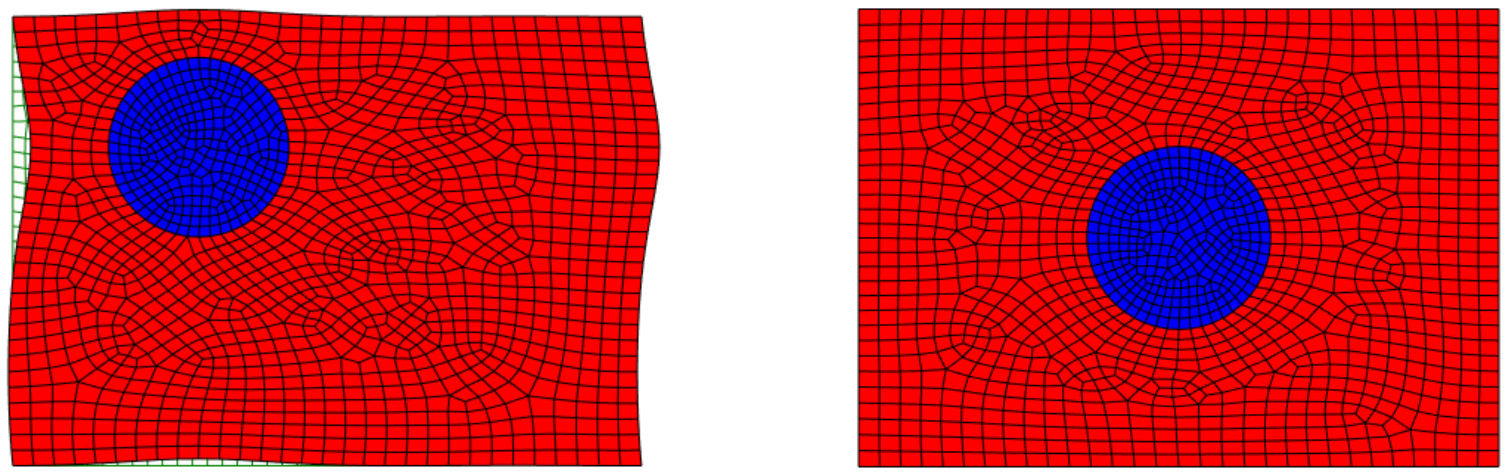

(a) Unit cell-1 (b) Unit cell-2

Figure 7. Deformed shapes of different unit cells under simple tension.

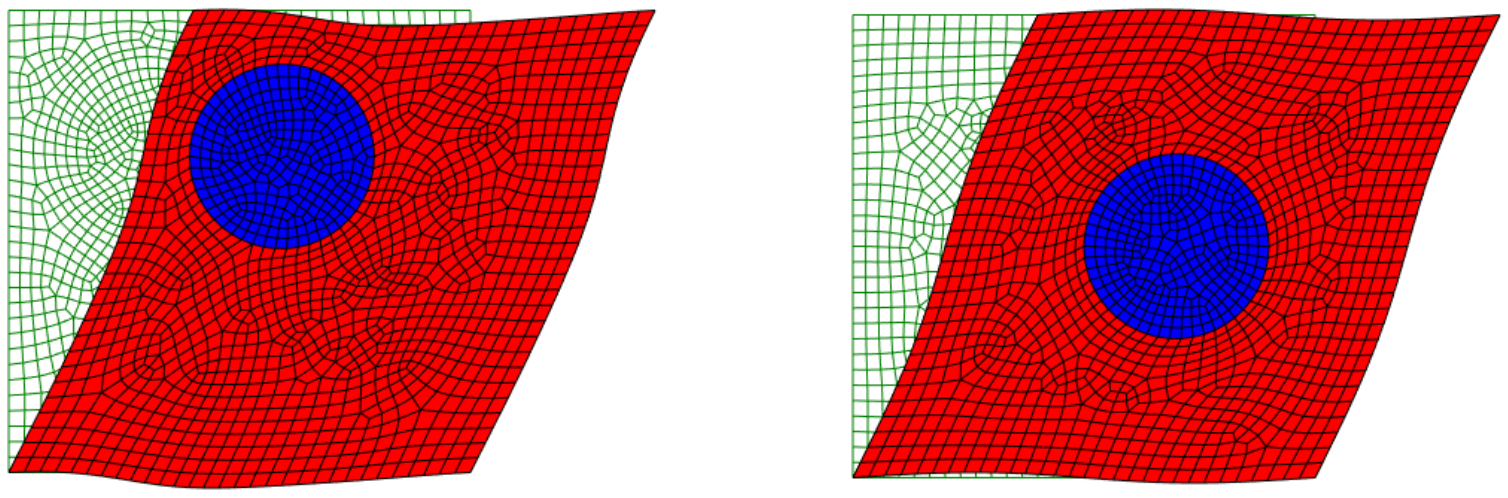

(a) Unit cell-1 (b) Unit cell-2

Figure 8. Deformed shapes of different unit cells under simple shear.

Table 3. Homogenized quantities from the RVEs of different unit cells under simple tension.

| RVE | Homogenized 1st PK stress $[\bar{\boldsymbol{P}}]$ | Homogenized tangent moduli $[\bar{\mathbb{A}}]$ | Homogenized stored energy $\bar{\psi}$ |
|---|---|---|---|
| 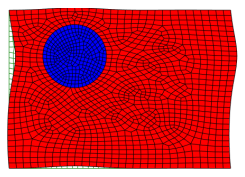 Unit cell-1 | $\begin{bmatrix} 11.6573 \\ 0 \\ 0 \\ 8.8865 \end{bmatrix}$ | $\begin{bmatrix} 26.0259 & -0.0002 & -0.0003 & 30.2150 \\ -0.0002 & 7.5260 & -1.1210 & 0 \\ -0.0003 & -1.1210 & 7.3172 & 0 \\ 30.2150 & 0 & 0 & 54.4927 \end{bmatrix}$ | 2.4347 |
| 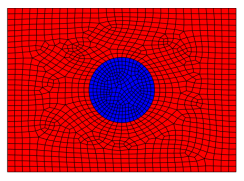 Unit cell-2 | $\begin{bmatrix} 11.6567 \\ 0 \\ 0 \\ 8.8866 \end{bmatrix}$ | $\begin{bmatrix} 26.0237 & 0 & -0.0001 & 30.2151 \\ 0 & 7.5253 & -1.1212 & 0.0001 \\ -0.0001 & -1.1212 & 7.3169 & 0.0002 \\ 30.2151 & 0.0001 & 0.0002 & 54.4909 \end{bmatrix}$ | 2.4346 |

Table 4. Homogenized quantities from the RVEs of different unit cells under simple shear.

| RVE | Homogenized 1st PK stress $[\bar{\boldsymbol{P}}]$ | Homogenized tangent moduli $[\bar{\mathbb{A}}]$ | Homogenized stored energy $\bar{\psi}$ |
|---|---|---|---|
| 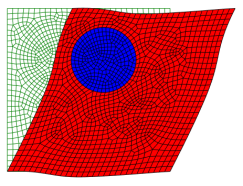 Unit cell-1 | $\begin{bmatrix} -0.5929 \\ 4.0980 \\ 3.8687 \\ -0.5731 \end{bmatrix}$ | $\begin{bmatrix} 35.1466 & -12.9755 & -2.9184 & 14.8978 \\ -12.9755 & 16.2037 & 11.5362 & -13.1511 \\ -2.9184 & 11.5362 & 9.7712 & -2.9799 \\ 14.8978 & -13.1511 & -2.9799 & 35.0999 \end{bmatrix}$ | 0.7710 |
| 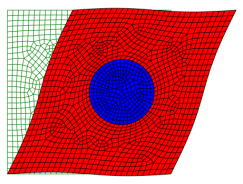 Unit cell-2 | $\begin{bmatrix} -0.5930 \\ 4.0976 \\ 3.8684 \\ -0.5731 \end{bmatrix}$ | $\begin{bmatrix} 35.1459 & -12.9752 & -2.9183 & 14.8981 \\ -12.9752 & 16.2023 & 11.5345 & -13.1520 \\ -2.9183 & 11.5345 & 9.7692 & -2.9804 \\ 14.8981 & -13.1520 & -2.9804 & 35.0999 \end{bmatrix}$ | 0.7709 |

## 3.2 Finite elastic-plastic deformation of a metamaterial

### 3.2.1 Invariance w.r.t number/choice of unit cells in RVE

The second example examines finite strain elastoplastic metamaterial that represents path-dependent material cases. A similar example has been considered in [51]. Both the matrix and inclusion materials are modeled using the finite strain $J_2$ elastoplasticity given in Table B 1, where the multiplicative decomposition of the deformation gradient is assumed, i.e. $\boldsymbol{F} = \boldsymbol{F}^e . \boldsymbol{F}^p$ and the internal variable can be taken as the elastic Finger tensor $\boldsymbol{b}^e$ or equivalently plastic right Cauchy-Green tensor $\boldsymbol{C}^p$ or its inverse $\boldsymbol{C}^{p-1}$. The matrix material is simulated as elastoplasticity with parameters: $\kappa_m$ = 17.5, $\mu_m$ = 8.0, $\sigma_y$ = 0.45 and $K^p$ = 0.1. The inclusion material is simulated as a hyperelastic phase (by choosing $\sigma_y \to \infty$, here $\sigma_y = 10^6$) with parameters: $\kappa_I = 100\kappa_m$ and $\mu_I = 100\mu_m$. Here, the metamaterial can be equivalently described by two unit cells, as seen in Figure 9, unit cell-1 and unit cell-2. The FE models of the two unit cells are shown in Figure 10. Both unit cells are of unit side length and unit thickness with the diameter of the hole and inclusions $d$ = 0.3. With the origin of a local coordinate system located at the center of the unit cells, for unit

cell-1 the inclusions are located at (-0.2, 0.2) and (-0.2, -0.2) and the hole is located at (0.2, 0), while for unit cell-2 the inclusions are at (0.3, 0.2) and (0.3, -0.2) and the hole is at (-0.3, 0).

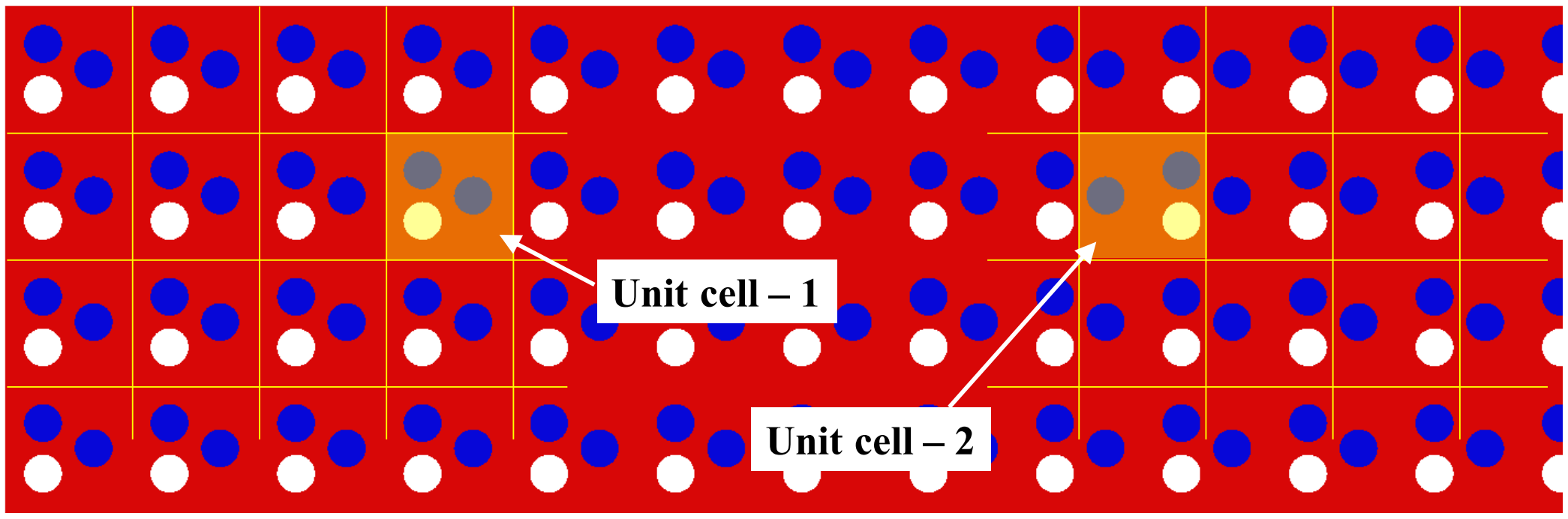

Figure 9. Illustration of two different ways of choosing a unit cell from the same periodic metamaterial.

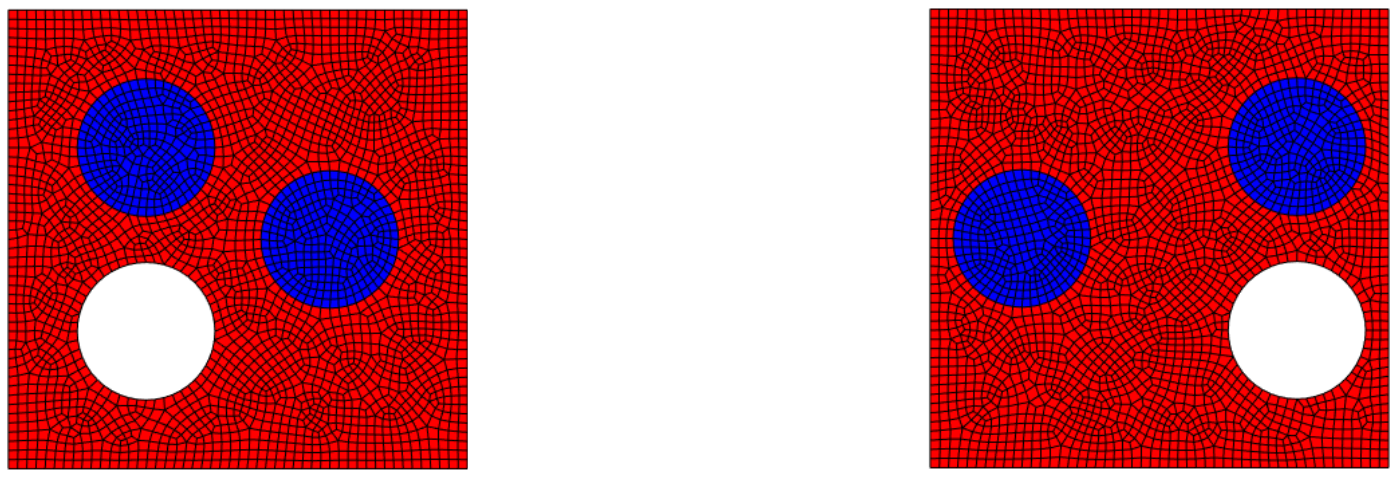

(a) Unit cell-1 ($n_{ele}$ = 2823) (b) Unit cell-2 ($n_{ele}$ = 2863)

Figure 10. FE meshes of two different unit cells of the elastoplastic metamaterial.

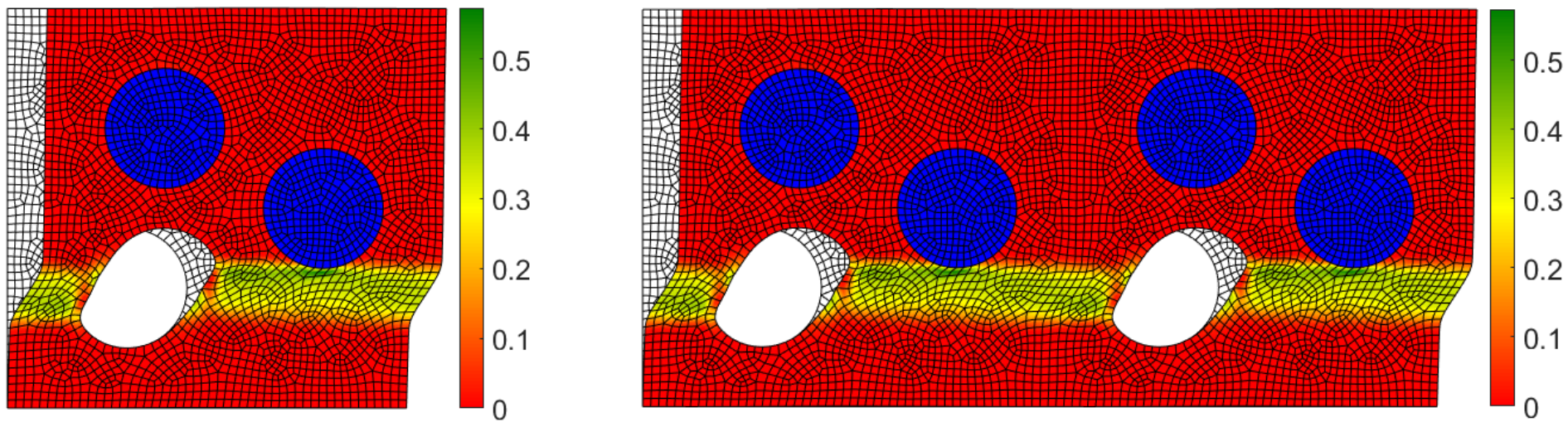

(a) 1×1 RVE (b) 2×1 RVE

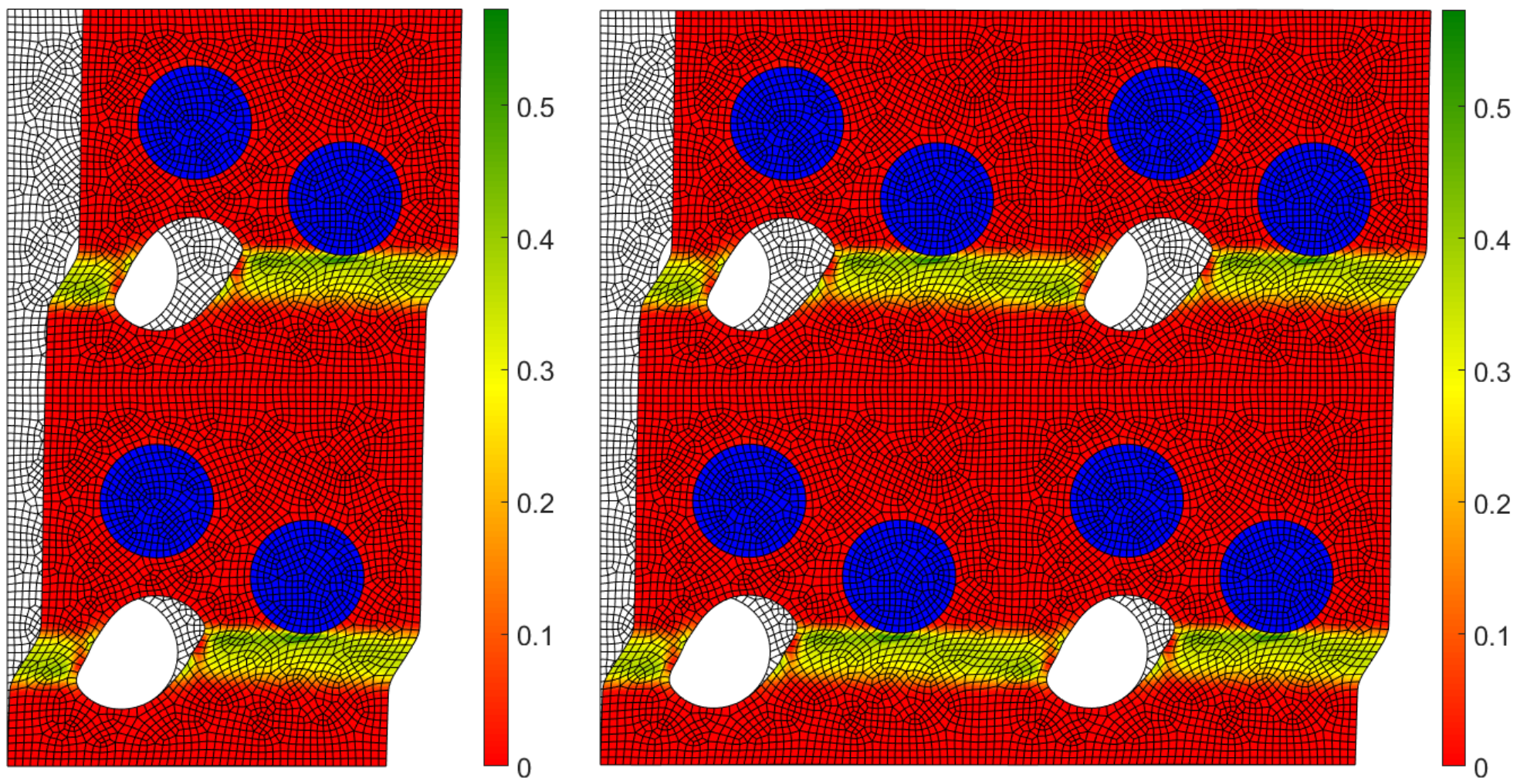

(c) 1×2 RVE (d) 2×2 RVE

Figure 11. Deformed shapes and equivalent plastic strain ($\alpha$) distributions of the RVE consisting of different number of unit cells (unit cell-1) under simple shear.

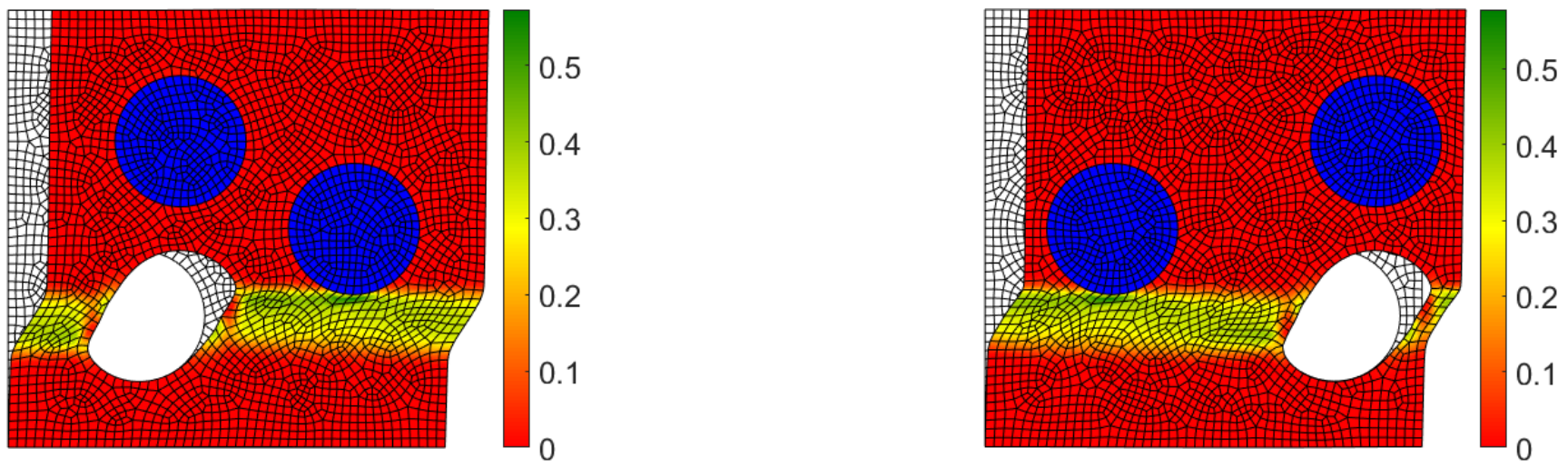

(a) Unit cell-1 (b) Unit cell-2

Figure 12. Deformed shapes and equivalent plastic strain ($\alpha$) distributions of the RVE with different unit cells under simple shear.

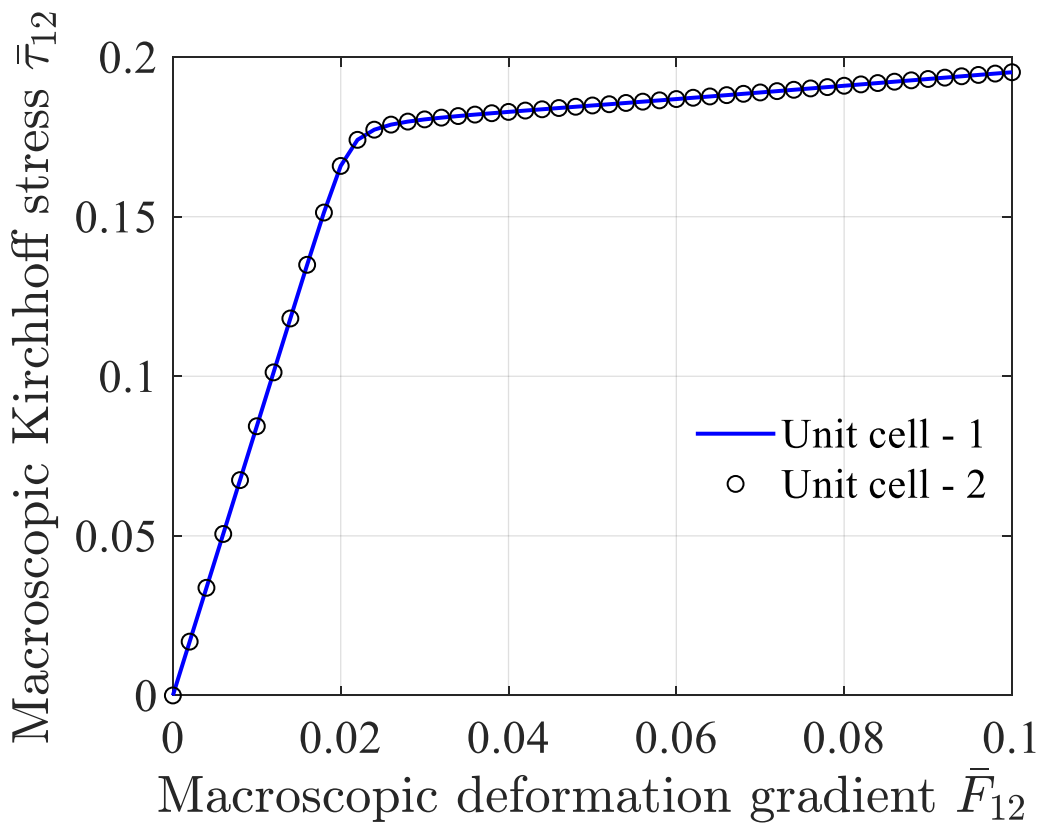

Figure 13. Macroscopic deformation gradient – stress curves of the RVE analysis with different unit cells under simple shear.

Table 5. Homogenized quantities from the elastoplastic RVE consisting of a different number of unit cells (Unit cell-1 in Figure 10a) under simple shear.

| RVE | Homogenized 1st PK stress $[\bar{\boldsymbol{P}}]$ | Homogenized tangent moduli $[\bar{\mathbb{A}}]$ | Homogenized stored energy $\bar{\psi}^e$ |
|---|---|---|---|
| 1×1<br>1×2<br>2×1<br>2×2 | $\begin{bmatrix} 0.0128 \\ 0.1893 \\ 0.1953 \\ 0.0598 \end{bmatrix}$ | $\begin{bmatrix} 26.1954 & -0.6689 & 0.3549 & 8.3450 \\ -0.6689 & 0.1601 & 0.0503 & -0.9698 \\ 0.3549 & 0.0503 & 0.2038 & 0.9365 \\ 8.3450 & -0.9698 & 0.9365 & 21.0161 \end{bmatrix}$ | $2.423\times10^{-3}$ |

Table 6. Homogenized quantities from the elastoplastic RVE of different unit cells under simple shear.

| RVE | Homogenized 1st PK stress $[\bar{\boldsymbol{P}}]$ | Homogenized tangent moduli $[\bar{\mathbb{A}}]$ | Homogenized stored energy $\bar{\psi}^e$ |
|---|---|---|---|
| 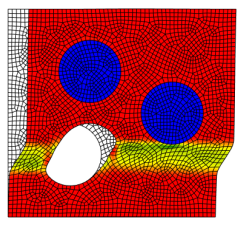 Unit cell-1 | $\begin{bmatrix} 0.0128 \\ 0.1893 \\ 0.1953 \\ 0.0598 \end{bmatrix}$ | $\begin{bmatrix} 26.1954 & -0.6689 & 0.3549 & 8.3450 \\ -0.6689 & 0.1601 & 0.0503 & -0.9698 \\ 0.3549 & 0.0503 & 0.2038 & 0.9365 \\ 8.3450 & -0.9698 & 0.9365 & 21.0161 \end{bmatrix}$ | $2.423\times10^{-3}$ |
| 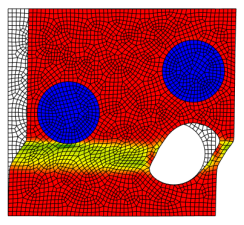 Unit cell-2 | $\begin{bmatrix} 0.0129 \\ 0.1893 \\ 0.1953 \\ 0.0600 \end{bmatrix}$ | $\begin{bmatrix} 26.1923 & -0.6768 & 0.3458 & 8.3341 \\ -0.6768 & 0.1608 & 0.0491 & -0.9884 \\ 0.3458 & 0.0491 & 0.2007 & 0.9157 \\ 8.3341 & -0.9884 & 0.9157 & 20.9943 \end{bmatrix}$ | $2.448\times10^{-3}$ |

To address the volumetric locking due to the incompressible plastic flow, the mixed u/p (9/3) element formulation [52] is adopted, where the independent pressure field $p$ is taken as $\tau_m$ defined by $\tau_m \overset{\text{def}}{=} (1/3)\boldsymbol{I}:\boldsymbol{\tau}$ . The macroscopic deformation gradient loading considers $[\bar{\boldsymbol{F}}] = [\bar{F}_{11} \quad \bar{F}_{21} \quad \bar{F}_{12} \quad \bar{F}_{22}]^T = [1 \quad 0 \quad 0.1 \quad 1]^T$. The sufficiency of RVE consisting of only one unit cell is again examined for representing materials with a path-dependent response. The analyses are carried out with unit cell-1 (Figure 10a). The sufficiency is confirmed by the results shown in Table 5 and Figure 11. Next, the equivalence of different choices of a unit cell is checked with the two different unit cells shown in Figure 10. The equivalent plastic strain ($\alpha$) distribution on the deformed shapes are shown in Figure 12(a) and (b), where similar shear band formations can be observed. Their homogenized deformation gradient versus stress curves (note that $\bar{\boldsymbol{\tau}} = \bar{\boldsymbol{P}}.\bar{\boldsymbol{F}}^T$) along the loading are compared in Figure 13, where a close match can be seen. A further comparison is made in Table 6 where the homogenized quantities including 1st PK stress, tangent moduli and stored elastic strain energy are compared for the two unit cells at the final loading step. Similar to the hyperelastic case in Section 3.1, the small differences herein can be attributed to the different FE mesh and inaccurate geometry modeling.

### *3.2.2 Invariance w.r.t shape of RVE*

To further illustrate the independence of the homogenization analysis results on the different choice of RVEs, consider the periodic metamaterial as shown in Figure 14 where the matrix material has the same properties as the elastoplastic matrix material in the previous example (see Figure 9). The elliptical holes are arranged in a periodic lattice as shown in Figure 14. The major and minor axes are of $\sqrt{3}/2$ and 0.5 units, respectively, and the major axis is at an angle 30° w.r.t. the horizontal direction. Three different RVEs are chosen, as shown in Figure 14. It should be noted that RVE-1 and RVE-2 both include one unit cell, i.e. they are the smallest periodic cells, while RVE-3 is not. The FE meshes of the three RVEs are shown in Figure 15. The deformation again considers simple shear, i.e. $[\bar{\boldsymbol{F}}] = [\bar{F}_{11} \quad \bar{F}_{21} \quad \bar{F}_{12} \quad \bar{F}_{22}]^T = [1 \quad 0 \quad 0.1 \quad 1]^T$. Figure 16 shows the deformed shape as well as the equivalent plastic strain ($\alpha$) distribution for different RVEs, where same shear band can be observed. Table 7 lists the homogenized quantities such as homogenized stress, tangent moduli, and stored elastic energy. Again, it can be seen that all the homogenized quantities match fairly well with each other.

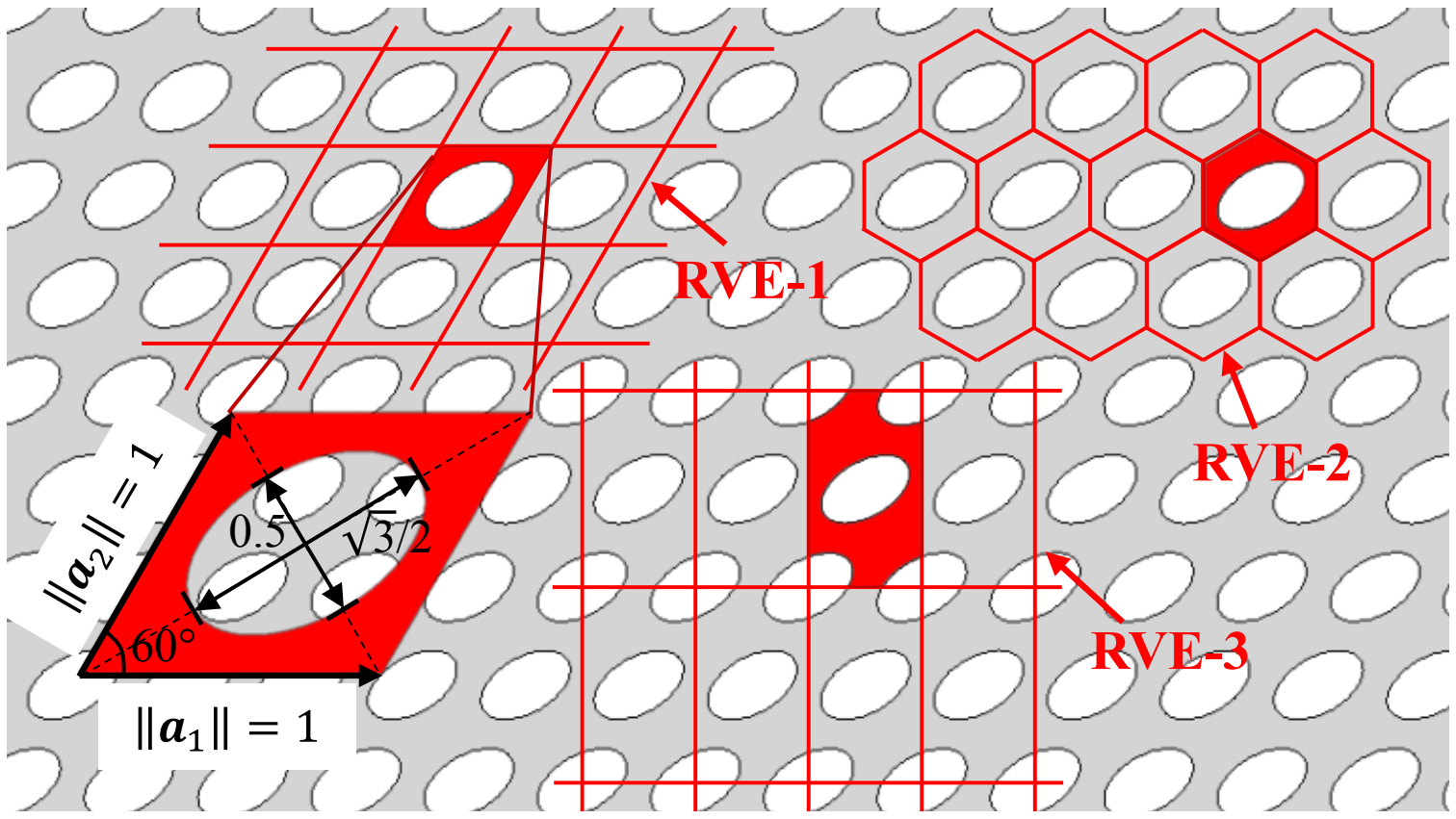

Figure 14. Illustration of three different ways of choosing RVE from the porous periodic elastoplastic metamaterial.

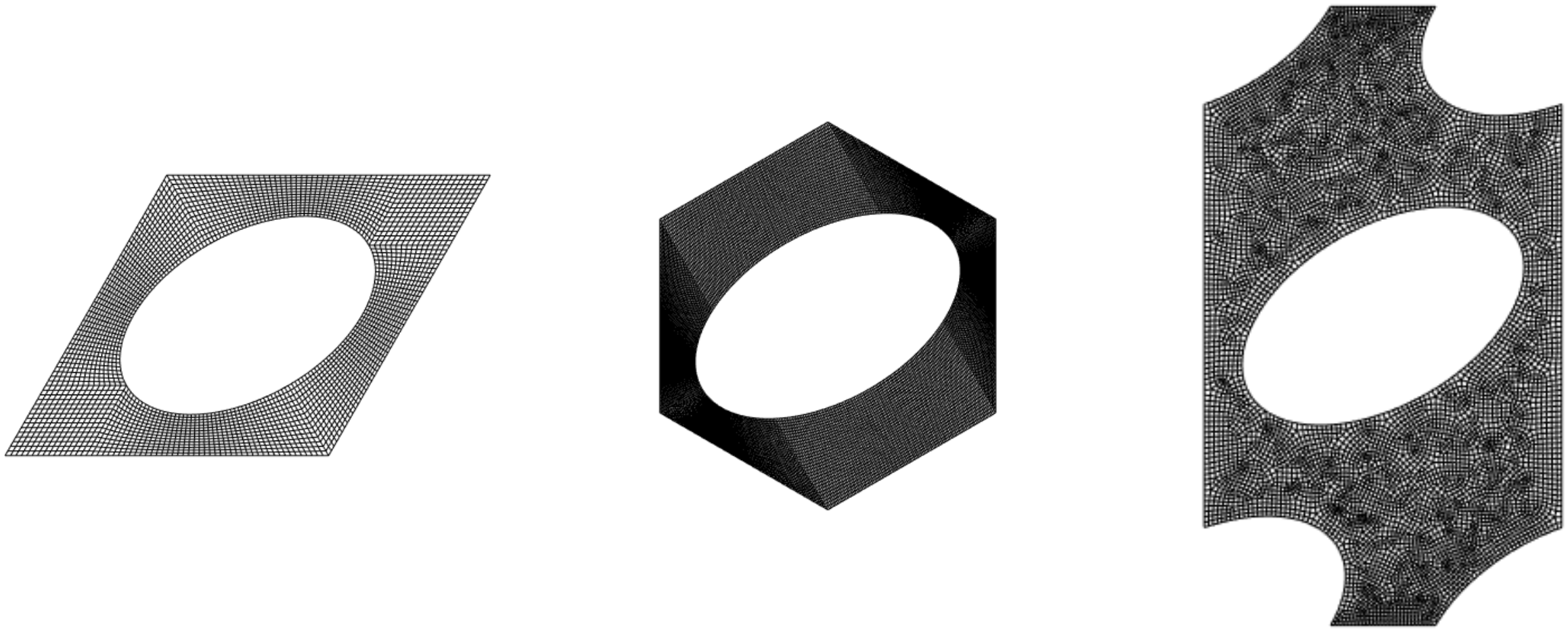

(a) RVE-1 ($n_{ele}$ = 3150) (b) RVE-2 ($n_{ele}$ = 13200) (c) RVE-3 ($n_{ele}$ = 7584)

Figure 15. FE meshes of three RVEs of the porous elastoplastic metamaterial.

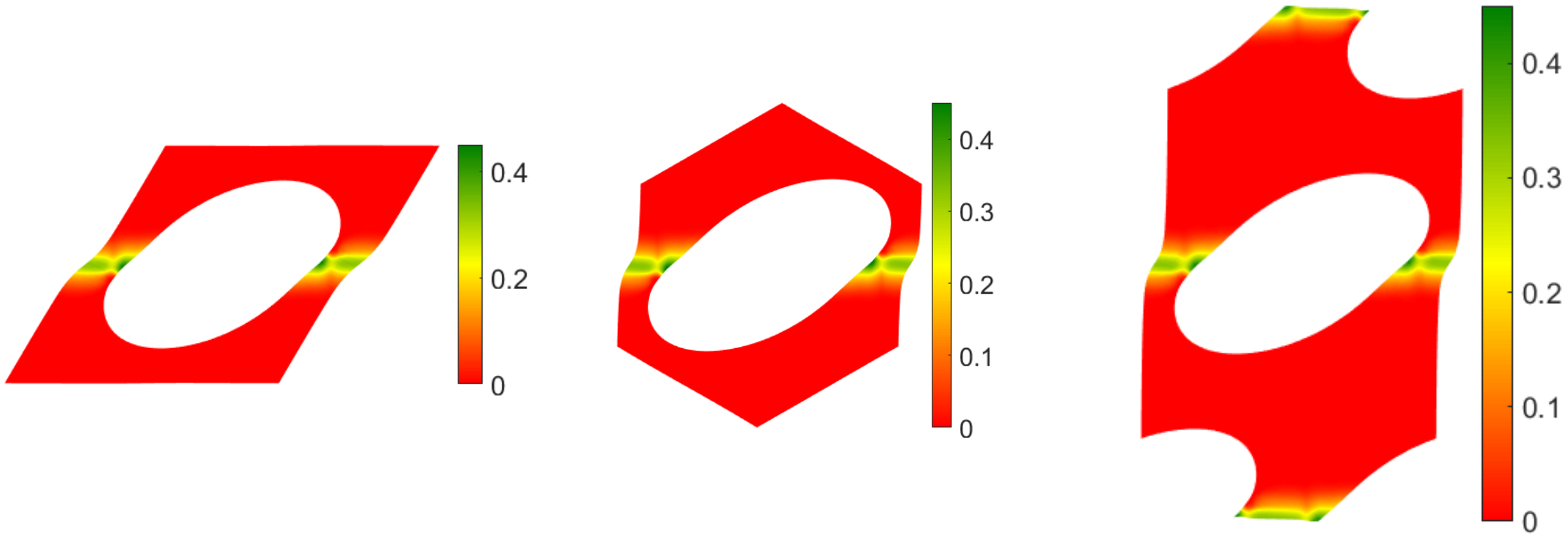

(a) RVE-1 (b) RVE-2 (c) RVE-3

Figure 16. Deformed shapes and equivalent plastic strain ($\alpha$) distributions of different RVEs under simple shear.

Table 7. Homogenized quantities from the different porous elastoplastic RVEs under simple shear.

| RVE | Homogenized 1st PK stress $[\bar{\boldsymbol{P}}]$ | Homogenized tangent moduli $[\bar{\mathbb{A}}]$ | Homogenized stored energy $\bar{\psi}^e$ |
|---|---|---|---|
| 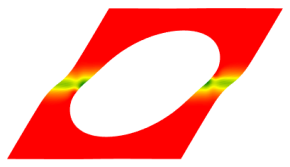 RVE-1 | $\begin{bmatrix}0.0393\\0.0748\\0.0817\\0.0692\end{bmatrix}$ | $\begin{bmatrix}9.5021 & -0.1584 & 0.0571 & 1.4065\\-0.1584 & 0.0372 & -0.0200 & -0.1788\\0.0571 & -0.0200 & 0.0607 & 0.1146\\1.4065 & -0.1788 & 0.1146 & 3.7513\end{bmatrix}$ | $1.389\times10^{-3}$ |
| 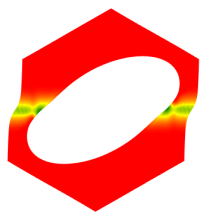 RVE-2 | $\begin{bmatrix}0.0393\\0.0748\\0.0817\\0.0692\end{bmatrix}$ | $\begin{bmatrix}9.5021 & -0.1582 & 0.0572 & 1.4064\\-0.1582 & 0.0370 & -0.0201 & -0.1783\\0.0572 & -0.0201 & 0.0606 & 0.1150\\1.4064 & -0.1783 & 0.1150 & 3.7504\end{bmatrix}$ | $1.388\times10^{-3}$ |
| 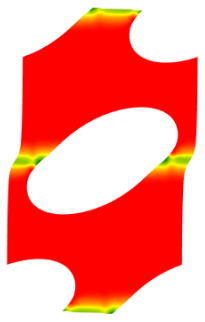 RVE-3 | $\begin{bmatrix}0.0393\\0.0748\\0.0817\\0.0692\end{bmatrix}$ | $\begin{bmatrix}9.5020 & -0.1586 & 0.0569 & 1.4064\\-0.1586 & 0.0373 & -0.0200 & -0.1796\\0.0569 & -0.0200 & 0.0607 & 0.1138\\1.4064 & -0.1796 & 0.1138 & 3.7513\end{bmatrix}$ | $1.389\times10^{-3}$ |

## 4 Multiscale Stability

A basic assumption in the homogenization analysis is that *one* unit cell can serve as the fundamental periodic cell during the entire loading process and can be taken as RVE. For finitely strained nonlinear metamaterials, this assumption, however, may be violated. Upon loading, for instance, buckling with wavelength possibly across arbitrary length can happen at the microscale [39], which will lead to the violation of the assumption that *one* unit cell can serve as the fundamental cell. If the buckling mode is periodic, a fundamental cell can still be found that may consist of more than one unit cell, while when the buckling mode is aperiodic, there is no fundamental periodic cell that can be further used [37]. From a macroscopic viewpoint, though polyconvexity in the sense of Ball [53] of the underlying material phases can be guaranteed by using appropriate constitutive models, which ensures the rank-1 convexity, the homogenized macroscopic metamaterial may still lose rank-1 convexity [33]. As shown in the previous studies, there exists a close connection between microscale buckling and macroscale loss of rank-1 convexity, i.e., long wavelength buckling on the microscale corresponds to the loss of rank-1 convexity of the homogenized incremental moduli at macroscale [36, 39]. It is also noted that the micro-instability (short wavelength buckling) occurs either before or simultaneously with the macro-instability (long wavelength buckling) [36, 39]. Compared to the macro-stability check, where a rank-1 convexity examination of the homogenized incremental moduli is only needed, the micro-stability check is much more computationally demanding, since the length scale of buckling mode is not a priori known.

### *4.1 Microscale stability*

For rate-independent solids, the stability is governed by Hill's stability criterion [54]. The principal solution branch ceases to be stable when the functional $\beta(\lambda)$ defined by

$$\beta(\lambda) = \min_{\boldsymbol{v}} Q(\boldsymbol{v}; \Omega) \quad \text{with} \quad Q(\boldsymbol{v}; \Omega) \stackrel{\text{def}}{=} \frac{\int_{\Omega} \nabla_X \widetilde{\boldsymbol{v}} : \mathbb{A} : \nabla_X \boldsymbol{v} \, dV}{\int_{\Omega} \nabla_X \widetilde{\boldsymbol{v}} : \nabla_X \boldsymbol{v} \, dV} \tag{44}$$

loses positive definiteness, where $\boldsymbol{v}$ is taken from the kinematically admissible displacement variation space $H_0^1(\Omega)$ for the corresponding macroscale boundary value problem. For periodic solids of infinite extent ($\Omega \to \mathbb{R}^d$), $\boldsymbol{v}$ is taken from locally integrable, bounded functions that ensure the finiteness of the ratio $Q$ [39], and correspondingly the minimum in Eq. (44) is taken as infimum. In Eq. (44) , $\widetilde{\boldsymbol{v}}$ denotes the complex conjugate of $\boldsymbol{v}$ and $\lambda$ stands for the loading parameter. The tensor $\mathbb{A}$ represents the tangent moduli (Eq. (25)) under the loading parameter $\lambda$ with the same periodicity as one unit cell. It was shown in [36] that this infimum $\beta(\lambda)$ can be computed through Bloch wave analysis, where the calculation is carried out within the RVE (e.g. one unit cell) $\Omega_0^{\mu}$ and is expressed as

$$\beta(\lambda) = \inf_{\boldsymbol{k}} \min_{\boldsymbol{u}} Q\left(\boldsymbol{v}_B(\boldsymbol{k}, \boldsymbol{u}); \Omega_0^{\mu}\right) \quad \text{with} \quad \boldsymbol{v}_B(\boldsymbol{k}, \boldsymbol{u}) = e^{i\boldsymbol{k}.\boldsymbol{X}} \boldsymbol{u} \tag{45}$$

where $\boldsymbol{v}_B$ is the Bloch wave representing the eigenmode in which $i = \sqrt{-1}$, $\boldsymbol{u}$ is a periodic function with the same periodicity as the RVE, i.e. $\boldsymbol{u}(\boldsymbol{X} + c_i \boldsymbol{a}_i) = \boldsymbol{u}(\boldsymbol{X})$ with $c_i$ arbitrary integers and $\boldsymbol{a}_i$ the $i^{th}$ periodic lattice vector ($i$ = 1, …, $d$), while the wavevector $\boldsymbol{k}$ is chosen in the reciprocal space spanned by the reciprocal bases $\boldsymbol{b}_i$ ($i$ = 1, …, $d$) defined by $\boldsymbol{a}_i . \boldsymbol{b}_j = 2\pi\delta_{ij}$ [55], i.e. $\boldsymbol{k} = \sum_{i=1}^{d} k_i \boldsymbol{b}_i$. For example, Figure 17(a) and (b) illustrate two different lattice spaces and their corresponding reciprocal lattice spaces.

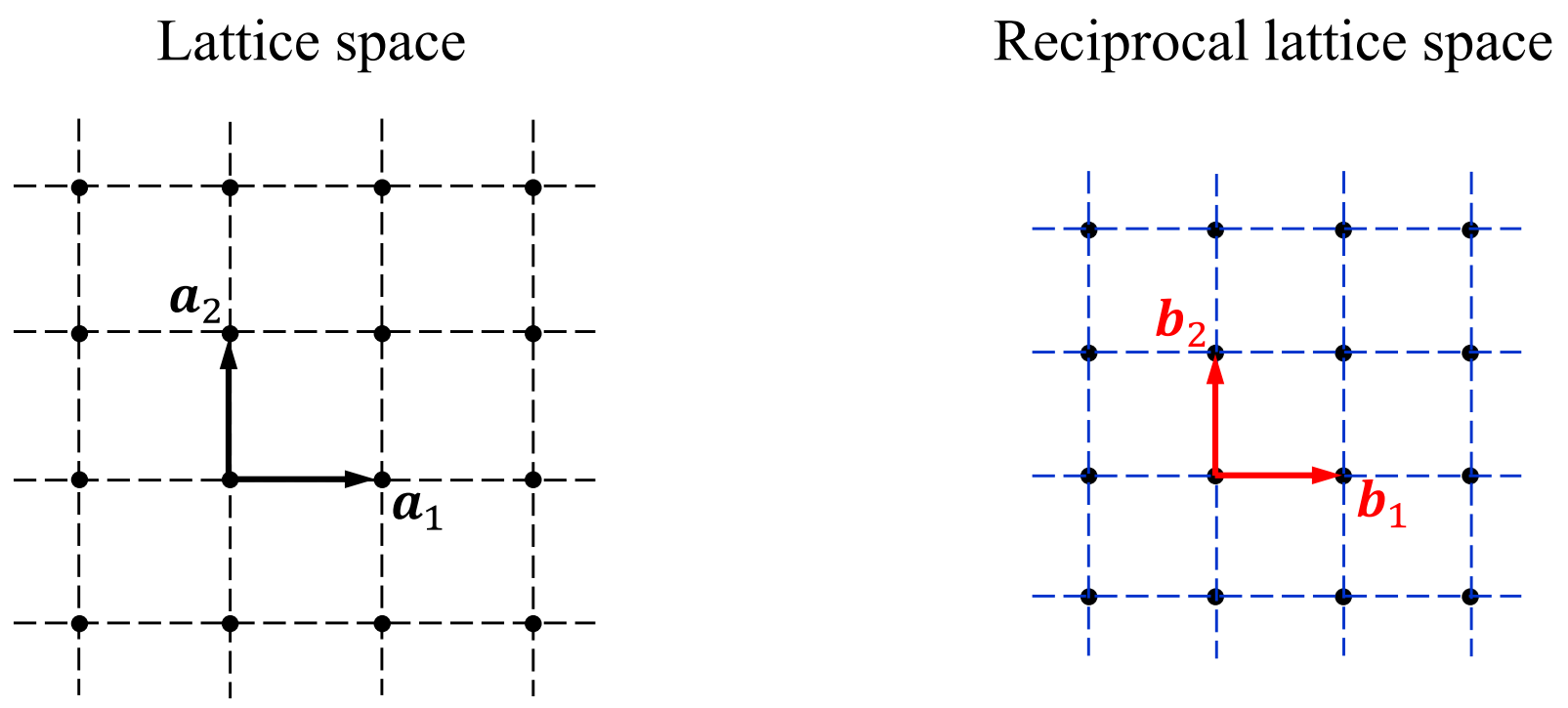

(a) Square-shape unit cell

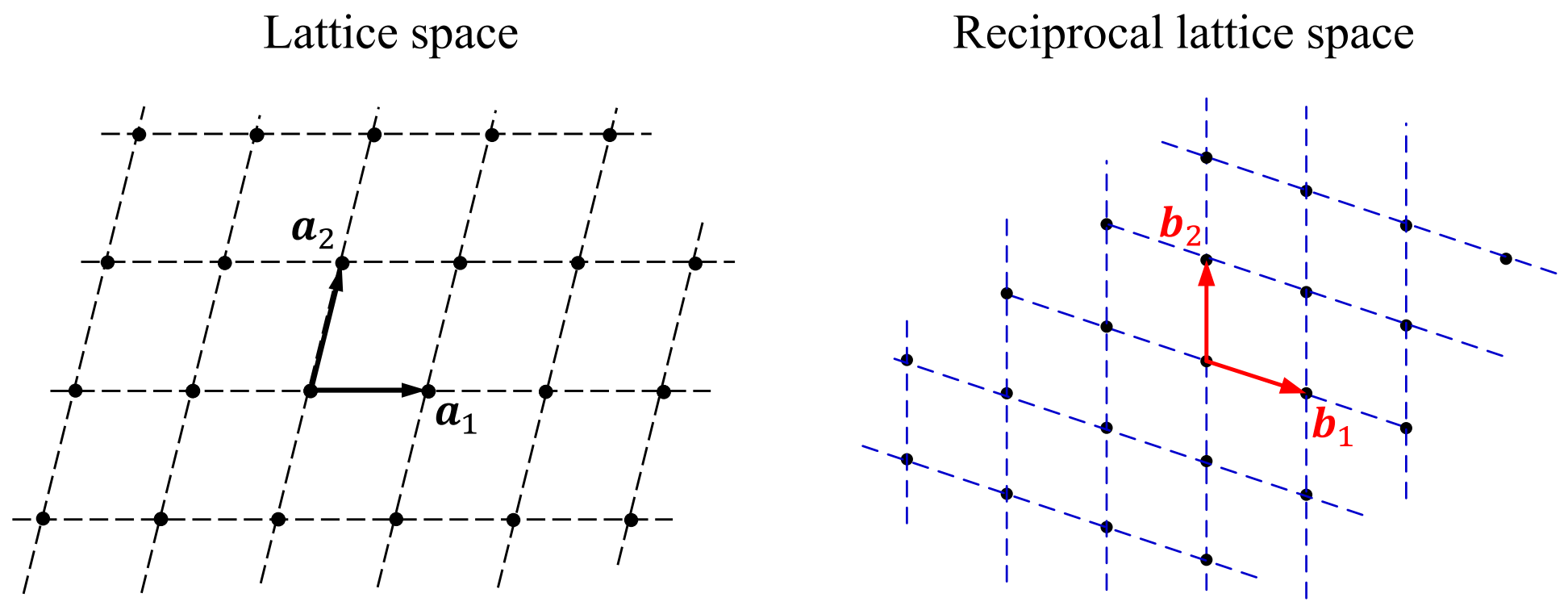

(b) Parallelogram-shape unit cell

Figure 17. Illustration of lattice and reciprocal spaces.

It is worth noting that two physically different types of buckling modes exist in the neighborhood of $\boldsymbol{k} = \mathbf{0}$, i.e., the long wavelength instability with $\boldsymbol{k} \to \mathbf{0}$ that leads to the loss of rank-1 convexity of the homogenized tangent moduli at the macroscale (see Section 4.2), and a short wavelength buckling mode with $\boldsymbol{k} = \mathbf{0}$ which has the same periodicity as the RVE. For short wavelength buckling, the infimum in Eq. (45) is achieved at $\boldsymbol{k} = \mathbf{0}$ and the microscale stability surface, plotted as $\beta(\lambda)$ against $k_1$ and $k_2$ (for 2D case), is continuous at origin ($\boldsymbol{k} = \mathbf{0}$) at critical loading parameter $\lambda_c$. However, for long wavelength buckling, the infimum in Eq. (45) is not reachable and can only be computed as the limit $\boldsymbol{k} \to \mathbf{0}$, and the microscale stability surface is discontinuous at origin (singular point). This is the reason for using infimum instead of minimum in Eq. (45). Interested readers are referred to Refs. [36, 39] for further theoretical details.

*4.1.1 Wavevector ($\boldsymbol{k}$) space*

For Bloch wave analysis, the sufficient wavevector space to be examined can be chosen as the primitive cell (smallest periodic cell or unit cell) in the reciprocal space, and the Bloch wave functions with wavevectors outside the primitive cell are all represented due to the periodicity. To see this, consider a 2D unit cell with periodic vectors $\boldsymbol{a}_1$ and $\boldsymbol{a}_2$. In the reciprocal lattice space, which is periodic with respect to periodic bases $\boldsymbol{b}_1$ and $\boldsymbol{b}_2$ (note that $\boldsymbol{a}_i.\boldsymbol{b}_j = 2\pi\delta_{ij}$), by definition, for any wavevector $\boldsymbol{k}$, there is a translational vector $\boldsymbol{T} = i_1\boldsymbol{b}_1 + i_2\boldsymbol{b}_2$ ($i_1, i_2 \in \mathbb{Z}$ = set of integer) such that $\boldsymbol{k} = \boldsymbol{k}_{PC} + \boldsymbol{T}$ with $\boldsymbol{k}_{PC}$ a wavevector in the primitive cell, see Figure 18a. As a result, the Bloch wave function with wavevector $\boldsymbol{k}$ is expressed as

$$\boldsymbol{v}_B = e^{i\boldsymbol{k}.\boldsymbol{X}}\boldsymbol{u}_{\boldsymbol{k}} \tag{46}$$

where a subscript $\boldsymbol{k}$ is used for the periodic function $\boldsymbol{u}_{\boldsymbol{k}}(\boldsymbol{X}) = \boldsymbol{u}(\boldsymbol{X})$ to make it clear that this periodic function is based on the wavevector $\boldsymbol{k}$. Note that the function $\boldsymbol{u}_{\boldsymbol{k}}(\boldsymbol{X})$ is periodic in the lattice space, i.e. $\boldsymbol{u}_{\boldsymbol{k}}(\boldsymbol{X} + q_1\boldsymbol{a}_1 + q_2\boldsymbol{a}_2) = \boldsymbol{u}_{\boldsymbol{k}}(\boldsymbol{X})$ for any $q_1, q_2 \in \mathbb{Z}$. Thus, it can be shown that

$$\boldsymbol{v}_B = e^{i\boldsymbol{k}.\boldsymbol{X}}\boldsymbol{u}_{\boldsymbol{k}} = e^{i\boldsymbol{k}_{PC}.\boldsymbol{X}}e^{i\boldsymbol{T}.\boldsymbol{X}}\boldsymbol{u}_{\boldsymbol{k}} = e^{i\boldsymbol{k}_{PC}.\boldsymbol{X}}\boldsymbol{u}_{\boldsymbol{k}_{PC}} \tag{47}$$

where

$$\boldsymbol{u}_{\boldsymbol{k}_{PC}}(\boldsymbol{X}) \stackrel{\text{def}}{=} e^{i\boldsymbol{T}.\boldsymbol{X}}\boldsymbol{u}_{\boldsymbol{k}}(\boldsymbol{X}) \tag{48}$$

is another periodic function in the lattice space, since $\forall\, q_1', q_2' \in \mathbb{Z}$

$$\begin{aligned}\boldsymbol{u}_{\boldsymbol{k}_{PC}}(\boldsymbol{X} + q_1'\boldsymbol{a}_1 + q_2'\boldsymbol{a}_2) &= e^{i\boldsymbol{T}.(\boldsymbol{X}+q_1'\boldsymbol{a}_1+q_2'\boldsymbol{a}_2)}\boldsymbol{u}_{\boldsymbol{k}}(\boldsymbol{X} + q_1'\boldsymbol{a}_1 + q_2'\boldsymbol{a}_2) = e^{i\boldsymbol{T}.\boldsymbol{X}}\boldsymbol{u}_{\boldsymbol{k}}(\boldsymbol{X}) \\ &= \boldsymbol{u}_{\boldsymbol{k}_{PC}}(\boldsymbol{X})\end{aligned} \tag{49}$$

Thus, Eq. (47) and Eq. (49) show that for any wave vector $\boldsymbol{k}$ in the reciprocal space, the Bloch wave function $\boldsymbol{v}_B$ can be equivalently expressed in terms of wavevector $\boldsymbol{k}_{PC}$ in the primitive cell and a corresponding periodic function that is still periodic w.r.t unit cell in the lattice space. From this discussion, it is clear that for a Bloch wave function representation, the wavevector $\boldsymbol{k}$ space can simply and sufficiently be chosen as *any* primitive cell in the reciprocal space. Thus, for any parallelogram-shaped RVE, the wavevector ($\boldsymbol{k}$) space can always be chosen as $k_1, k_2 \in [0,1)$. It is noted that as a special case, the 1st Brillouin zone, which is mostly used in solid-state physics [55], is also a primitive cell (see Figure 18b) in the reciprocal space, and thus can also be used as the wavevector space for evaluating $\beta(\lambda)$ in Eq. (45). In this study, the primitive cell with $k_1, k_2 \in [0,1)$ is used, as shown in Figure 18a.

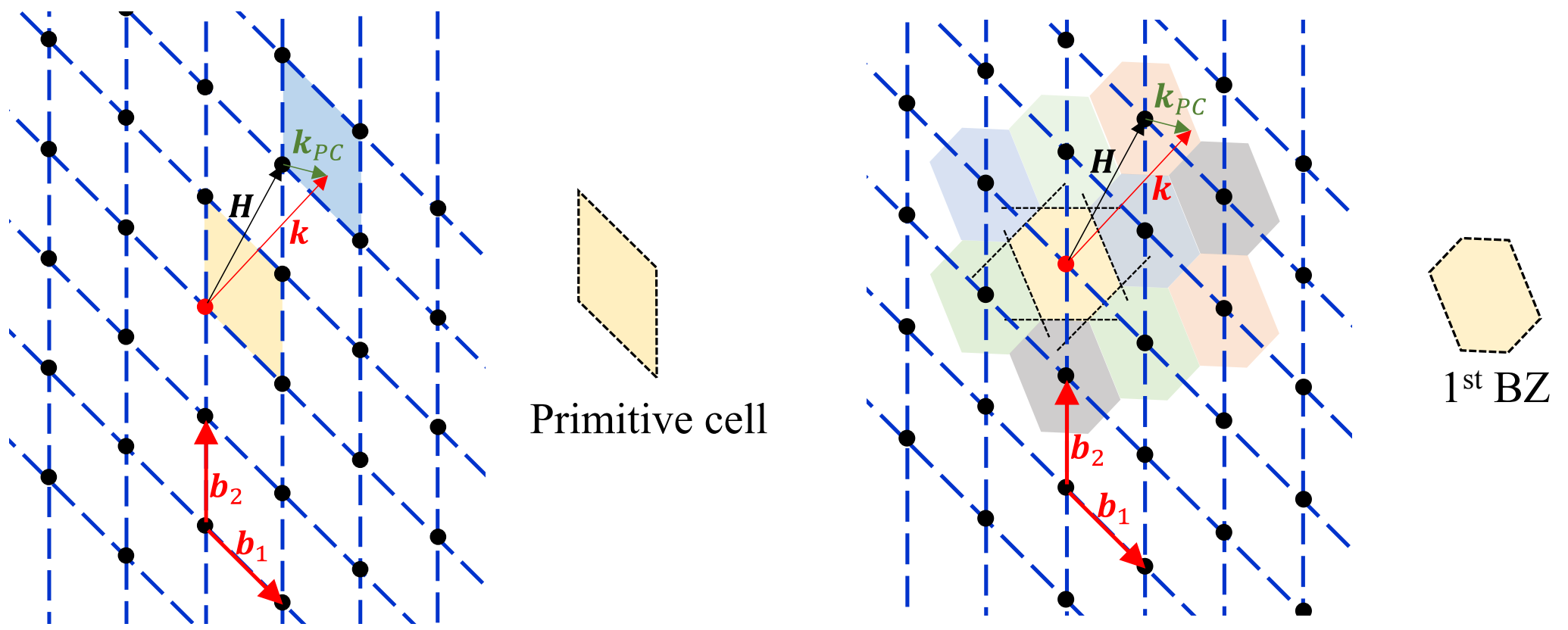

(a) Parallelogram shaped primitive cell (PC) (b) 1st Brillouin zone (BZ) as primitive cell

Figure 18. Wavevector ($\boldsymbol{k}$) space for Bloch wave analysis.

### *4.1.2 Interpretation of the Bloch wave function as buckling mode*

The equivalence between Eq. (44) and Eq. (45) has been established in [36]. However, since the buckling mode in Eq. (44) is in real physical space, it is not clear how the complex-valued Bloch wave function can be treated as the buckling mode or how a real-valued mode can be extracted from the Bloch wave function $\boldsymbol{v}_B$. As shown in [36] that if X is a subsequence in $H_0^1(\Omega)$ and $\mathrm{X} \oplus i\mathrm{X}$ is its complexification, then

$$\min_{\boldsymbol{v} \in \mathrm{X} \oplus i\mathrm{X}} Q(\boldsymbol{v}; \Omega) = \min_{\boldsymbol{v} \in \mathrm{X}} Q(\boldsymbol{v}; \Omega) \tag{50}$$

For further exposition, the proof of this assertion is detailed below.

***Proof***

First, it is clear that $\min_{\boldsymbol{v} \in \mathrm{X} \oplus i\mathrm{X}} Q(\boldsymbol{v}; \Omega) \leq \min_{\boldsymbol{v} \in \mathrm{X}} Q(\boldsymbol{v}; \Omega)$ since X is a subset of $\mathrm{X} \oplus i\mathrm{X}$. Thus, it is enough to show that $\min_{\boldsymbol{v} \in \mathrm{X}} Q(\boldsymbol{v}; \Omega) \leq \min_{\boldsymbol{v} \in \mathrm{X} \oplus i\mathrm{X}} Q(\boldsymbol{v}; \Omega)$. To this end, for any $\boldsymbol{v} \in \mathrm{X} \oplus i\mathrm{X}$, write

$$\boldsymbol{v} = \mathrm{Re}(\boldsymbol{v}) + i\,\mathrm{Im}(\boldsymbol{v}) = \boldsymbol{w}_1 + i\boldsymbol{w}_2 \tag{51}$$

Substituting Eq. (51) into $Q(\boldsymbol{v}; \Omega)$ defined in Eq. (44), using the major symmetry of $\mathbb{A}$, gives

$$Q(\boldsymbol{v}; \Omega) = \frac{\int_\Omega \boldsymbol{\nabla w}_1 : \mathbb{A} : \boldsymbol{\nabla w}_1 \, dV + \int_\Omega \boldsymbol{\nabla w}_2 : \mathbb{A} : \boldsymbol{\nabla w}_2 \, dV}{\int_\Omega \boldsymbol{\nabla w}_1 : \boldsymbol{\nabla w}_1 \, dV + \int_\Omega \boldsymbol{\nabla w}_2 : \boldsymbol{\nabla w}_2 \, dV} \tag{52}$$

Since $\boldsymbol{w}_1, \boldsymbol{w}_2 \in \mathrm{X}$, assuming that $\beta = \min_{\boldsymbol{v} \in \mathrm{X}} Q(\boldsymbol{v}; \Omega)$ and $\beta_1 \stackrel{\text{def}}{=} Q(\boldsymbol{w}_1; \Omega)$ and $\beta_2 \stackrel{\text{def}}{=} Q(\boldsymbol{w}_2; \Omega)$ gives $\beta_1 \geq \beta$ and $\beta_2 \geq \beta$. Without loss of generality, assume that $\beta_1 \leq \beta_2$. Thus,

$$\beta \leq \beta_1 \leq \frac{\int_\Omega \boldsymbol{\nabla w}_1 : \mathbb{A} : \boldsymbol{\nabla w}_1 \, dV + \int_\Omega \boldsymbol{\nabla w}_2 : \mathbb{A} : \boldsymbol{\nabla w}_2 \, dV}{\int_\Omega \boldsymbol{\nabla w}_1 : \boldsymbol{\nabla w}_1 \, dV + \int_\Omega \boldsymbol{\nabla w}_2 : \boldsymbol{\nabla w}_2 \, dV} \leq \beta_2 \tag{53}$$

which results in $\min_{\boldsymbol{v} \in \mathrm{X}} Q(\boldsymbol{v}; \Omega) \leq \min_{\boldsymbol{v} \in \mathrm{X} \oplus i\mathrm{X}} Q(\boldsymbol{v}; \Omega)$ with the equality holding only when $\beta_1 = \beta_2 = \beta$. That means, if the minimum $\min_{\boldsymbol{v} \in \mathrm{X}} Q(\boldsymbol{v}; \Omega)$ is reached by a set of vectors $\mathrm{Z} \subseteq \mathrm{X}$, the minimum $\min_{\boldsymbol{v} \in \mathrm{X} \oplus i\mathrm{X}} Q(\boldsymbol{v}; \Omega)$ is reached by the complexification of the same particular set, i.e. $\boldsymbol{v} \in \mathrm{Z} \oplus i\mathrm{Z}$. ■

The above proof demonstrates that if the minimum of $Q(\boldsymbol{v}; \Omega)$ is reached in the complex domain, both the real part and the imaginary part belong to the same set which leads to the minimum value of $Q(\boldsymbol{v}; \Omega)$ in the real domain. In terms of the Bloch wave analysis, this means that if zero value of the function $\inf_{\boldsymbol{k}} \min_{\boldsymbol{u}} Q\left(\boldsymbol{v}_B(\boldsymbol{k}, \boldsymbol{u}); \Omega_0^\mu\right)$ is detected by a complex-valued Bloch wave function $\boldsymbol{v}_B$, then the buckling mode can be either obtained from the real part or the imaginary part, since both are from the same set and differ only by a phase shift.

### *4.1.3 Remarks*

As mentioned above, the buckling modes can be of any finite wavelength or even aperiodic. This poses a significant challenge in the stability analysis of periodic metamaterials since the solution belongs to the entire (infinite) lattice space, and FE model including multiple unit cells, has to be employed if stability is directly investigated, as proposed in Ref. [45]. However, expressing the buckling modes in terms of the Bloch wave functions simplifies the solution process, since instead of searching the entire lattice space, the Bloch wave solutions are sought using FE model with only one unit cell [36]. However, in this case, all $\boldsymbol{k}$ values belonging to the primitive cell in the reciprocal space that theoretically consists of infinite points have to be examined. From a finite element analysis viewpoint, the Bloch function method makes the problem computationally tractable, as it makes it possible to determine the stability of periodic metamaterials while employing only one unit cell as RVE, which would otherwise become computationally intractable if the direct method presented in Ref. [45] is employed. Furthermore, if the metamaterial microscale stability is examined by structural stability analysis with an increasing number of unit cells in RVE as proposed in Ref. [45], this strategy will still fail to capture *aperiodic* buckling modes. Thus, this direct stability examination is both computationally expensive and inaccurate. Indeed, as shown in [36], it is only when the search space includes RVE that consists of an infinite number of unit cells will the direct method in [45] give the same result as the Bloch wave analysis on a unit cell.

*4.2 Macroscale stability*

As a measure of the macroscopic stability, the rank-1 convexity of the homogenized tangent moduli ensures the absence of discontinuities in the deformation gradient field on the macroscale. Macroscopic stability can be assessed by examining the positive definiteness of the ellipticity indicator $B(\lambda)$ defined by

$$B(\lambda) = \min_{\bar{\boldsymbol{m}},\bar{\boldsymbol{M}}} (\bar{\boldsymbol{m}} \otimes \bar{\boldsymbol{M}}):\bar{\mathbb{A}}:(\bar{\boldsymbol{m}} \otimes \bar{\boldsymbol{M}}) \tag{54}$$

where $\bar{\boldsymbol{m}}$ and $\bar{\boldsymbol{M}}$ span over all possible directions with $\|\bar{\boldsymbol{m}}\| = \|\bar{\boldsymbol{M}}\| = 1$. A recent study has also shown that upon the loss of rank-one convexity there is *not* always a discontinuous/localized deformation pattern on the bifurcated branch [56]. The presence or absence of localized deformation depends on the stability of the bifurcated branch. When there is a discontinuous deformation corresponds to the loss of rank-1 convexity, i.e. $B(\lambda) = 0$, the corresponding minimizing vector $\bar{\boldsymbol{M}}$ represents the normal to the curves across which the jump discontinuities appear and $\bar{\boldsymbol{m}}$ determines the nature of the discontinuous mode (simple shear if $\bar{\boldsymbol{m}}$ is orthogonal to $\bar{\boldsymbol{M}}$ or pure splitting if $\bar{\boldsymbol{m}}$ is parallel to $\bar{\boldsymbol{M}}$ or mixture otherwise) [57]. As has been proved in [36], the following inequality holds

$$\beta(\lambda) \leq \beta(\lambda; \boldsymbol{k} \to \boldsymbol{0}) \leq B(\lambda) \tag{55}$$

Eq. (55) implies that if the microscopic stability is preserved, i.e. $\beta(\lambda) > 0$, the homogenized macroscopic material is rank-1 convex. It is noted that the first inequality in Eq. (55) is straightforward since the $\boldsymbol{k}$ space in $\beta(\lambda)$ includes the case $\boldsymbol{k} \to \boldsymbol{0}$. Besides, it is also shown in [36] that the loss of rank-1 convexity corresponds to a long wavelength microscale buckling, i.e.

$$B(\lambda) = 0 \quad \text{if} \quad \beta(\lambda; \boldsymbol{k} \to \boldsymbol{0}) = 0 \tag{56}$$

As a result, the detection of long wavelength buckling can be more efficiently carried out by rank-1 convexity examination compared to the Bloch wave analysis.

### *4.3 Implementation details – Bloch analysis*

In this section, three numerical methods are presented that can be used for microscale stability analysis described in Section 4.1 using Bloch waves. Since only zero values of the stability indicator $\beta(\lambda)$ are of interest, the normalization by the denominator in Eq. (44) can be ignored. To this end, the discrete form of Eq. (45) can be expressed as

$$\beta(\lambda) = \inf_{\boldsymbol{k}} \min_{\boldsymbol{u}} F(\boldsymbol{v}_B) \tag{57}$$

with

$$F(\boldsymbol{v}) = \boldsymbol{v}^* \boldsymbol{K}_T \boldsymbol{v} \quad \text{and} \quad \boldsymbol{v}_B(\boldsymbol{k}, \boldsymbol{u}) = e^{i\boldsymbol{k}.\boldsymbol{X}} \boldsymbol{u} \tag{58}$$

where the symbol $*$ denotes the complex conjugate transpose. All the presented methods aim at finding the lowest value of loading parameter $\lambda = \lambda_c$, also known as critical load/point, at which the quadratic stability functional in Eq. (57) loses its positive definiteness. The first two methods are based on condensation schemes and the last method is based on the null-space projection scheme. Note that this is a case of a constrained eigenvalue problem, where constraints are determined by the Bloch conditions.

#### *4.3.1 Condensation method – I*

In this method, the DOFs (microscale displacement field) of the underlying RVE are partitioned into three groups, $\boldsymbol{v}_a$ on the negative side, $\boldsymbol{v}_b$ on the positive side and $\boldsymbol{v}_i$ of the interior nodes, i.e.

$$\boldsymbol{v} = \begin{bmatrix} \boldsymbol{v}_a \\ \boldsymbol{v}_b \\ \boldsymbol{v}_i \end{bmatrix} \tag{59}$$

and the tangent stiffness matrix $\boldsymbol{K}_T$ (Eq. (25)) at the loading step ($\lambda$) of interest is accordingly partitioned as

$$\boldsymbol{K}_T = \begin{bmatrix} \boldsymbol{K}_{aa} & \boldsymbol{K}_{ab} & \boldsymbol{K}_{ai} \\ \boldsymbol{K}_{ba} & \boldsymbol{K}_{bb} & \boldsymbol{K}_{bi} \\ \boldsymbol{K}_{ia} & \boldsymbol{K}_{ib} & \boldsymbol{K}_{ii} \end{bmatrix} \tag{60}$$

Since the buckling mode can be represented by the Bloch wave function, i.e. Eq. (45), it can be shown that the DOFs on the positive side are related to the DOFs on the negative side by

$$\boldsymbol{v}_b = \boldsymbol{M}(\boldsymbol{k})\boldsymbol{v}_a \tag{61}$$

where, for instance, in the case of parallelogram RVE (see Figure 2a or Figure 3a) the matrix $\boldsymbol{M} = \boldsymbol{M}(\boldsymbol{k})$ contains nonzero elements of the form $e^{i\boldsymbol{k}.\boldsymbol{a}_1}$, $e^{i\boldsymbol{k}.\boldsymbol{a}_2}$ and $e^{i\boldsymbol{k}.(\boldsymbol{a}_1+\boldsymbol{a}_2)}$ where $\boldsymbol{a}_1$ and $\boldsymbol{a}_2$ are two periodic vectors, i.e.,

$$\boldsymbol{M}(\boldsymbol{k}) = \begin{bmatrix} \begin{bmatrix} e^{i\boldsymbol{k}.\boldsymbol{a}_1} & & \\ & \ddots & \\ & & e^{i\boldsymbol{k}.\boldsymbol{a}_1} \end{bmatrix} & & \\ & \begin{bmatrix} e^{i\boldsymbol{k}.\boldsymbol{a}_2} & & \\ & \ddots & \\ & & e^{i\boldsymbol{k}.\boldsymbol{a}_2} \end{bmatrix} & \\ & & \begin{bmatrix} e^{i\boldsymbol{k}.\boldsymbol{a}_1} & 0 \\ 0 & e^{i\boldsymbol{k}.\boldsymbol{a}_1} \\ e^{i\boldsymbol{k}.(\boldsymbol{a}_1+\boldsymbol{a}_2)} & 0 \\ 0 & e^{i\boldsymbol{k}.(\boldsymbol{a}_1+\boldsymbol{a}_2)} \\ e^{i\boldsymbol{k}.\boldsymbol{a}_2} & 0 \\ 0 & e^{i\boldsymbol{k}.\boldsymbol{a}_2} \end{bmatrix} \end{bmatrix} \tag{62}$$

where the first matrix block represents the Bloch periodic boundary condition on the left and right pairing sides and the second matrix block represents the bottom and top pairing sides, while the third block represents the boundary conditions on the corner nodes, see Eq. (45) and Figure 3a, where corner node 1 is taken as on the negative side while corner nodes 2, 3, 4 are taken as on the positive side and are related to the corner node 1 by the Bloch periodic conditions. The Bloch wave analysis for microscale stability then seeks the first point during the loading process where the eigenvalue(s) $\beta$ of the tangent stiffness matrix $\boldsymbol{K}_T$ with eigenmode satisfying Eq. (45)$_2$ is (are) zero, i.e.

*Find the lowest value of the loading parameter $\lambda_c$ such that the minimum of the eigenvalues $\beta(\lambda_c) = 0$ with $\beta$ defined as*

$$\begin{bmatrix} \boldsymbol{K}_{aa} & \boldsymbol{K}_{ab} & \boldsymbol{K}_{ai} \\ \boldsymbol{K}_{ba} & \boldsymbol{K}_{bb} & \boldsymbol{K}_{bi} \\ \boldsymbol{K}_{ia} & \boldsymbol{K}_{ib} & \boldsymbol{K}_{ii} \end{bmatrix} \begin{bmatrix} \boldsymbol{v}_a \\ \boldsymbol{v}_b \\ \boldsymbol{v}_i \end{bmatrix} = \beta \begin{bmatrix} \boldsymbol{v}_a \\ \boldsymbol{v}_b \\ \boldsymbol{v}_i \end{bmatrix} \quad \text{with} \quad \boldsymbol{v}_b = \boldsymbol{M}\boldsymbol{v}_a \tag{63}$$

which leads to

$$\begin{bmatrix} \boldsymbol{K}_{aa} & \boldsymbol{K}_{ab} & \boldsymbol{K}_{ai} \\ \boldsymbol{K}_{ba} & \boldsymbol{K}_{bb} & \boldsymbol{K}_{bi} \\ \boldsymbol{K}_{ia} & \boldsymbol{K}_{ib} & \boldsymbol{K}_{ii} \end{bmatrix} \boldsymbol{H} \begin{bmatrix} \boldsymbol{v}_a \\ \boldsymbol{v}_i \end{bmatrix} = \beta \boldsymbol{H} \begin{bmatrix} \boldsymbol{v}_a \\ \boldsymbol{v}_i \end{bmatrix} \quad \text{with} \quad \boldsymbol{H} = \begin{bmatrix} \boldsymbol{I} & \boldsymbol{0} \\ \boldsymbol{M} & \boldsymbol{0} \\ \boldsymbol{0} & \boldsymbol{I} \end{bmatrix} \tag{64}$$

where $\boldsymbol{I}$ is an identity matrix of appropriate size. Due to the linear independency of the columns in matrix $\boldsymbol{H}$, the system of equations in Eq. (64) is equivalent to

$$\begin{bmatrix} \widehat{\boldsymbol{K}}_{aa} & \widehat{\boldsymbol{K}}_{ai} \\ \widehat{\boldsymbol{K}}_{ia} & \widehat{\boldsymbol{K}}_{ii} \end{bmatrix} \begin{bmatrix} \boldsymbol{v}_a \\ \boldsymbol{v}_i \end{bmatrix} = \beta \boldsymbol{G} \begin{bmatrix} \boldsymbol{v}_a \\ \boldsymbol{v}_i \end{bmatrix} \quad \text{with} \quad \boldsymbol{G} = \boldsymbol{H}^* \boldsymbol{H} \tag{65}$$

where

$$\begin{gathered} \widehat{\boldsymbol{K}}_{aa} = \boldsymbol{K}_{aa} + \boldsymbol{K}_{ab}\boldsymbol{M} + \boldsymbol{M}^*\boldsymbol{K}_{ba} + \boldsymbol{M}^*\boldsymbol{K}_{bb}\boldsymbol{M} \\ \widehat{\boldsymbol{K}}_{ai} = \boldsymbol{K}_{ai} + \boldsymbol{M}^*\boldsymbol{K}_{bi} \\ \widehat{\boldsymbol{K}}_{ia} = \boldsymbol{K}_{ia} + \boldsymbol{K}_{ib}\boldsymbol{M} \\ \widehat{\boldsymbol{K}}_{ii} = \boldsymbol{K}_{ii} \end{gathered} \tag{66}$$

and $*$ is the complex conjugate transpose since $\boldsymbol{M}$ is a complex-valued matrix. It should be noted that since the non-zero entries of the matrix $\boldsymbol{M}$ are $e^{i\boldsymbol{k}.\boldsymbol{L}}$ with $\boldsymbol{L}$ some periodic translation vectors (e.g. $\boldsymbol{a}_1$, $\boldsymbol{a}_2$ for the nodes on the side while $\boldsymbol{a}_1 + \boldsymbol{a}_2$ for some corner nodes), the matrix $\boldsymbol{M}^*\boldsymbol{M}$ is a diagonal matrix with most of the entries equal to 1 and a few entries corresponding to slave corner node DOFs to be 2 or 3 depending on different RVE shapes (see Figure 3 and remarks in Section 2.1.1). Hence, the matrix $\boldsymbol{G}$ is a diagonal matrix, as well. As a result, the original eigenvalue problem, Eq. (63), is equivalently transformed to the following condensed eigenvalue problem

$$\boldsymbol{G}^{-1}\begin{bmatrix}\widehat{\boldsymbol{K}}_{aa} & \widehat{\boldsymbol{K}}_{ai}\\ \widehat{\boldsymbol{K}}_{ia} & \widehat{\boldsymbol{K}}_{ii}\end{bmatrix}\begin{bmatrix}\boldsymbol{v}_a\\ \boldsymbol{v}_i\end{bmatrix} = \beta\begin{bmatrix}\boldsymbol{v}_a\\ \boldsymbol{v}_i\end{bmatrix} \tag{67}$$

where the invertibility of the matrix $\boldsymbol{G}$ is obvious.

***Remark****:* Since the 1st bifurcation point is of interest, i.e. the loading parameter $\lambda$ at which the minimum eigenvalue $\beta = 0$, the existence of zero (or negative) eigenvalue of the matrix

$$\boldsymbol{G}^{-1}\begin{bmatrix}\widehat{\boldsymbol{K}}_{aa} & \widehat{\boldsymbol{K}}_{ai}\\ \widehat{\boldsymbol{K}}_{ia} & \widehat{\boldsymbol{K}}_{ii}\end{bmatrix} \tag{68}$$

is equivalent to the existence of zero (or negative) eigenvalue of the matrix

$$\begin{bmatrix}\widehat{\boldsymbol{K}}_{aa} & \widehat{\boldsymbol{K}}_{ai}\\ \widehat{\boldsymbol{K}}_{ia} & \widehat{\boldsymbol{K}}_{ii}\end{bmatrix} \tag{69}$$

due to the positive definiteness of the matrix $\boldsymbol{G}$ or equivalently $\boldsymbol{G}^{-1}$. Moreover, even though in general the eigenvectors of the matrix in Eq. (68) are different from those of the matrix in Eq. (69), the eigenvectors corresponding to the buckling mode, i.e. the eigenvector corresponding to zero eigenvalues, are the same for the systems in Eq. (68) and Eq. (69). As a result, eigen analysis can be carried out for the matrix in Eq. (69).

### *4.3.2 Condensation method – II*

The condensation method in Section 4.3.1 implements the Bloch wave analysis by condensing out $\boldsymbol{v}_b$, i.e. the degrees of freedom corresponding to the positive side. To further reduce the computational cost, Triantafyllidis et al. [39] proposed another condensation method, where the eigen analysis is carried out for a matrix of the same size as $\boldsymbol{K}_{aa}$. From Eq. (63)₁, assuming that the analysis is carried out at the zero eigenvalue, i.e. at $\beta = 0$, results in

$$\boldsymbol{K}_{ia}\boldsymbol{v}_a + \boldsymbol{K}_{ib}\boldsymbol{v}_b + \boldsymbol{K}_{ii}\boldsymbol{v}_i = \boldsymbol{0} \tag{70}$$

From Eq. (70), one part of the eigenvector ($\boldsymbol{v}_i$) can be expressed in terms of $\boldsymbol{v}_a$ as

$$\boldsymbol{v}_i = \boldsymbol{W}\boldsymbol{v}_a \quad \text{with} \quad \boldsymbol{W} \overset{\text{def}}{=} -\boldsymbol{K}_{ii}^{-1}(\boldsymbol{K}_{ia} + \boldsymbol{K}_{ib}\boldsymbol{M}) \tag{71}$$

where the relation $\boldsymbol{v}_b = \boldsymbol{M}\boldsymbol{v}_a$ is used. Thus, to use this condensation method the inverse $\boldsymbol{K}_{ii}^{-1}$ have to be calculated. Note that this may lead to increase memory requirements during the solution process, as the inverse of sparse matrices is, in general, not sparse. With eigenvector components $\boldsymbol{v}_b$ and $\boldsymbol{v}_i$ expressed in terms of $\boldsymbol{v}_a$, the original eigenvalue equation in Eq. (63) can be rewritten as

$$\begin{bmatrix} \boldsymbol{K}_{aa} & \boldsymbol{K}_{ab} & \boldsymbol{K}_{ai} \\ \boldsymbol{K}_{ba} & \boldsymbol{K}_{bb} & \boldsymbol{K}_{bi} \\ \boldsymbol{K}_{ia} & \boldsymbol{K}_{ib} & \boldsymbol{K}_{ii} \end{bmatrix} \widehat{\boldsymbol{H}}\boldsymbol{v}_a = \beta\widehat{\boldsymbol{H}}\boldsymbol{v}_a \quad \text{with} \quad \widehat{\boldsymbol{H}} = \begin{bmatrix} \boldsymbol{I} \\ \boldsymbol{M} \\ \boldsymbol{W} \end{bmatrix} \tag{72}$$

which, due to the linear independency of the column vectors in matrix $\widehat{\boldsymbol{H}}$, is equivalent to

$$\widehat{\boldsymbol{K}}\boldsymbol{v}_a = \beta\boldsymbol{D}\boldsymbol{v}_a \quad \text{with} \quad \boldsymbol{D} = \widehat{\boldsymbol{H}}^*\widehat{\boldsymbol{H}} \tag{73}$$

or

$$\boldsymbol{D}^{-1}\widehat{\boldsymbol{K}}\boldsymbol{v}_a = \beta\boldsymbol{v}_a \tag{74}$$

where

$$\begin{aligned} \widehat{\boldsymbol{K}} = \widehat{\boldsymbol{H}}^*\boldsymbol{K}_T\widehat{\boldsymbol{H}} &= \boldsymbol{K}_{aa} + \boldsymbol{K}_{ab}\boldsymbol{M} + \boldsymbol{K}_{ai}\boldsymbol{W} + \boldsymbol{M}^*(\boldsymbol{K}_{ba} + \boldsymbol{K}_{bb}\boldsymbol{M} + \boldsymbol{K}_{bi}\boldsymbol{W}) \\ &= \boldsymbol{K}_{aa} + \boldsymbol{K}_{ab}\boldsymbol{M} + \boldsymbol{M}^*\boldsymbol{K}_{ba} + \boldsymbol{M}^*\boldsymbol{K}_{bb}\boldsymbol{M} \\ &- (\boldsymbol{K}_{ai} + \boldsymbol{M}^*\boldsymbol{K}_{bi})\boldsymbol{K}_{ii}^{-1}(\boldsymbol{K}_{ia} + \boldsymbol{K}_{ib}\boldsymbol{M}) \end{aligned} \tag{75}$$

As can be seen, the size of the matrix in the eigen analysis is significantly reduced by using the conditions in Eq. (70). Also, the inversion of $\boldsymbol{K}_{ii}$ is only needed once at each loading step. Thus, this method can be more efficient than the one in Section 4.3.1, especially when the sampling grid in the $\boldsymbol{k}$-space is large. Moreover, when the 1st bifurcation point is of interest, the eigen analysis of matrix $\widehat{\boldsymbol{K}}$ can be carried out to search for a zero (or negative) eigenvalue instead of the original matrix $\boldsymbol{D}^{-1}\widehat{\boldsymbol{K}}$ due to the positive definiteness of matrix $\boldsymbol{D}$ or $\boldsymbol{D}^{-1}$. The positive definiteness of $\boldsymbol{D}$ can be established, as for any vector $\boldsymbol{v}_a \neq \boldsymbol{0}$

$$\boldsymbol{v}_a^* \boldsymbol{D} \boldsymbol{v}_a = \boldsymbol{v}_a^T (\boldsymbol{I} + \boldsymbol{M}^* \boldsymbol{M} + \boldsymbol{W}^* \boldsymbol{W}) \boldsymbol{v}_a = \|\boldsymbol{v}_a\|^2 + \|\boldsymbol{v}_b\|^2 + \|\boldsymbol{v}_i\|^2 > 0 \tag{76}$$

Similarly, the buckling mode, i.e. the eigenvector corresponding to zero eigenvalue, can also be obtained by eigen analysis on matrix $\widehat{\boldsymbol{K}}$ instead of $\boldsymbol{D}^{-1}\widehat{\boldsymbol{K}}$.

### *4.3.3 Null-space projection method*

In this method, the constraints on the eigenmode, i.e. Eq. $(45)_2$, are expressed as

$$\boldsymbol{C}\boldsymbol{v} = \boldsymbol{0} \tag{77}$$

where matrix $\boldsymbol{C}$ is of size $N_c \times N$ such that $\mathrm{rank}(\boldsymbol{C}) = N_c$, with $N_c < N$ the number of constraints, and $N$ the total number of DOFs. From Eq. (77), it is clear that the eigenvectors should belong to the null space $\mathcal{N}(\boldsymbol{C})$ of the constraint matrix $\boldsymbol{C}$ with $\dim\left(\mathcal{N}(\boldsymbol{C})\right) = N - N_c$. As a result, the eigenvector/buckling mode $\boldsymbol{v}$ should be spanned in a space consisting of $(N - N_c)$ bases vectors $\boldsymbol{y}_1, \boldsymbol{y}_2, \ldots, \boldsymbol{y}_{N-N_c}$ of $\mathcal{N}(\boldsymbol{C})$, i.e.

$$\boldsymbol{v} = \boldsymbol{Y}\boldsymbol{z} \quad \text{with} \quad \boldsymbol{Y} = \begin{bmatrix} | & & | \\ \boldsymbol{y}_1 & \cdots & \boldsymbol{y}_{N-N_c} \\ | & & | \end{bmatrix}_{N \times (N-N_c)} \tag{78}$$

where $\boldsymbol{z} \in \mathbb{R}^{N-N_c}$. Hence, with the null-space based representation of the eigenvector $\boldsymbol{v}$, the constrained eigen analysis (Eq. (63)) becomes

$$\boldsymbol{K}_T \boldsymbol{Y} \boldsymbol{z} = \beta \boldsymbol{Y} \boldsymbol{z} \tag{79}$$

To further proceed, the basis matrix $\boldsymbol{Y}$ of $\mathcal{N}(\boldsymbol{C})$ is needed. To this end, the QR decomposition of $\boldsymbol{C}^*$ matrix is carried out, which gives

$$\boldsymbol{C}^* = \boldsymbol{Q}\boldsymbol{R} \quad \text{with} \quad \boldsymbol{Q} = [[\boldsymbol{Q}_1]_{N \times N_c} \quad [\boldsymbol{Q}_2]_{N \times (N-N_c)}] \quad \text{and} \quad \boldsymbol{R} = \begin{bmatrix} [\boldsymbol{R}_1]_{N_c \times N_c} \\ [\boldsymbol{0}]_{(N-N_c) \times N_c} \end{bmatrix} \tag{80}$$

with an illustration shown in Figure 19. From Figure 19 and by the definition of QR decomposition, it is known that the columns in $\boldsymbol{Q}$ matrix are orthonormal vectors and columns in $\boldsymbol{Q}_1$ form the basis for $\mathcal{R}(\boldsymbol{C}^*)$, i.e. the range space of $\boldsymbol{C}^*$, while the columns in $\boldsymbol{Q}_2$ form the basis for $\mathcal{N}(\boldsymbol{C})$, i.e. the null space of $\boldsymbol{C}$. Therefore, $\boldsymbol{Y}$ can be chosen as $\boldsymbol{Q}_2$ and as a result, Eq. (79) can be equivalently expressed as

$$\boldsymbol{Q}_2^* \boldsymbol{K}_T \boldsymbol{Q}_2 \boldsymbol{z} = \beta \boldsymbol{z} \tag{81}$$

where the orthonormality of the column vectors in $\boldsymbol{Q}_2$ is used. This projected eigenvalue problem is then solved and the eigenvector $\boldsymbol{v}$ (buckling mode) can be simply recovered by $\boldsymbol{v} = \boldsymbol{Q}_2 \boldsymbol{z}$.

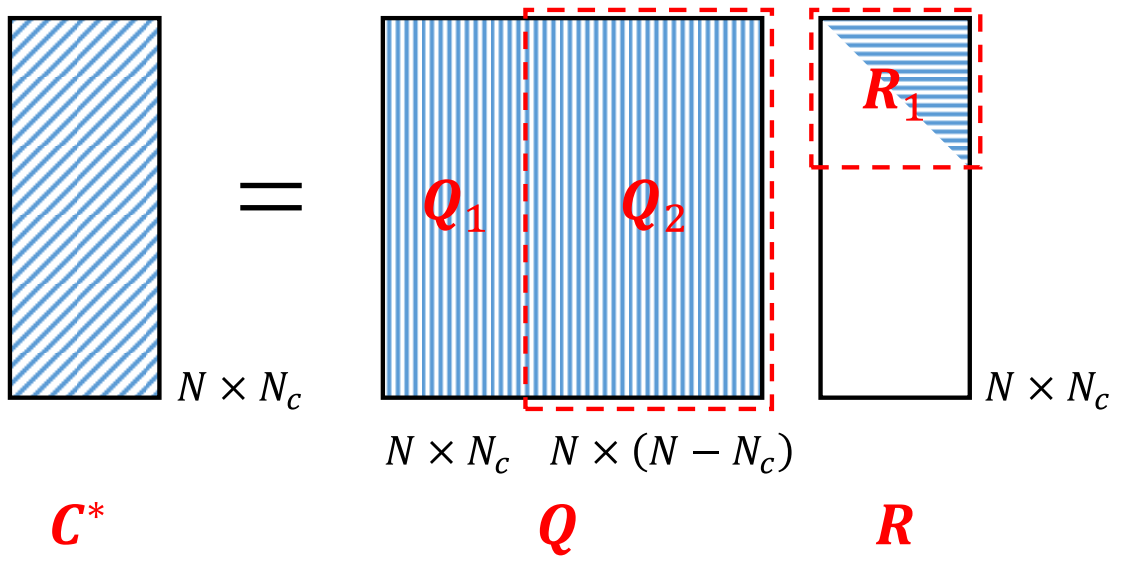

Figure 19. Illustration of QR decomposition of $\boldsymbol{C}^*$ matrix.

### *4.3.4 Remarks*

With the Bloch type boundary conditions rephrased in a constraints matrix in the null-space projection method, it has been shown that this method is particularly useful for the case where the boundary nodal degree of freedoms are not properly aligned, e.g. in isogeometric analysis [49]. In all the three methods, for buckling modes that are larger than one unit cell, i.e. $\boldsymbol{k} \neq \mathbf{0}$, the rigid-body motion is implicitly suppressed by the Bloch boundary conditions (Eq. $(45)_2$), so there is no need for adding extra constraints. However, for buckling mode that is periodic with respect to one unit cell, i.e. $\boldsymbol{k} = \mathbf{0}$, extra constraints are needed for constraining the rigid-body translation, e.g. fixing one arbitrary point. For instance, when $\boldsymbol{k} \neq \mathbf{0}$, $N_c = 2m$ for 2D case in Eq. (78) where $m$ denotes the number of pairs of nodes lying on the negative and positive boundary sides (see Eq. (15)), while when $\boldsymbol{k} = \mathbf{0}$, $N_c = 2m + 2$ where an arbitrary node in the domain is fixed.

## 5 Numerical Examples – Multiscale Stability

The purpose of this section is to demonstrate the efficacy of the proposed framework for the multiscale stability analysis using representative test cases. In the first example, the three methods presented in Section 4.3 are used and compared in detecting the multiscale instability point, while in the rest of the following examples, the null-space projection method is used for the Bloch wave analysis. Both strain driven and stress driven homogenization frameworks are employed for studying the multiscale stabilities of different microstructures. The goal of multiscale stability analysis is to determine the 1st bifurcation point and the buckling mode at the 1st bifurcation point. For the Bloch wave analyses, the $\boldsymbol{k}$-space mesh with $(k_1, k_2) \in [0,1) \times [0,1)$ consists of a 100×100 uniform mesh over $[0,1) \times [0,1)$ and 100×100 uniform meshes in the three refined zones $(0, 0.01) \times (0.01, 1]$, $(0.01, 1] \times (0, 0.01)$ and $(0, 0.01) \times (0, 0.01)$. For the rank-1 convexity check, the positive definiteness of the indicator $B(\lambda)$ in Eq. (54) is examined at each loading step at every $\pi/720$ radian increment in both $\overline{\boldsymbol{m}}$ and $\overline{\boldsymbol{M}}$ space. After the critical load is approximately located, bisection analyses are further carried out to further refine the accuracy of the detected 1st bifurcation point ($\lambda_c$) based on the selected $\boldsymbol{k}$-mesh.

## *5.1 Example-1: Comparison of the three methods*

The first example compares the three methods presented in Section 4.3. Specifically, a square unit cell of unit size with a central circular hole of radius $r$ = 0.4 is considered with matrix material following neo-Hookean model in Eq. (43) with $\kappa$ = 166.67 and $\mu$ = 35.71. Uniaxial load is considered where $\theta$ = 0° and $\phi = \pi/2$ in Eqns. (32) and (35), respectively, under stress driven framework. For the verification of the stress driven loop, Figure 20 plots the homogenized Kirchhoff stress in the principal direction, i.e. $\bar{\boldsymbol{\tau}}'$ $(= \bar{\boldsymbol{\tau}})$ in Eq. (32), where it can be seen that both $\bar{\tau}_{12} = \bar{\tau}_{21} = 0$ and $\bar{\tau}_{11} = 0$ in the loading process. The microscale stability surfaces in Figure 21 and Figure 22 are plotted where the stability indicator $\beta_{\boldsymbol{k}}$ is defined as the minimum eigenvalues ($\beta$) in Eq. (63) with wavevector ($\boldsymbol{k}$) at the 1st bifurcation load from different methods. As discussed in Section 4.3.1 and Section 4.3.2, the multiscale stability analysis (seeking 1st bifurcation load and buckling mode) is not affected by either including $\boldsymbol{G}^{-1}$ (or $\boldsymbol{D}^{-1}$) in Eq. (67) (or Eq. (74)) or excluding these matrices in eigen analysis. Excluding $\boldsymbol{G}^{-1}$ and $\boldsymbol{D}^{-1}$ in the two condensation methods, the 1st bifurcation loads detected using all the three methods are the same (with loading step size $\Delta\lambda$ = 7.5×10$^{-5}$) and the three buckling modes that are also the same, see Figure 21. However, the microscale stability surfaces can be different, as expected, see Figure 21(a) and (b). To demonstrate the equivalency of the three methods, the matrices $\boldsymbol{G}^{-1}$ and $\boldsymbol{D}^{-1}$ are included in the two condensation methods and the Bloch wave analyses are carried out again with the results shown in Figure 22. This result shows that not only the same 1st bifurcation load and buckling mode are obtained from the three methods, the microscale stability surfaces are also identical. It is also noted that a clear discontinuity at the origin $(k_1, k_2)$ = (0, 0) can be observed in the microscale stability surfaces in Figure 22. This discontinuity implies that the buckling mode with periodicity of one unit cell is not present at the 1st bifurcation load.

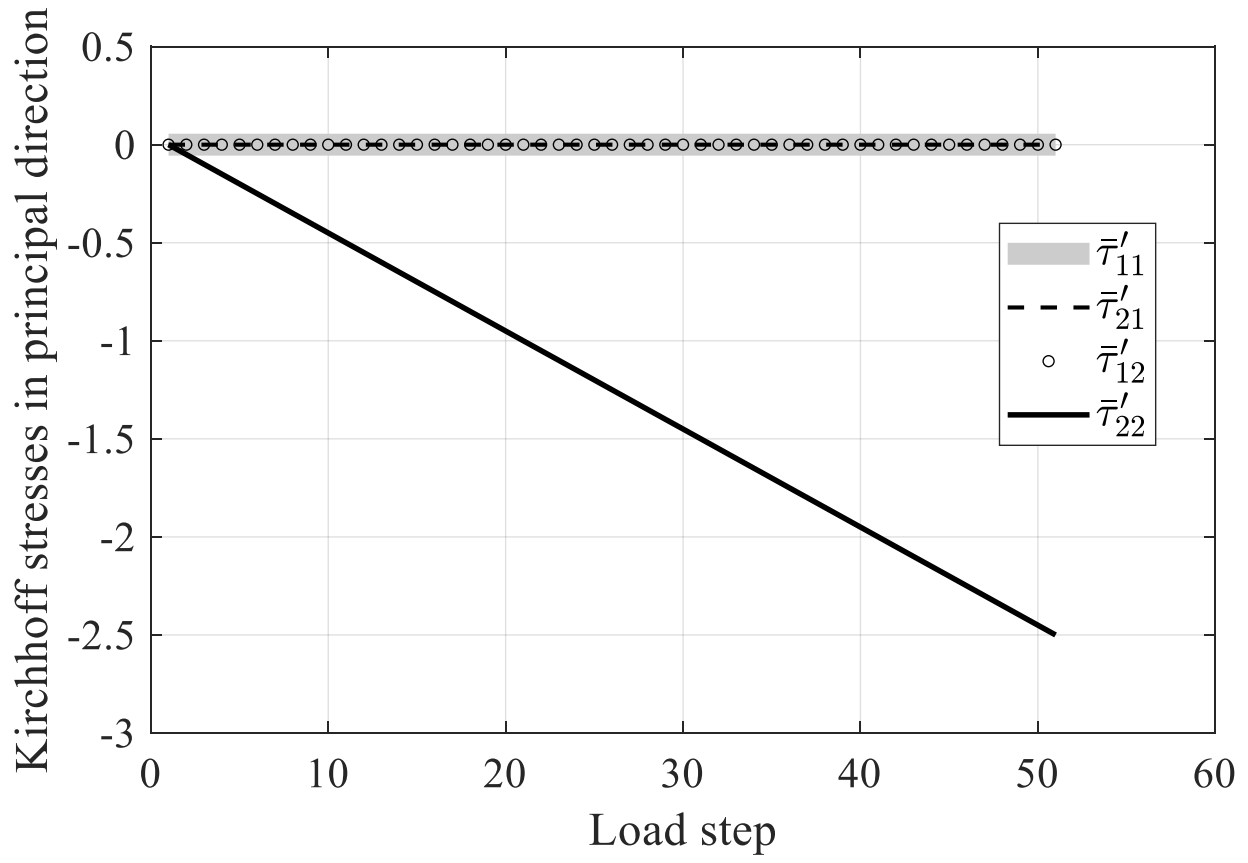

Figure 20. Homogenized Kirchhoff stresses in the principal direction at load steps.

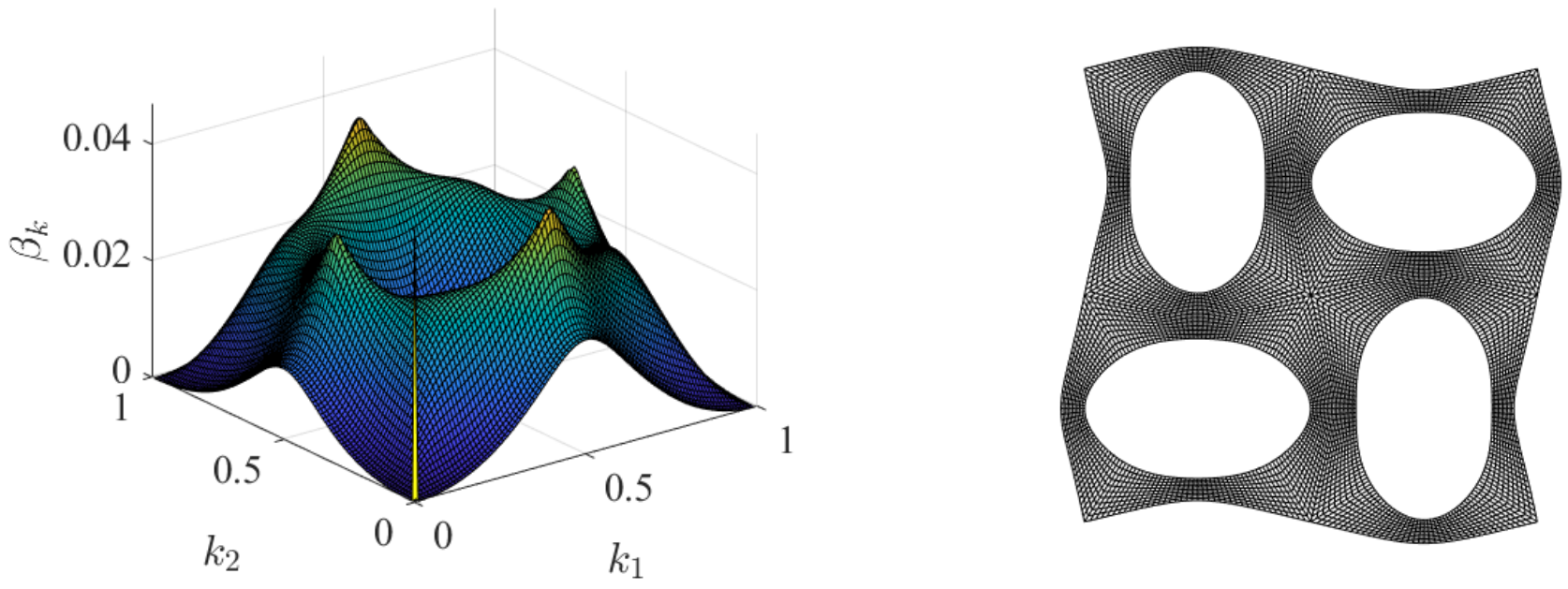

(a) Condensation method – I: $\lambda_c$ = 2.4620

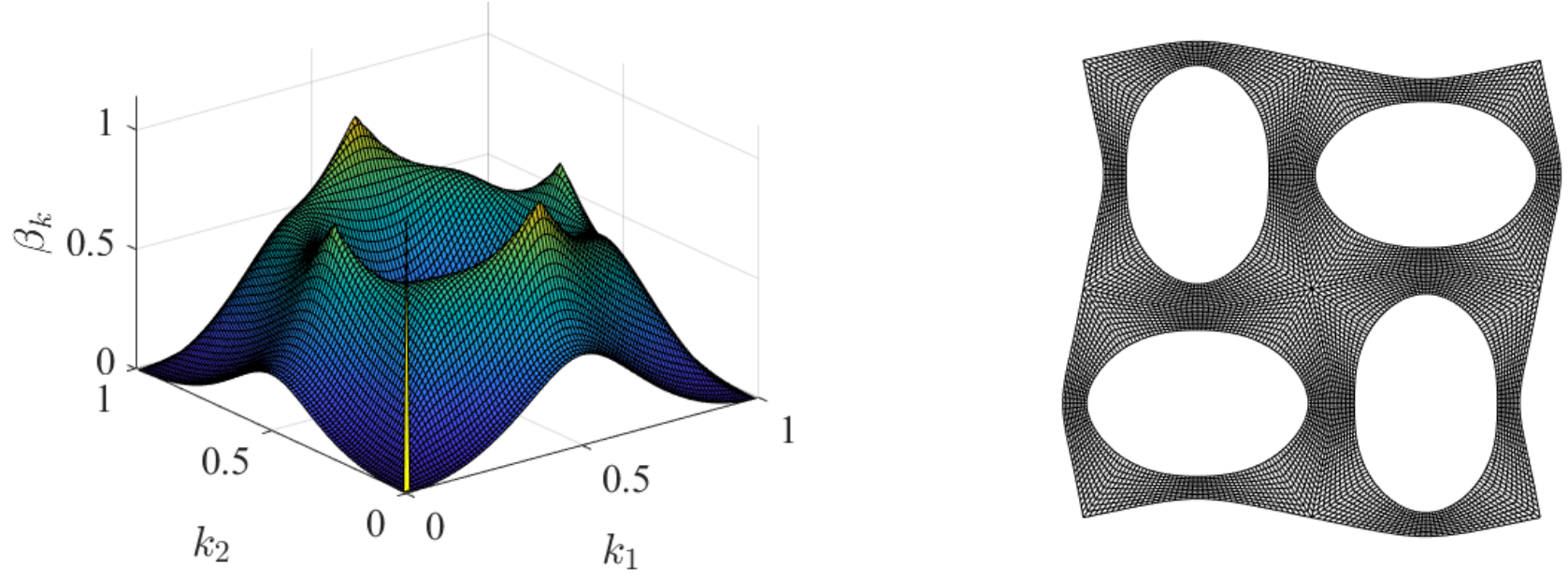

(b) Condensation method – II: $\lambda_c$ = 2.4620

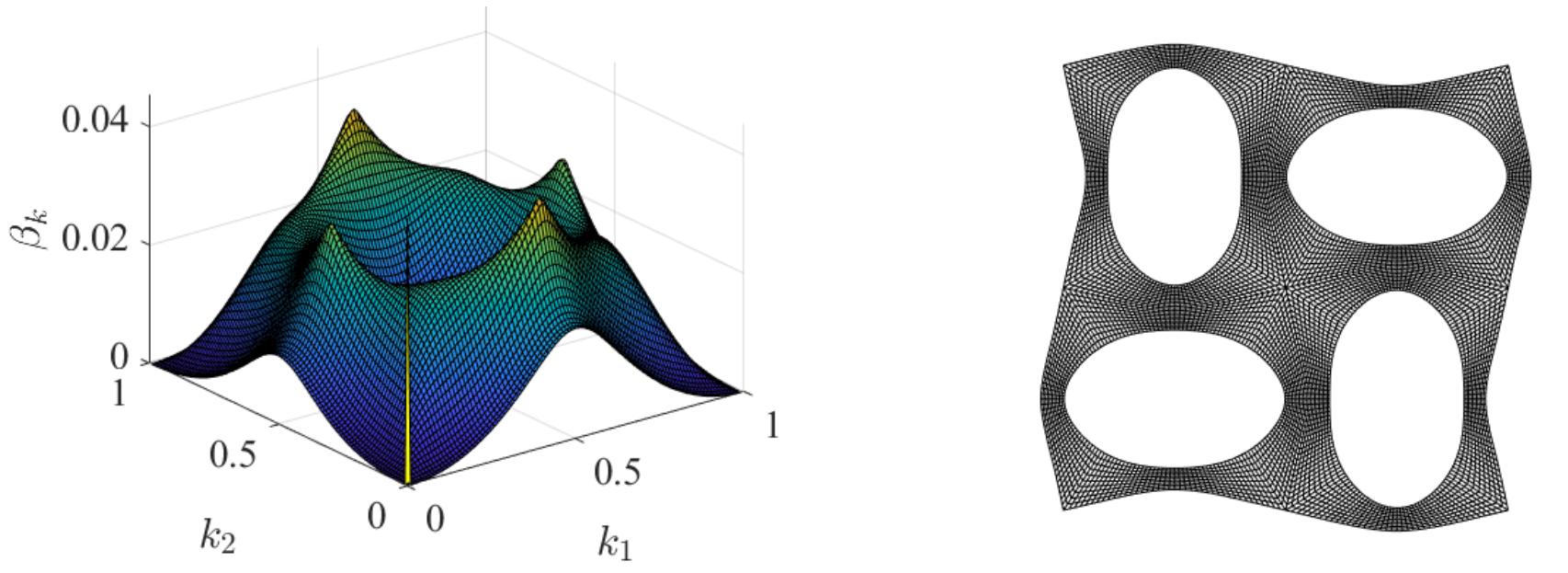

(c) Null-space projection method: $\lambda_c$ = 2.4620

Figure 21. Microscale stability surfaces and buckling modes at 1st bifurcation points ($\lambda_c$) using different methods with $\boldsymbol{G}^{-1}$ and $\boldsymbol{D}^{-1}$ *excluded* in Eq. (67) and Eq. (74). (*left column:* microscale surface, *right column:* buckling mode).

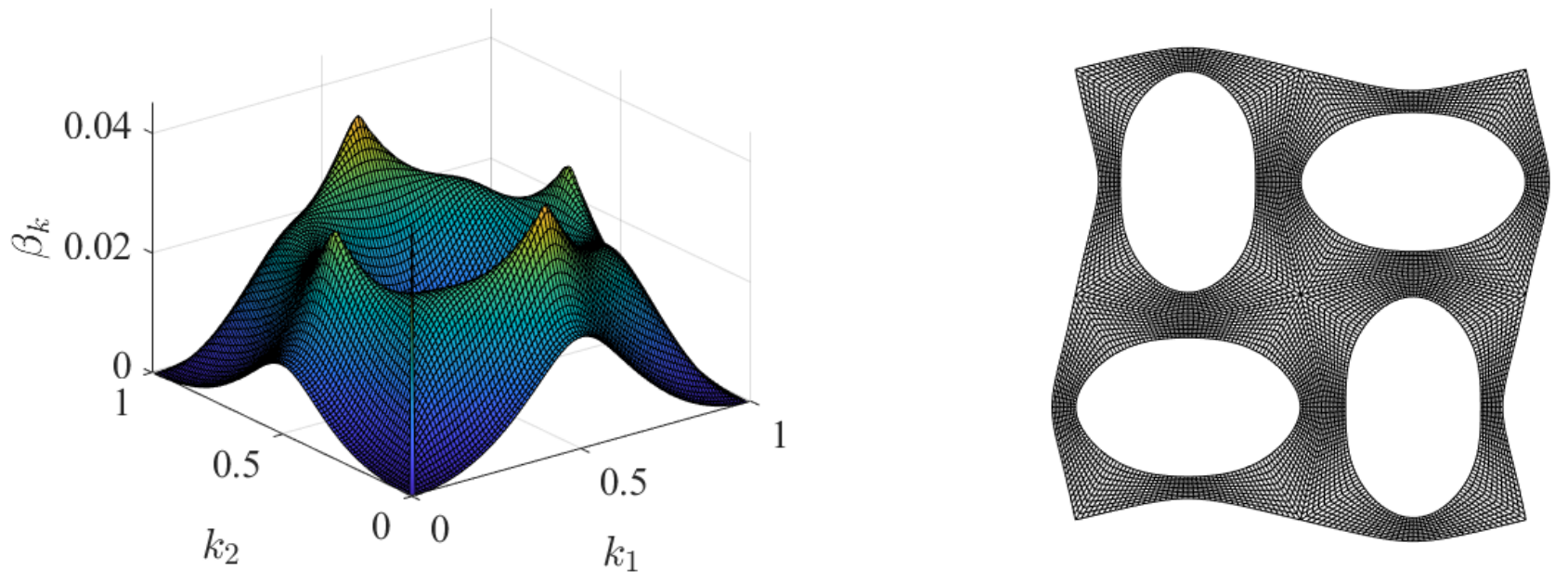

(a) Condensation method – I: $\lambda_c$ = 2.4620

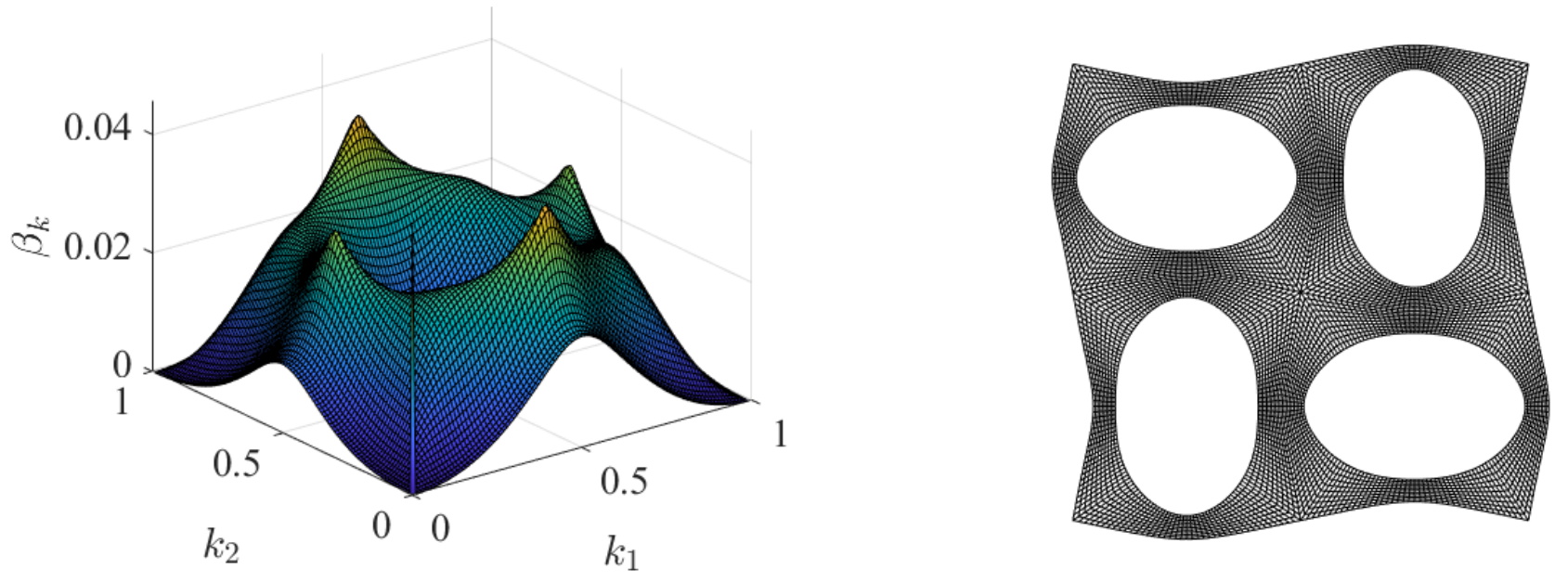

(b) Condensation method – II: $\lambda_c$ = 2.4620

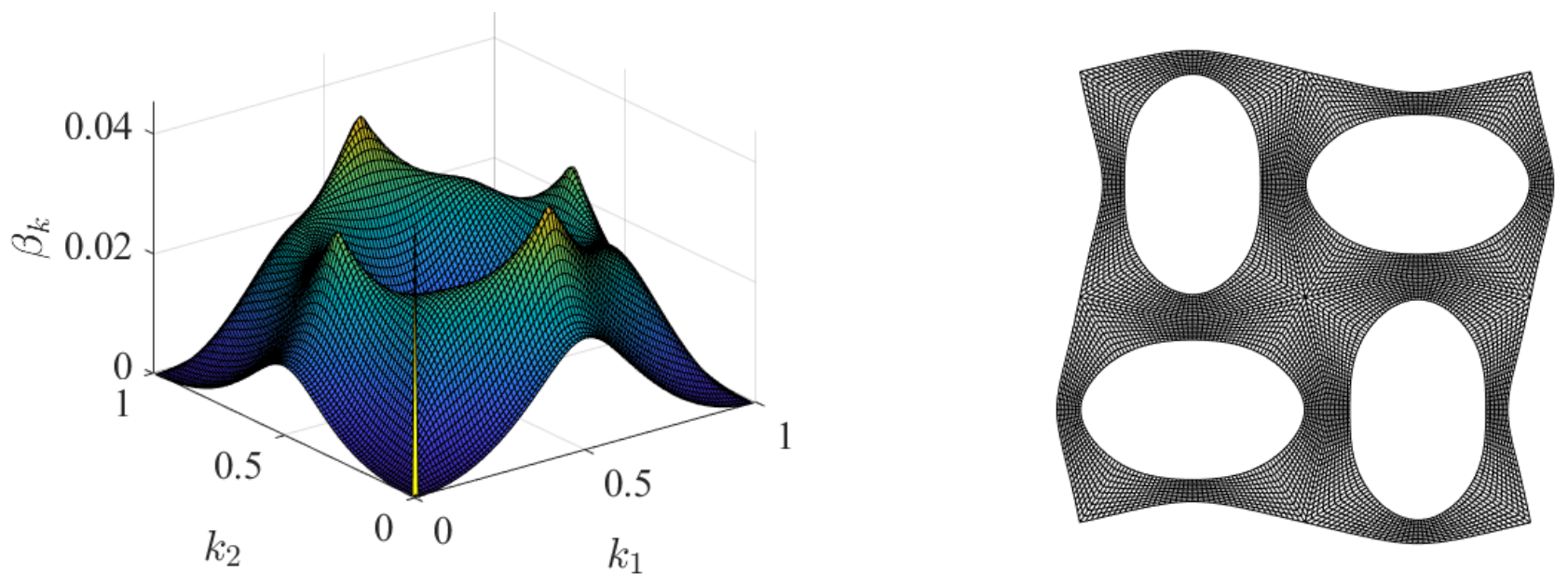

(c) Null-space projection method: $\lambda_c$ = 2.4620

Figure 22. Microscale stability surfaces and buckling modes at 1st bifurcation points ($\lambda_c$) using different methods with $\boldsymbol{G}^{-1}$ and $\boldsymbol{D}^{-1}$ *included* in Eq. (67) and Eq. (74). (*left column:* microscale surface, *right column:* buckling mode)

### *5.2 Example-2: Equivalency of different RVEs in multiscale stability analysis*

The second example considers periodic metamaterial with inclined circular holes of radius $r$ = 0.4, see Figure 23, where the overall metamaterial can be represented by different periodic cells. The matrix material again follows neo-Hookean model (Eq. (43)) with $\kappa$ = 166.67 and $\mu$ = 35.71. A constrained compression along an inclined angle (30°) is considered under strain driven framework. The macroscopic deformation gradient $\overline{\boldsymbol{F}}$ is applied as

$$\overline{\boldsymbol{F}} = \boldsymbol{Q}\begin{bmatrix}1 & 0\\0 & \lambda\end{bmatrix}\boldsymbol{Q}^T \tag{82}$$

where the bases transformation matrix $\boldsymbol{Q}$ is given in Eq. $(32)_2$ with $\theta$ = 30° and $\lambda$ represents loading parameter that decreases from 1 indicating compression.

Among the three RVEs (parallelogram, hexagon, and rectangle), the parallelogram and hexagon are both the (smallest) fundamental unit cells of the same size, while the rectangle-shaped RVE is twice the size as compared to other RVEs. For parallelogram and hexagon-shaped RVEs, the periodic vectors $\boldsymbol{a}_1$ and $\boldsymbol{a}_2$ are both unit vectors with the angle between them equal to 60°. The FE meshes of the three RVEs are shown in Figure 24. The buckling mode together with the calculated 1st bifurcation load factor ($\lambda_c$) for each case are shown in Figure 25. As can be seen, the same buckling mode is detected using different RVEs with close values for the 1st bifurcation load. However, as can be seen from Figure 26, the microscale stability contours can be different depending on the different choices of RVEs. In addition, the macroscale stability (absence of long wavelength buckling) is checked by the rank-1 convexity analysis of the homogenized tangent moduli and the results are presented in Table 8. As $B(\lambda_c) > 0$ for all cases, the rank-1 convexity of the homogenized tangent moduli is preserved and consequently there is no macroscale, i.e. long wavelength, instability at the critical points associated with short wavelength buckling. As already shown in Table 3, Table 4 and Table 6, the calculation of the homogenized tangent moduli can be slightly affected by the different choices of RVEs due to numerical errors. This influence can also be seen in the macroscale stability indicator $B(\lambda)$ calculation. Despite the small numerical differences brought by the errors in geometry and FE modeling and other approximation errors, this example demonstrates that different RVEs can be equivalently used for homogenization as well as multiscale stability analysis.

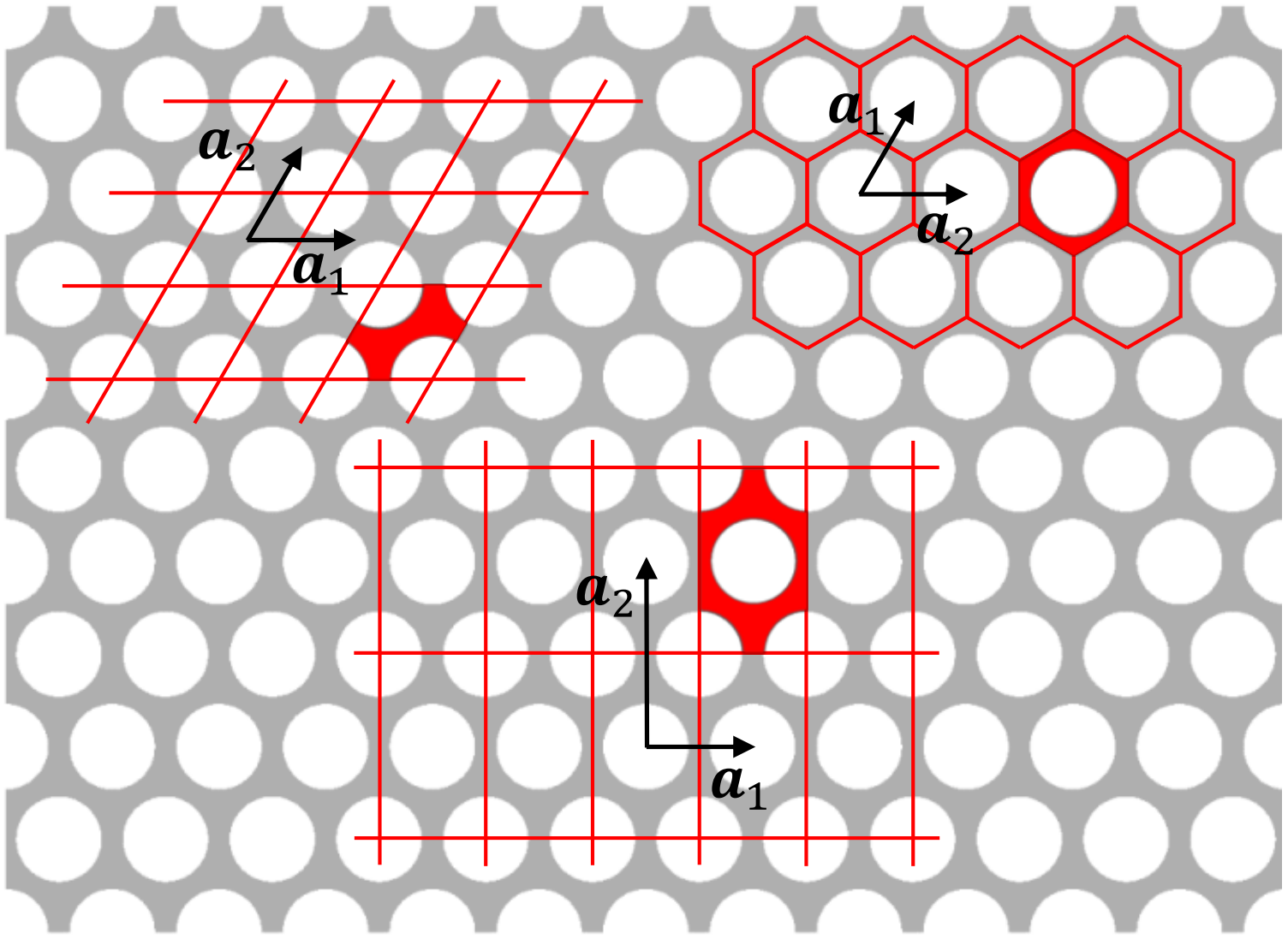

Figure 23. Illustration of different choices of RVE for composite with inclined circular holes

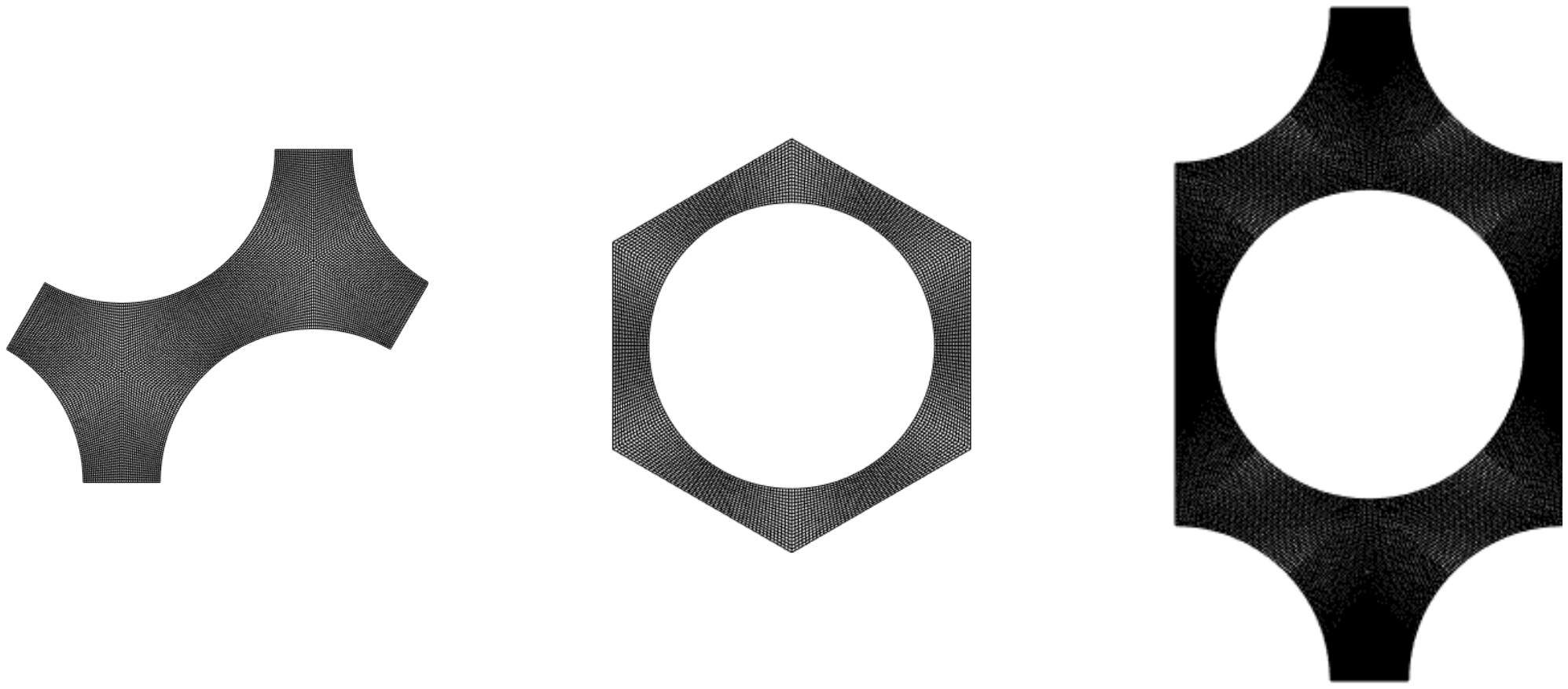

(a) Parallelogram (b) Hexagon (c) Rectangle

Figure 24. FE meshes of the three different RVEs.

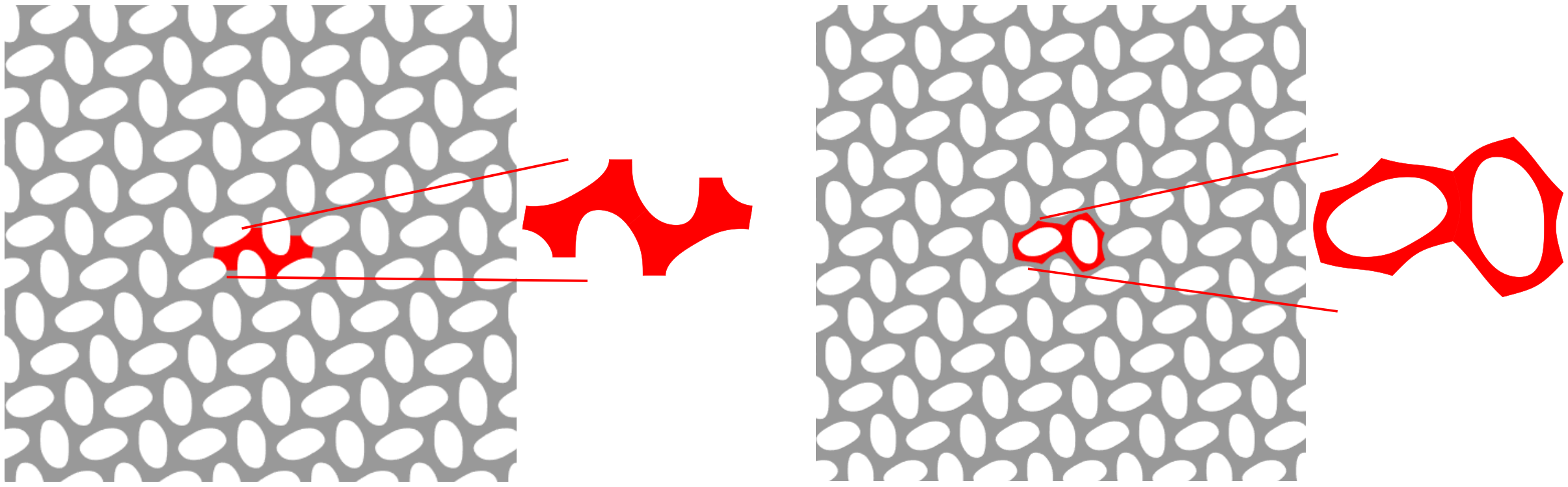

(a) Parallelogram: $\lambda_c$ = 0.9271, $(k_1, k_2)$ = (0.5, 0) (b) Hexagon: $\lambda_c$ = 0.9270, $(k_1, k_2)$ = (0, 0.5)

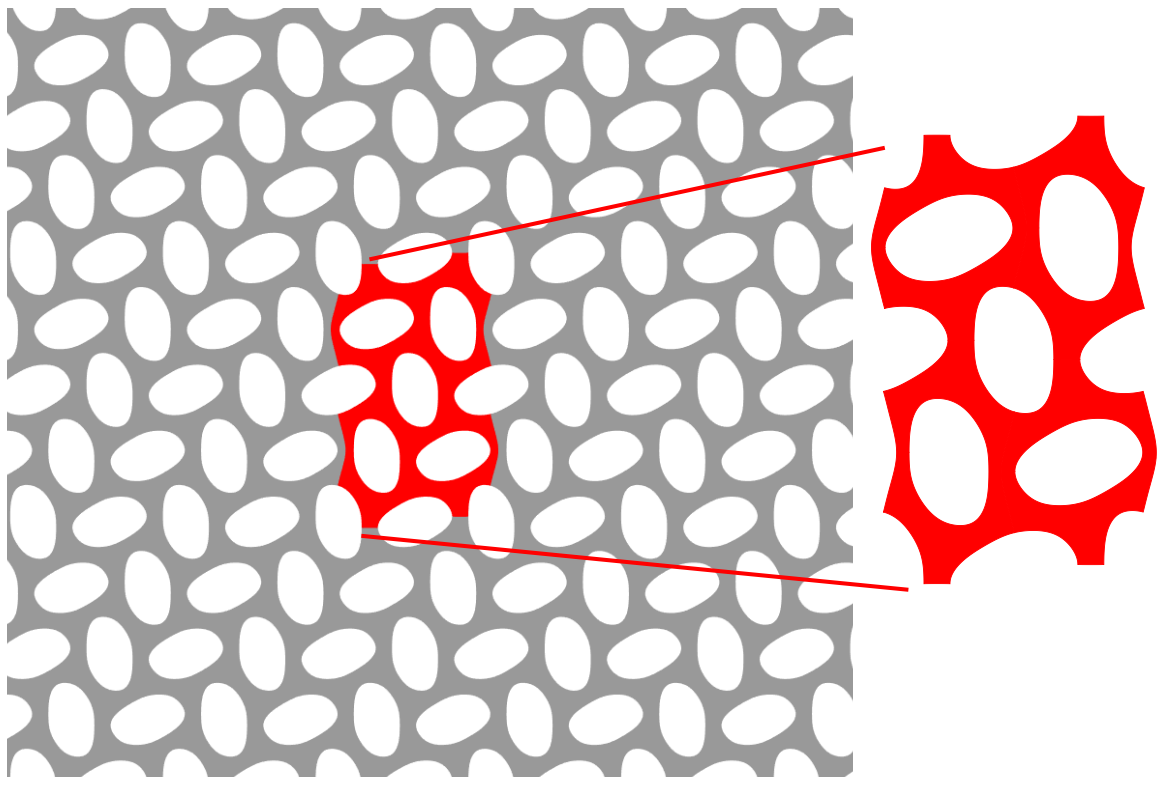

(c) Rectangle: $\lambda_c$ = 0.9270, $(k_1, k_2)$ = (0.5, 0.5)

Figure 25. 1st bifurcation load and the corresponding buckling mode obtained using different RVEs.

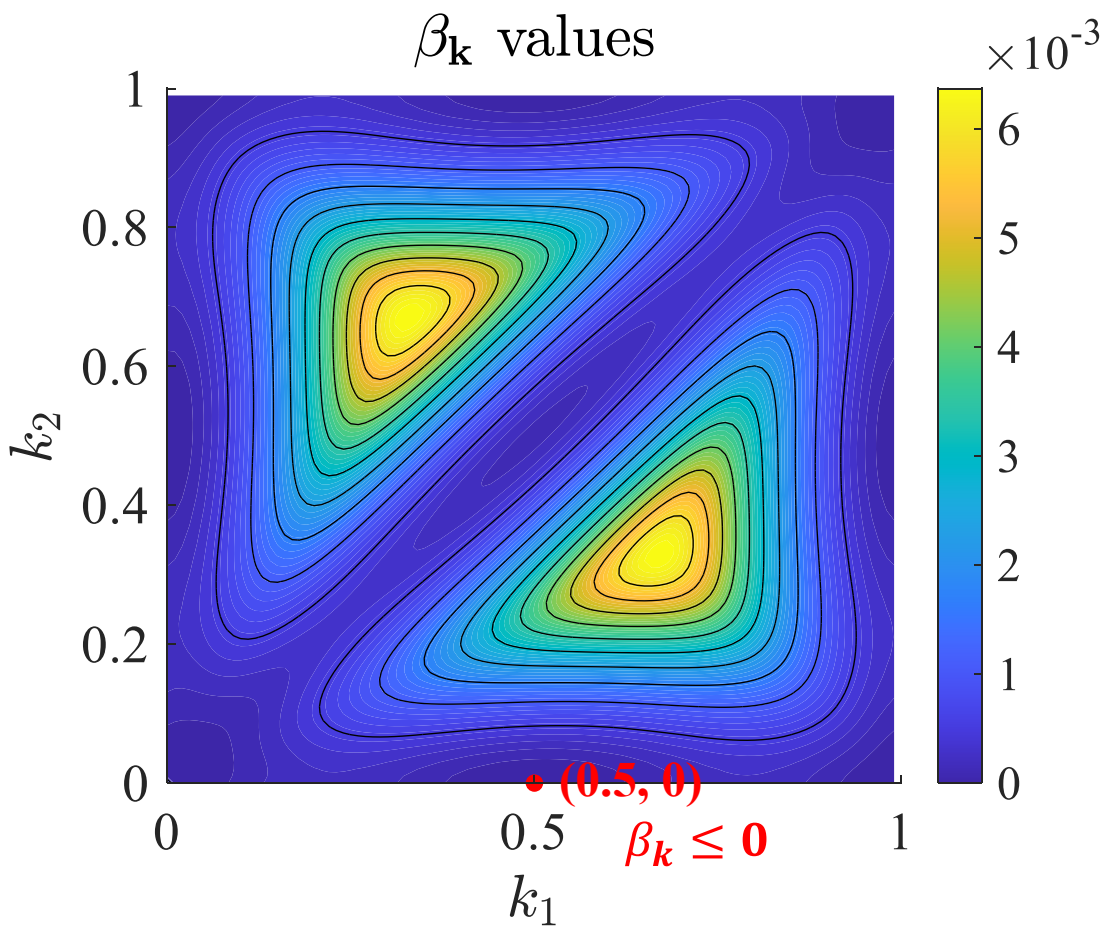

(a) Parallelogram: $\lambda_c$ = 0.9271

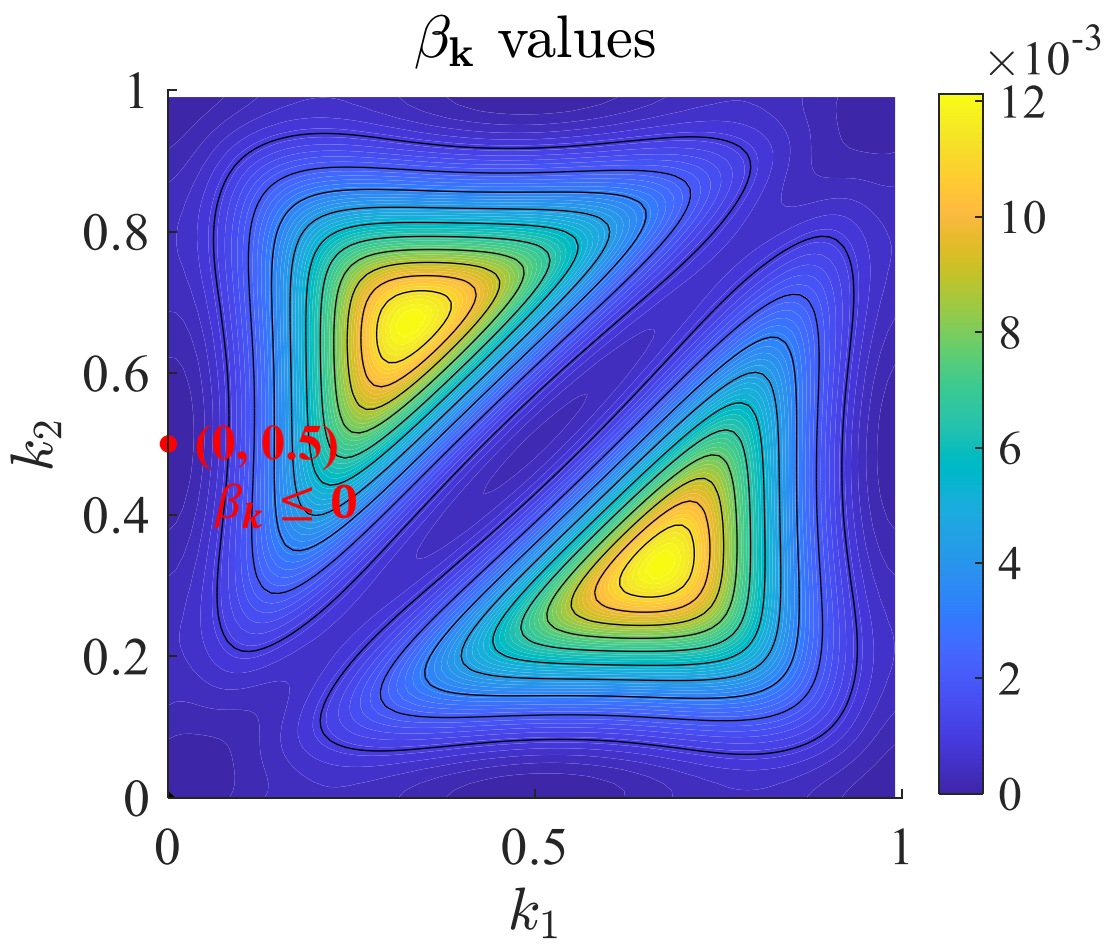

(b) Hexagon: $\lambda_c$ = 0.9270

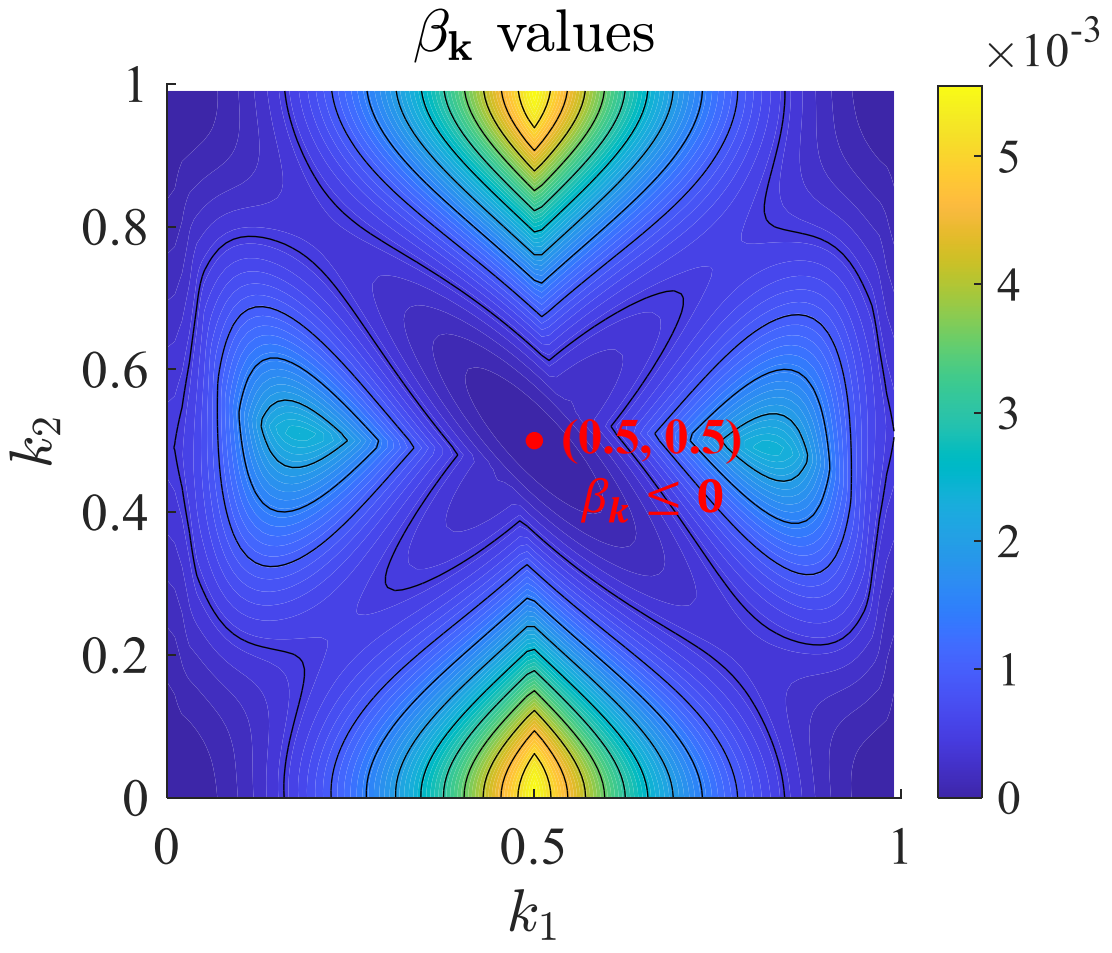

(c) Rectangle: $\lambda_c$ = 0.9270

Figure 26. Microscale stability contours at the 1st bifurcation point obtained using different RVEs.

Table 8. Rank-1 convexity analysis results at 1st bifurcation point of the metamaterial with inclined circular holes.

| RVEs | 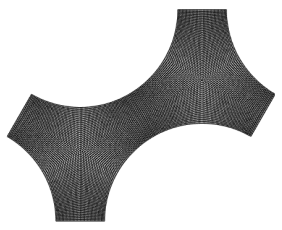 | 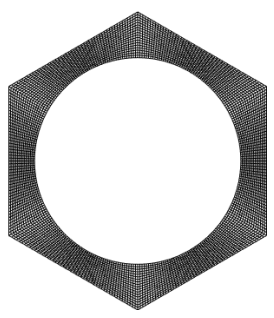 | 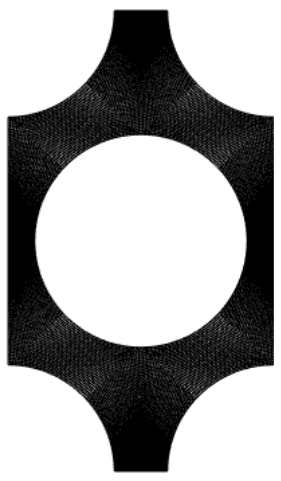 |
|---|---|---|---|
| Macroscale stability indicator $B(\lambda_c)$ | 4.0684 | 4.0743 | 4.0748 |

### 5.3 Example-3: Hyperelastic honeycomb under different stress states

A hyperelastic honeycomb metamaterial is considered in this example with the geometry and FE model details shown in Figure 27 and matrix material properties (neo-Hookean): $\kappa$ = 833.33, $\mu$ = 384.62. Similar honeycomb has been examined in [47]. The stress driven framework is adopted to study the stability performance of the honeycomb under different stress states. Three cases are examined: (1) $\phi = \pi/2$ (uniaxial compression); (2) $\phi$ = arctan(0.5); (3) $\phi = \pi/4$ (equi-biaxial compression). The principal stress orientation angle $\theta = 0°$ and the stress amplitude (also loading parameter) $\lambda$ is increasing from zero until 1st bifurcation point is identified. The second case represents that the ratio of the two principle stresses is 0.5, i.e. $\bar{\tau}_1/\bar{\tau}_2 = 0.5$. The results are shown in Figure 28. It can be seen that the three loading cases lead to different types of bifurcations – simple, double and triple bifurcations, which is consistent with the results in [47]. Next, the rank-1 convexity of the homogenized tangent moduli is checked at the 1st bifurcation point and the results are presented in Table 9, which confirms the short wavelength type buckling as $B(\lambda_c) > 0$ for these cases.

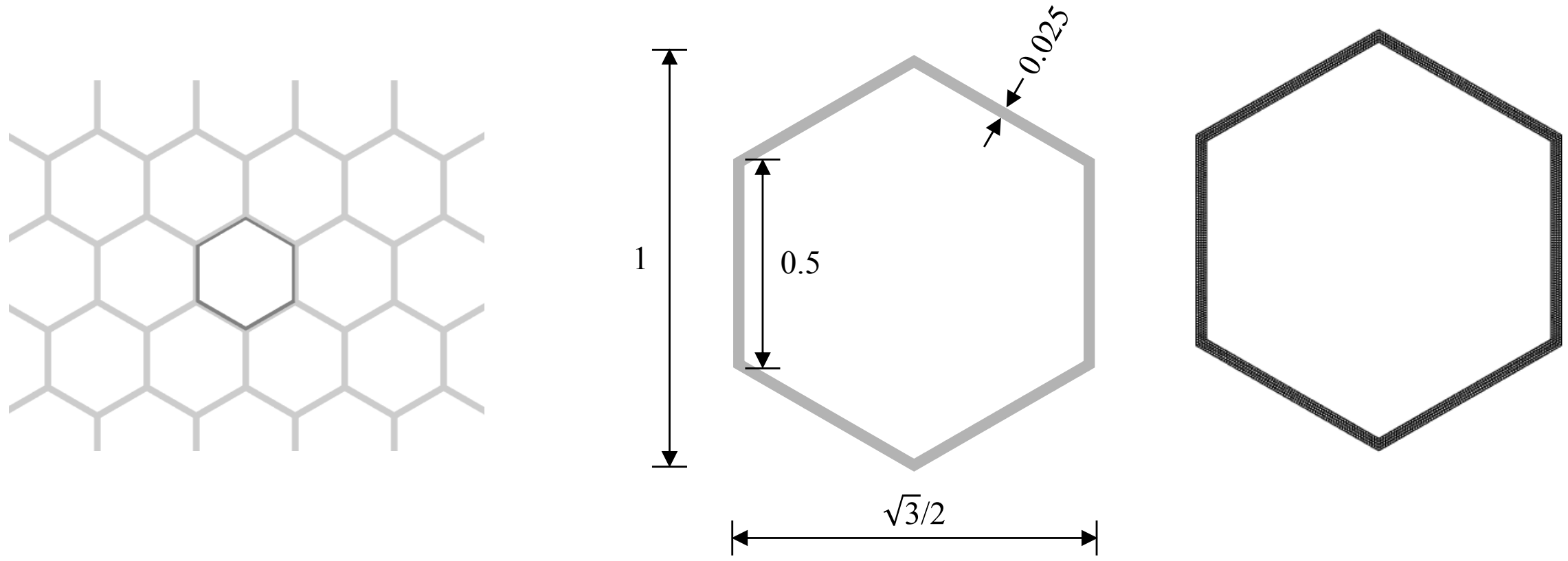

(a) Periodic composite (b) Unit cell (RVE) (c) FE mesh

Figure 27. Geometric sketch of a hyperelastic honeycomb: (a) periodic composite; (b) unit cell; (c) FE mesh.

Micro instability contour    Buckling mode(s)

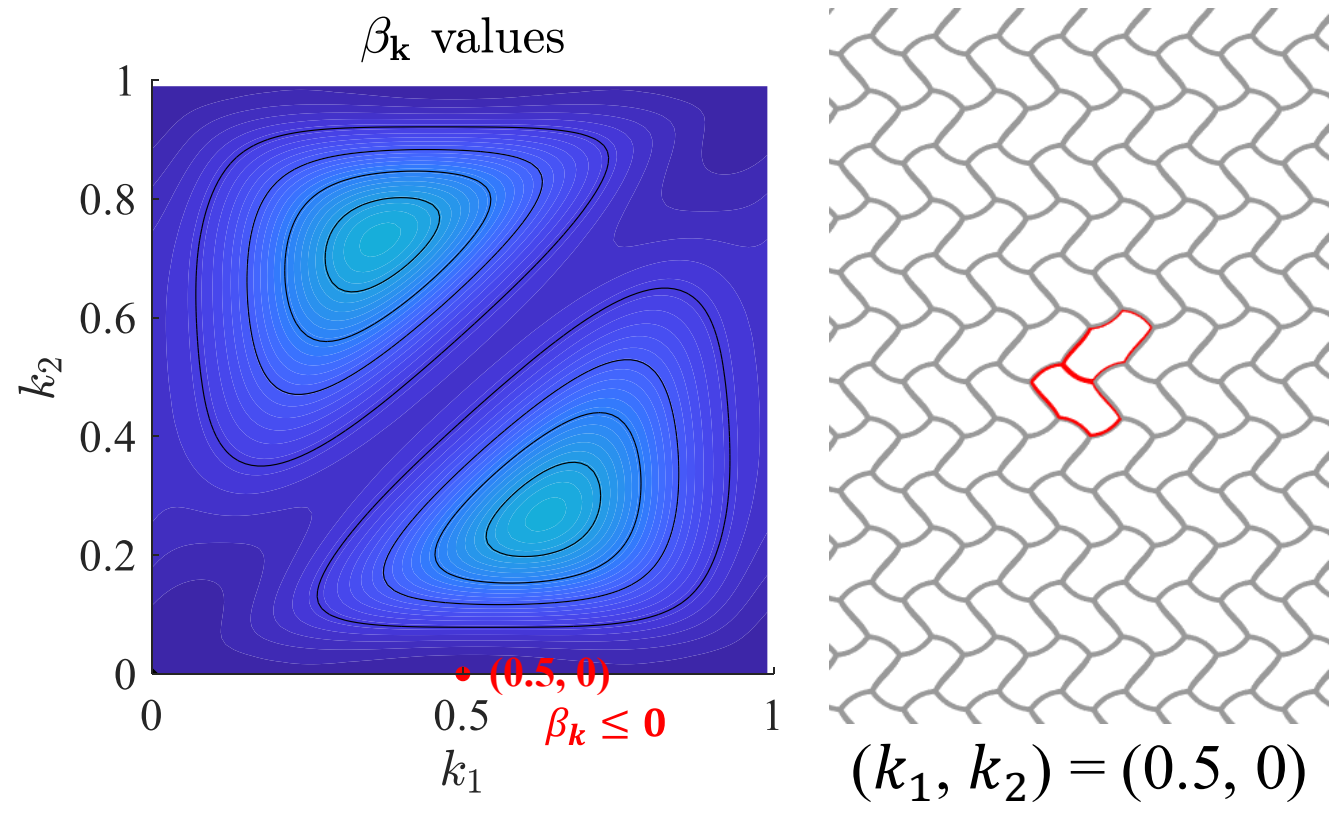

(a) $\phi = \pi/2$ (uniaxial compression), single bifurcation, $\lambda_c = 0.2177$.

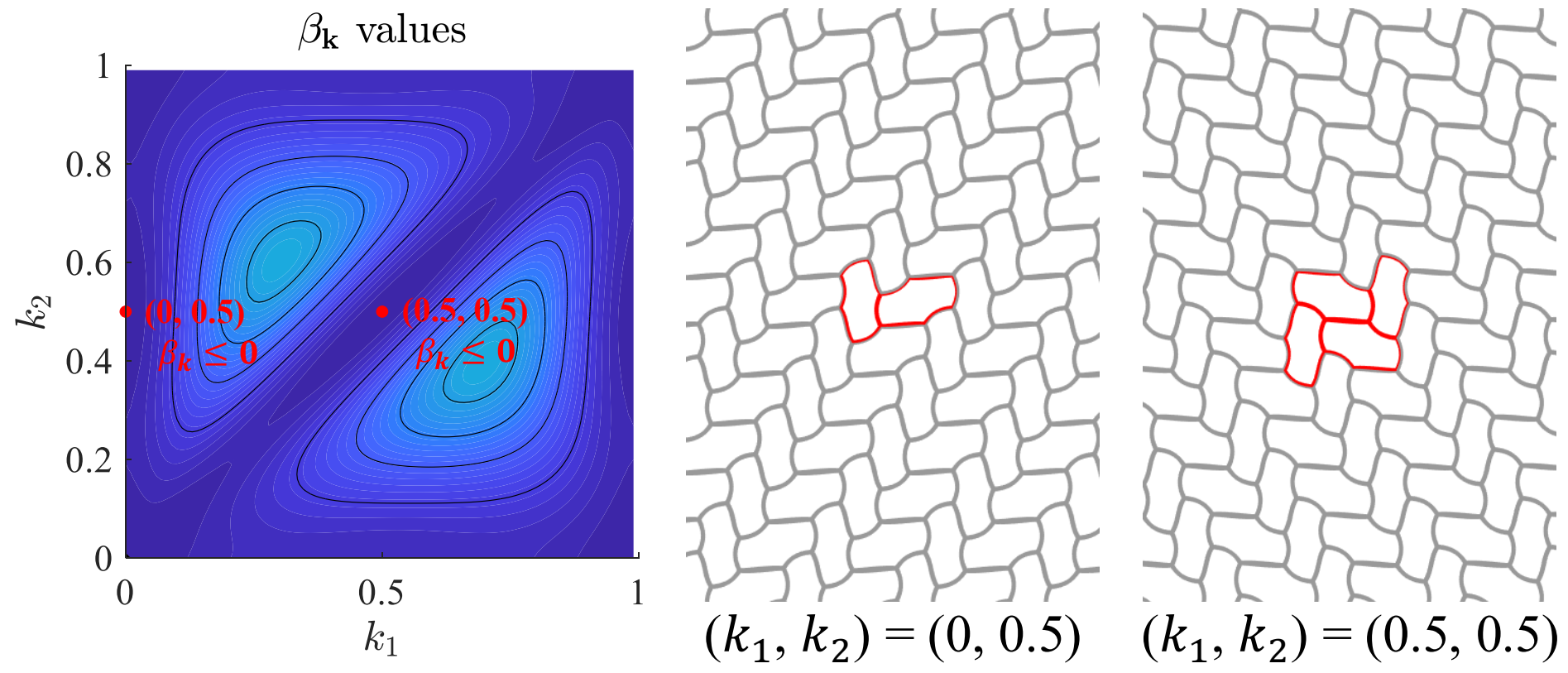

(b) $\phi = \arctan(0.5)$, double bifurcation, $\lambda_c = 0.2650$.

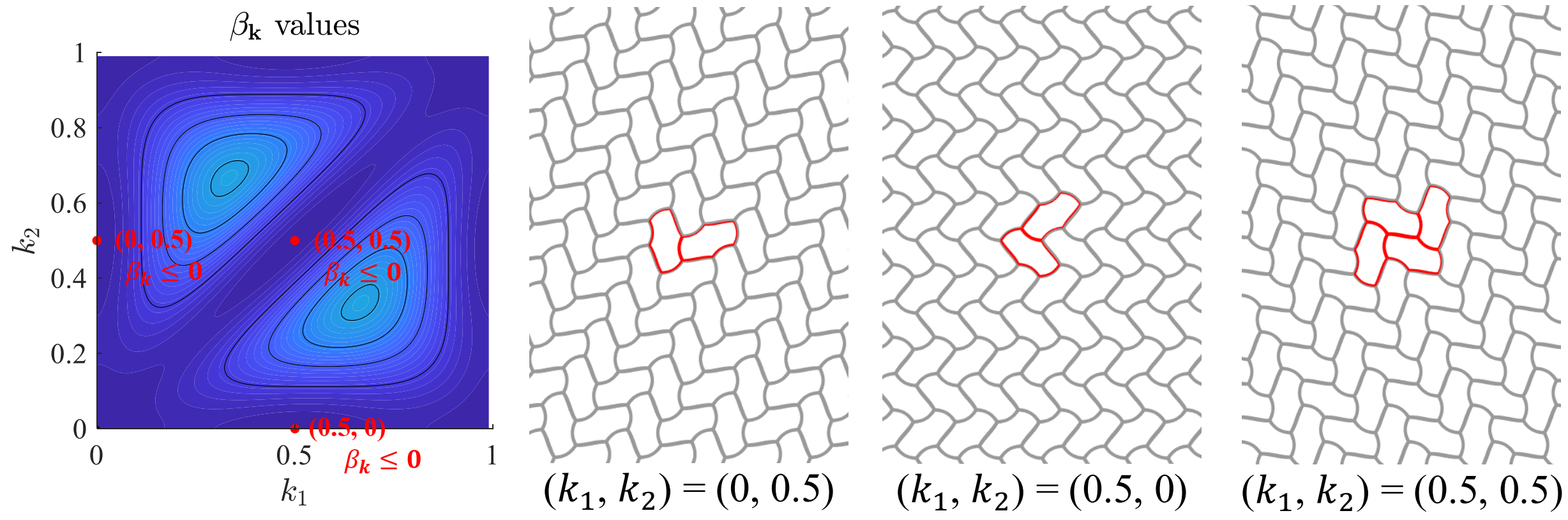

(c) $\phi = \pi/4$ (equi-biaxial compression), triple bifurcation, $\lambda_c = 0.2820$.

Figure 28. Microscale instability contour plots and buckling modes of the three cases under stress driven condition. Note: The principal Kirchhoff stresses $\bar{\tau}_i$ are $\lambda_c \sin\phi$ and $\lambda_c \cos\phi$.

Table 9. Rank-1 convexity analysis results at 1st bifurcation point of the hyperelastic honeycomb metamaterial.

| 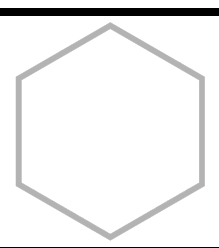 | $\phi = \pi/2$ | $\phi = \arctan(0.5)$ | $\phi = \pi/4$ |
|---|---|---|---|
| Macroscale stability indicator $B(\lambda_c)$ | 0.2551 | 0.2559 | 0.3353 |

### 5.4 Example-4: Buckling mode switching in multimaterial triangular lattice

This example is used to demonstrate how the material constituents' properties can affect the multiscale buckling behavior. A multimaterial triangular lattice-like metamaterial is used with geometry and FE mesh details shown in Figure 29. Uniaxial compression along an inclined angle (-30°) is considered with stress driven approach, i.e. $\theta$ = -30° and $\phi = \pi/2$ in Eqns. (32) and (35). The metamaterial comprises two materials with material-1 plotted as red color and material-2 plotted as blue color. The parameters of material-1 are fixed as $\kappa_1 = 833.33$ and $\mu_1 = 384.62$. Three cases are considered: $(\kappa_2, \mu_2) = 0.1(\kappa_1, \mu_1)$, $(\kappa_2, \mu_2) = (\kappa_1, \mu_1)$ and $(\kappa_2, \mu_2) = 10(\kappa_1, \mu_1)$, and the Bloch wave analysis results are shown in Figure 30. As can be seen, all the buckling wavelengths are short. However, increasing material-2 properties from $0.1(\kappa_1, \mu_1)$ to $10(\kappa_1, \mu_1)$ changes the microscale buckling mode from 2×2 to 1×2. For $(\kappa_2, \mu_2) = (\kappa_1, \mu_1)$, two aperiodic buckling modes (Figure 30b) are found based on the employed $\boldsymbol{k}$-space mesh, i.e. double bifurcation point. The rank-1 convexity results shown in Table 10 again confirms the short wavelength type buckling for all these cases as $B(\lambda_c) > 0$.

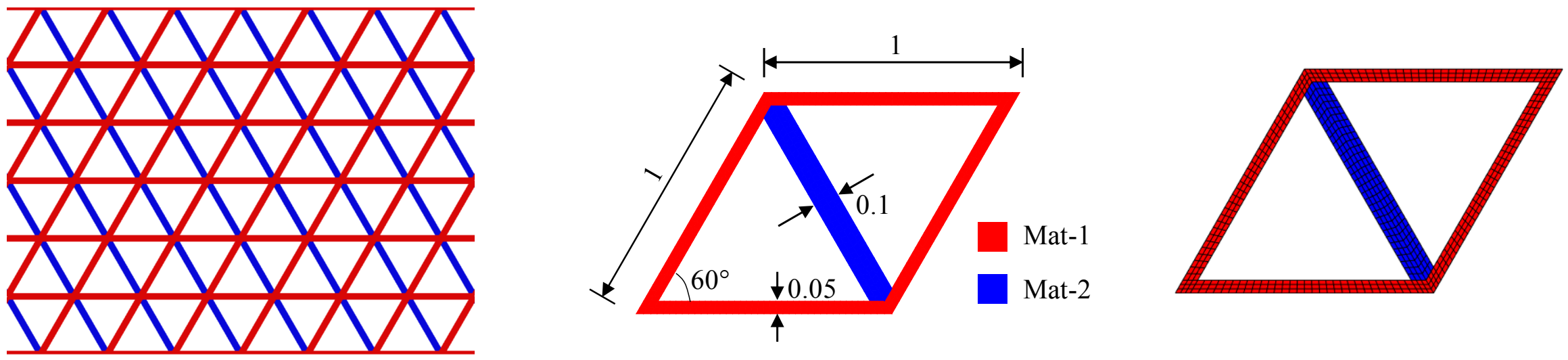

(a) Periodic composite (b) Unit cell (RVE) (c) FE mesh

Figure 29. Geometric sketch of a multimaterial equilateral triangular lattice metamaterial: (a) periodic composite; (b) unit cell; (c) FE mesh.

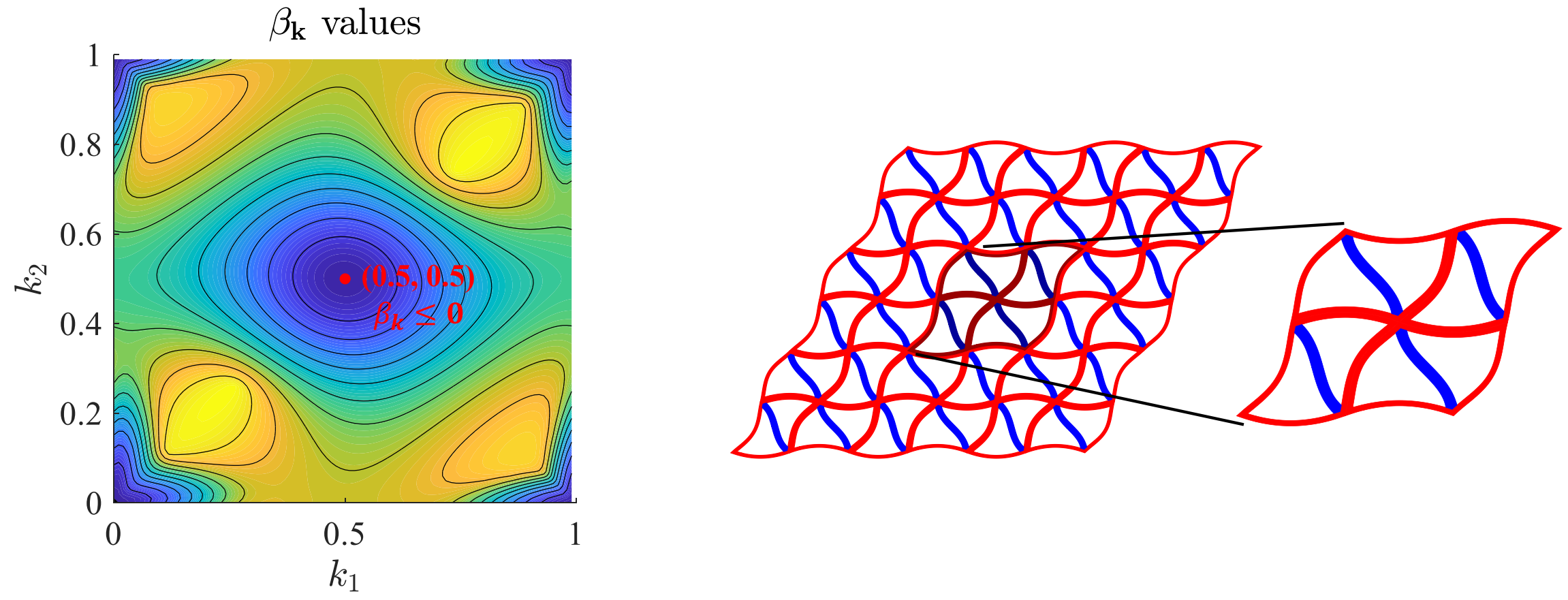

(a) $(\kappa_2, \mu_2) = 0.1(\kappa_1, \mu_1)$: $\lambda_c = 2.2464$, $(k_1, k_2) = (0.5, 0.5)$, periodic buckling mode (2×2)

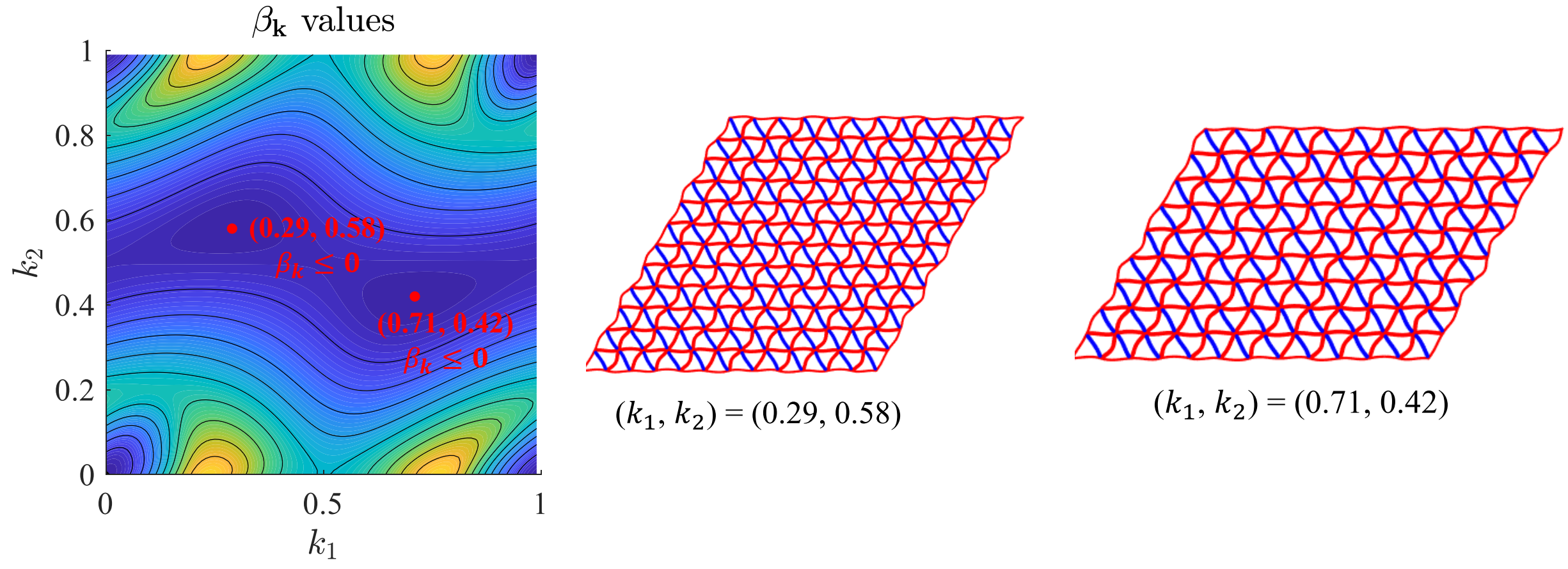

(b) $(\kappa_2, \mu_2) = (\kappa_1, \mu_1)$: $\lambda_c = 3.6160$ , two aperiodic buckling modes (only a small portion of buckling mode are plotted)

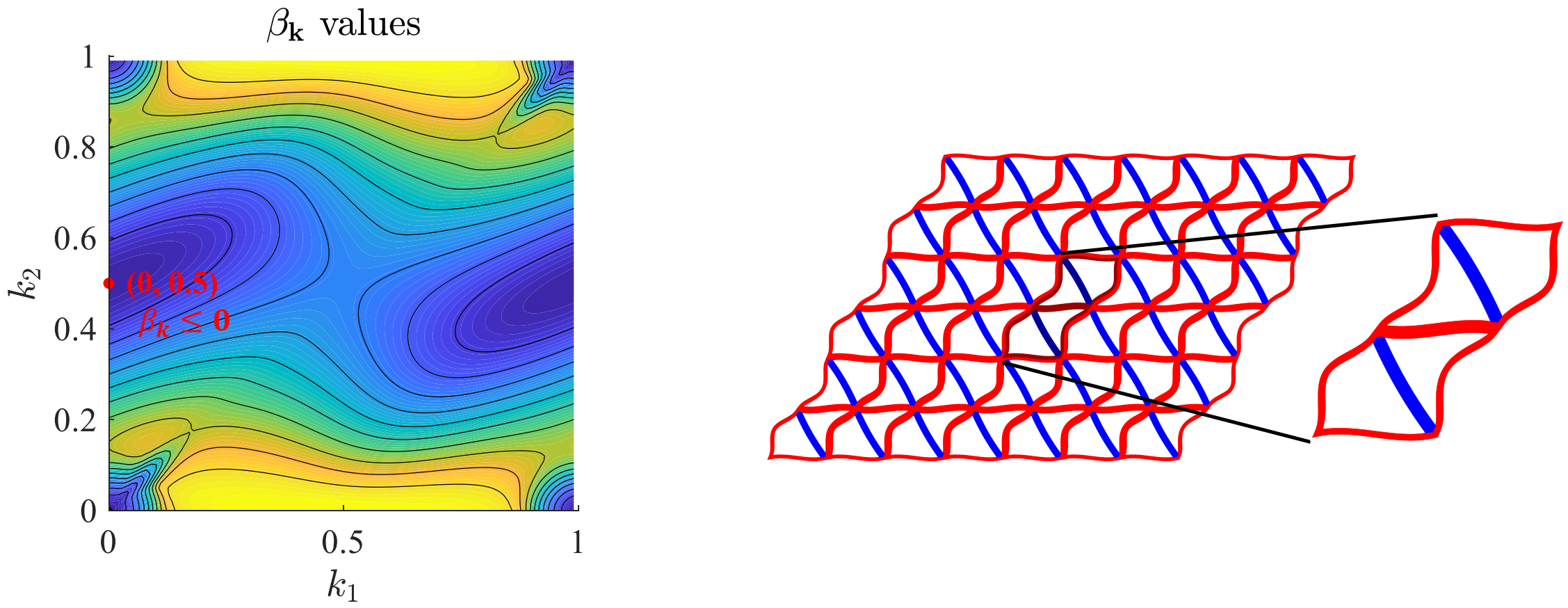

(c) $(\kappa_2, \mu_2) = 10(\kappa_1, \mu_1)$: $\lambda_c = 4.7109$, $(k_1, k_2) = (0, 0.5)$, periodic buckling mode (1×2)

Figure 30. Buckling modes of the triangular lattice obtained with different properties of material-2 (blue color).

Table 10. Rank-1 convexity analysis results at 1st bifurcation point of a triangular lattice with different properties of material-2 (blue color).

| 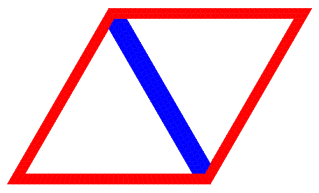 | $(\kappa_2, \mu_2) = 0.1(\kappa_1, \mu_1)$ | $(\kappa_2, \mu_2) = (\kappa_1, \mu_1)$ | $(\kappa_2, \mu_2) = 10(\kappa_1, \mu_1)$ |
|---|---|---|---|
| Macroscale stability indicator $B(\lambda_c)$ | 6.7276 | 46.7195 | 51.7206 |

### *5.5 Example-5: Highly localized buckling mode*

This example intends to show a different short wavelength buckling mode, i.e. a mode which has the same periodicity as the original microstructure. The geometry of the metamaterial's microstructure and the FE mesh of the RVE including one unit cell are shown in Figure 31. The matrix material follows neo-Hookean hyperelasticity with $\kappa$ = 166.67 and $\mu$ = 35.71. In this example, a strain driven loading is considered (Eq. (21)) where the macroscopic deformation gradient is parameterized by $\overline{\boldsymbol{F}} = \lambda \cdot [1 \quad 0 \quad 0 \quad 1]^T$ in which the loading factor $\lambda$ is decreasing from 1, representing biaxial compression. Figure 32 shows the microscopic stability surface at the 1st bifurcation point ($\lambda_c$ = 0.9934) where the stability indicator $\beta_{\boldsymbol{k}}$ is plotted against the wavevectors $\boldsymbol{k}$ in the primitive cell in reciprocal space. As can be seen, the surface at the origin $(k_1, k_2) = (0, 0)$ is continuous, i.e. the origin point is *not singular*, and the minimum value is indeed achieved at the origin. With the chosen $\boldsymbol{k}$ mesh, $\beta_{\boldsymbol{k}}$ is negative only at origin while remaining positive at other points on the surface (Figure 32b). This result is in contrast with that in Figure 22, where origin was a discontinuous point in the $\boldsymbol{k}$ space. This means that the buckling mode at the 1st bifurcation point is 1×1 mode, i.e. periodic with respect to one unit cell. The buckling mode at the 1st bifurcation load is plotted in Figure 33. The rank-1 convexity analysis result at $\lambda_c$, $B(\lambda_c)$ = 5.9469, which indicates macroscale stability.

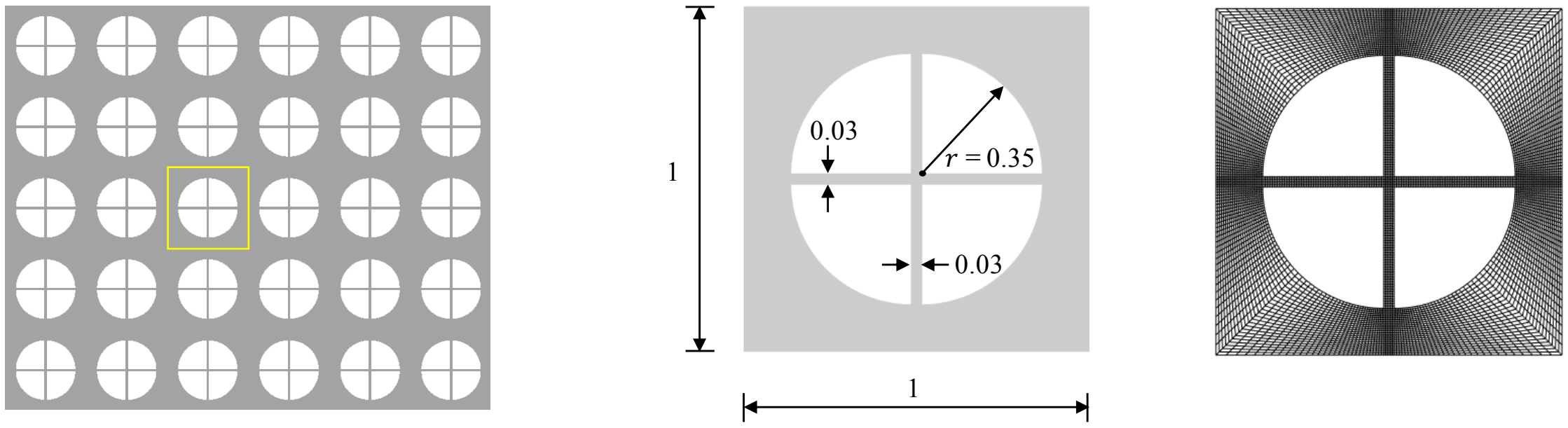

(a) Periodic composite (b) Unit cell (RVE) (c) FE mesh

Figure 31. Geometric sketch of a metamaterial with circular holes and inner hole cross-bars: (a) periodic metamaterial; (b) unit cell; (c) FE mesh.

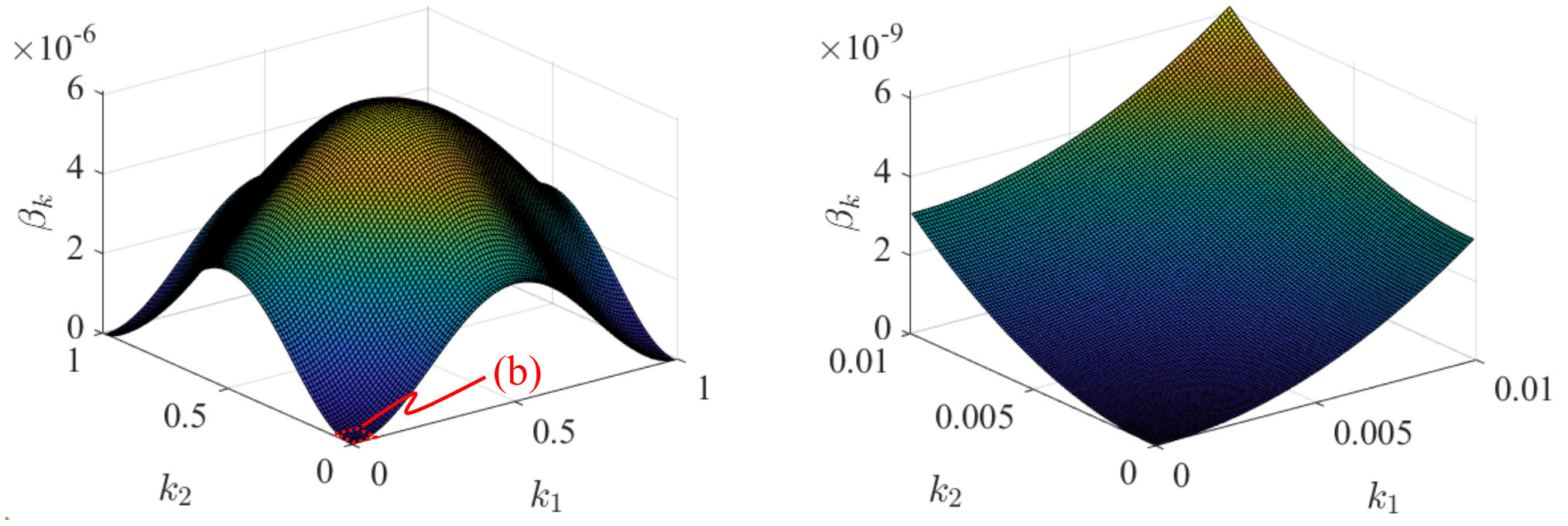

(a) Uniform mesh in [0,1) × [0,1) (b) Uniform mesh in [0,0.01) × [0,0.01)

Figure 32. Microscale stability surface at the 1st bifurcation point $\lambda_c$ = 0.9934.

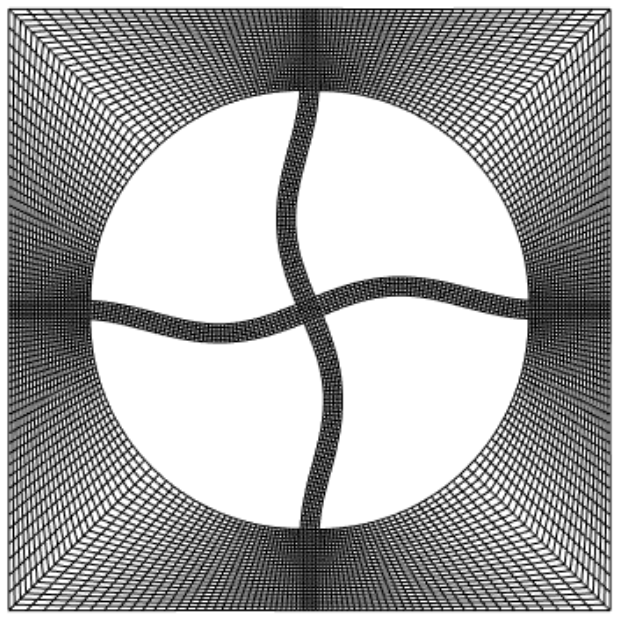

Figure 33. Buckling mode (1×1) at the 1st bifurcation point $\lambda_c$ = 0.9934.

### *5.6 Example-6: Long wavelength buckling*

This example serves the purpose of demonstrating a long wavelength buckling case. The periodic metamaterial under examination is shown in Figure 34 with FE mesh of the RVE. The material's properties follow neo-Hookean model with $\kappa$ = 166.67 and $\mu$ = 35.71. The metamaterial is under a constrained compression with macroscopic deformation gradient $\bar{\boldsymbol{F}} = [1 \quad 0 \quad 0 \quad \lambda]^T$ where $\lambda$ is decreasing from one. With strain driven approach, the 1st bifurcation point is detected with both Bloch wave analysis and rank-1 convexity check. From stretch ratio ($\lambda$) 0.98244 to 0.98243, both the Bloch functional indicator $\beta(\lambda)$ in Eq. (44) and the macroscopic instability indicator $B(\lambda)$ in Eq. (54) changes sign from positive to negative indicating the presence of a bifurcation point. The Bloch functional surface $\beta_{\boldsymbol{k}}$ at $\lambda_c$ = 0.98243 is plotted in Figure 35, while the macroscale stability curves by rank-1 convexity analysis are shown in Figure 36 for the two adjacent loading steps where $B_\alpha \overset{\text{def}}{=} \min_{\bar{\boldsymbol{m}}}(\bar{\boldsymbol{m}} \otimes \bar{\boldsymbol{M}}): \bar{\mathbb{A}}: (\bar{\boldsymbol{m}} \otimes \bar{\boldsymbol{M}})$ with $\bar{\boldsymbol{M}} = [\cos\alpha \quad \sin\alpha]^T$ and $\alpha \in [0, \pi)$. It can be seen that long wavelength buckling can be equivalently detected from both the Bloch wave analysis and the examination of rank-1 convexity of the homogenized tangent moduli. This result is in contrast with other cases where rank-1 convexity (macro stability) is preserved at the onset of first bifurcation associated with microscale buckling with finite wavelengths.

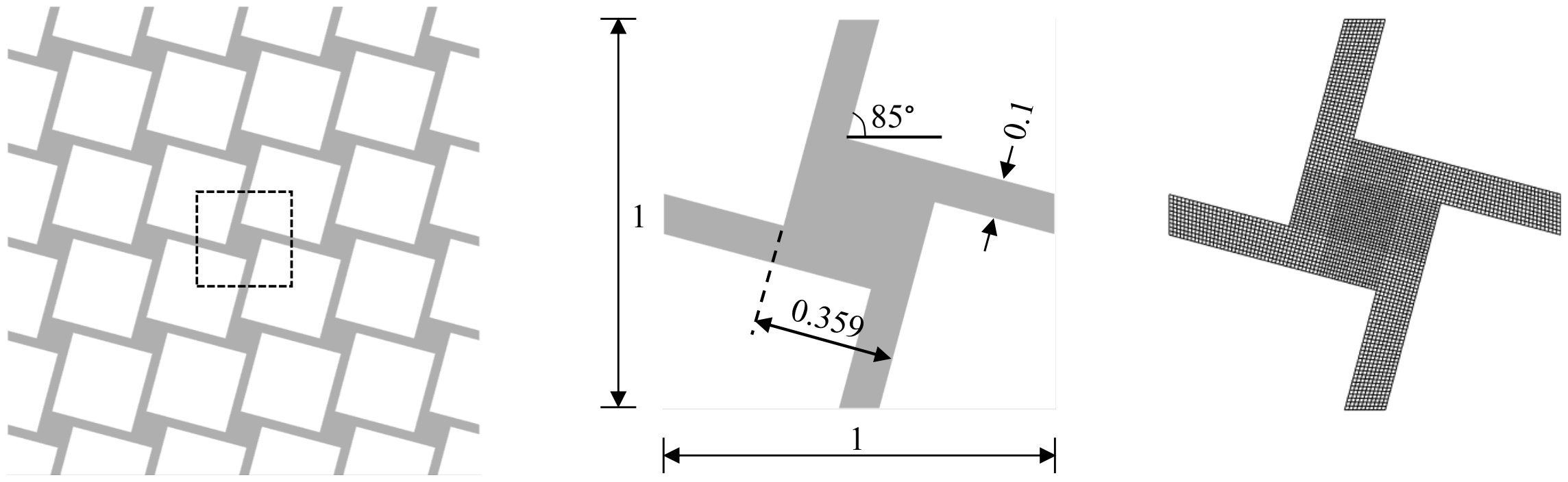

(a) Periodic metamaterial (b) Unit cell (RVE) (c) FE mesh

Figure 34. Geometric sketch of a metamaterial with rotated square-shaped voids: (a) periodic metamaterial; (b) unit cell; (c) FE mesh.

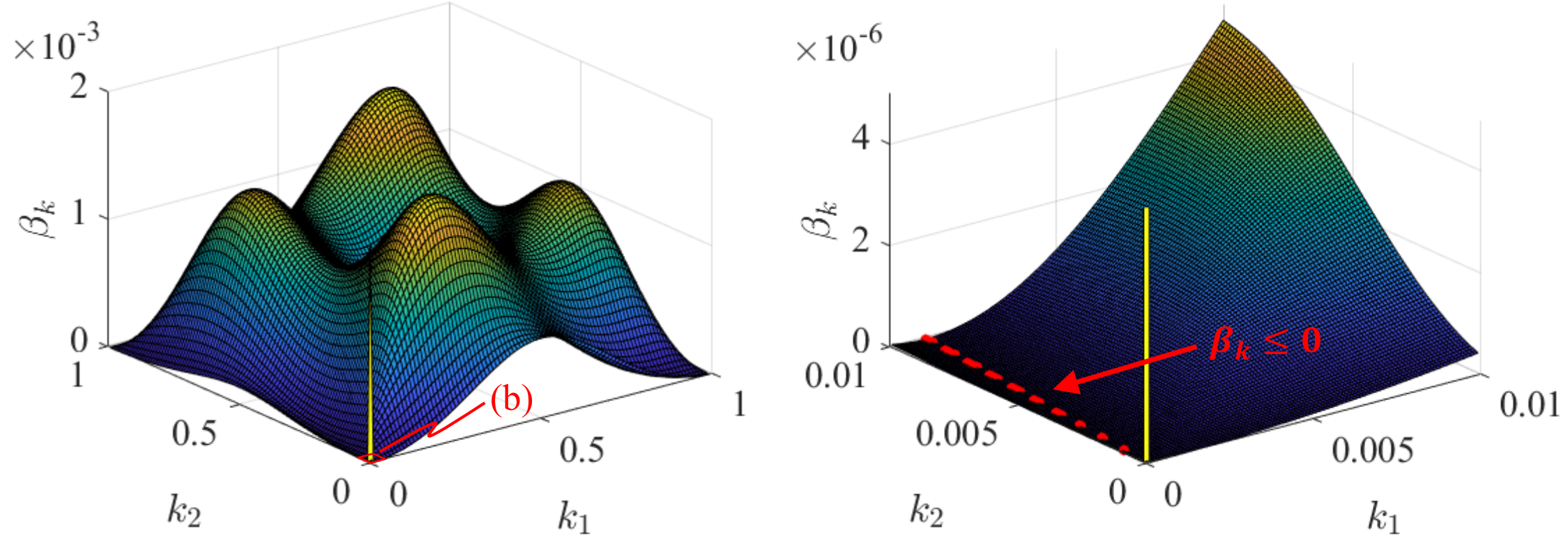

(a) Uniform mesh in [0,1) × [0,1) (b) Uniform mesh in [0,0.01) × [0,0.01)

Figure 35. Microscale stability surface at the 1st bifurcation point $\lambda_c$ = 0.98243.

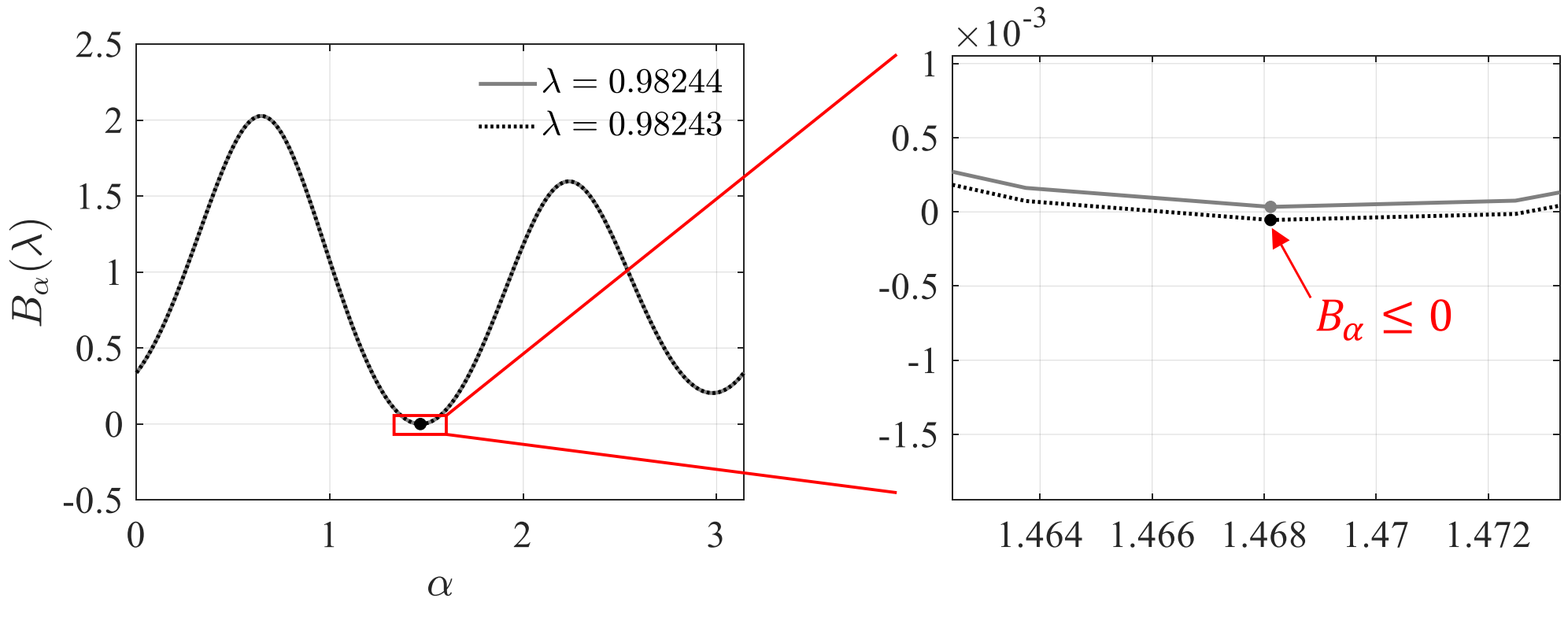

Figure 36. Macroscale stability curves from rank-1 convexity analysis at before and after 1st bifurcation points.

### 5.7 Example-7: Elastoplastic honeycomb under different loading orientations

The last example considers a honeycomb metamaterial with the underlying material constituent following the finite strain $J_2$ plasticity model with parameters $\kappa$ = 17.5, $\mu$ = 8.0, $\sigma_y$ = 0.45 and $K^p$ = 0.1 (Appendix B). Constrained compression under strain driven loading is considered with three loading orientations: $\theta$ = 0° (vertical), 45° (inclined) and 90° (horizontal), see Eq. (82). The geometry and FE mesh details are shown in Figure 37. Same as the example in Section 3.2, the mixed u/p (9/3) element formulation is used to address the locking issue. As has been studied in [37] where the honeycomb lattice is discretized using elastoplastic beam elements, the 1st bifurcation point depends on the loading conditions. With the 2D plane strain formulation, the Bloch stability analysis results for different macroscopic stretch orientations are given in Figure 38, where different short wavelength buckling modes are present. All are single bifurcation points. In addition, the critical load magnitude also varies according to the direction of applied loading. Figure 39 shows the deformed shapes and equivalent plastic strain ($\alpha$) distributions in the RVE at the 1st bifurcation point under different loading conditions. These results demonstrate that the inelasticity occurs before bifurcation in all these cases. Finally, the rank-1 convexity results are given in Table 11, which indicates that the macroscale stability is preserved.

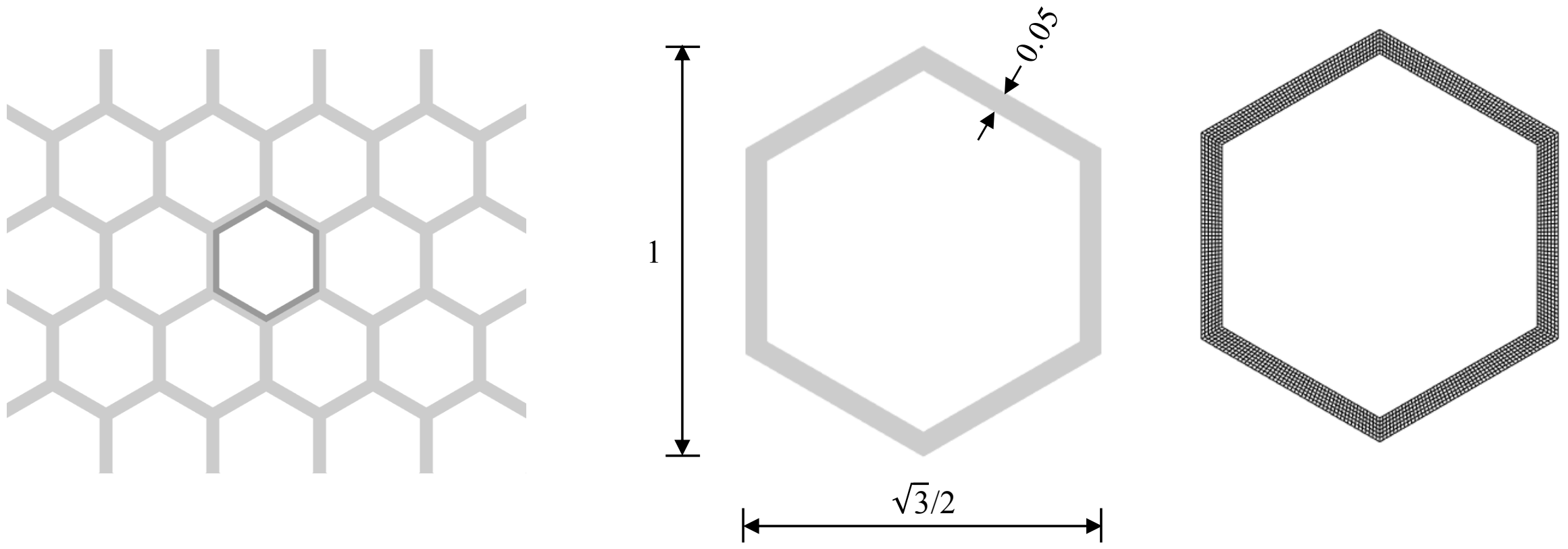

(a) Periodic composite (b) Unit cell (RVE) (c) FE mesh

Figure 37. Geometric sketch of an elastoplastic honeycomb: (a) periodic composite; (b) unit cell; (c) FE mesh.

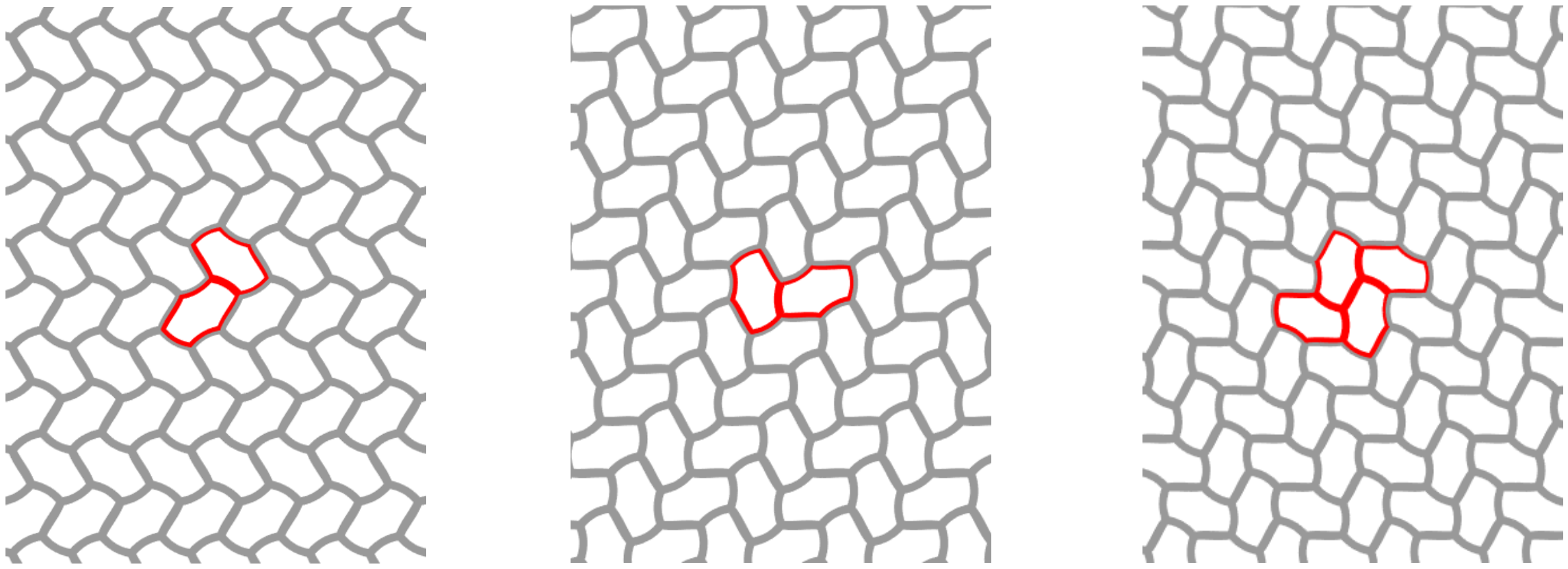

(a) $\theta$ = 0°, $\lambda_c$ = 0.9783, ($k_1$, $k_2$) = (0.5, 0) (a) $\theta$ = 45°, $\lambda_c$ = 0.9783 ($k_1$, $k_2$) = (0, 0.5) (a) $\theta$ = 90°, $\lambda_c$ = 0.9780 ($k_1$, $k_2$) = (0.5, 0.5)

Figure 38. First bifurcation loads and their corresponding buckling modes of elastoplastic honeycomb under different stretch orientations.

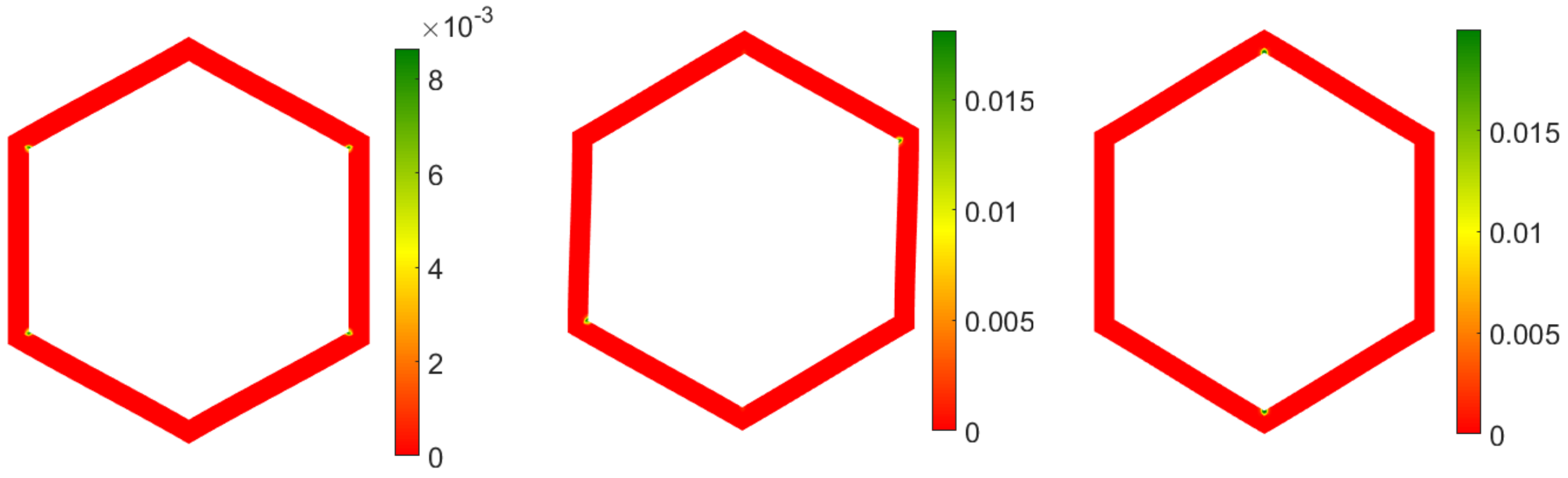

(a) $\theta$ = 0° (vertical)　(b) $\theta$ = 45° (inclined)　(c) $\theta$ = 90° (horizontal)

Figure 39. Deformed shapes and equivalent plastic strain ($\alpha$) distributions of the RVE at the 1st bifurcation point (along the principal branch) under different stretch orientations: (a) $\theta$ = 0°; (b) $\theta$ = 45°; (c) $\theta$ = 90°.

Table 11. Rank-1 convexity analysis results at 1st bifurcation point of elastoplastic honeycomb under different loading orientations.

| Principal macro stretch orientation | $\theta$ = 0° (vertical) | $\theta$ = 45° (inclined) | $\theta$ = 90° (horizontal) |
|---|---|---|---|
| Macroscale stability indicator $B(\lambda_c)$ | 0.0472 | 0.0437 | 0.0421 |

## 6 Conclusions

In this study, a consistent computational framework is proposed for both multiscale homogenization and micro/macro stability analyses. The homogenization method is verified through hyperelastic and elastoplastic periodic metamaterials test cases where the invariance of the homogenization results with respect to RVEs of different sizes and shapes is shown. The multiscale stability analysis with Bloch wave formulation is detailed which includes the selection of wave vector space and retrieval of the real-valued buckling mode from the complex-valued Bloch wave representations. Three methods for the resulted constrained eigenvalue problem are laid out with implementation details and the equivalence of the three methods is demonstrated. Besides, the validity of the stability analysis framework is illustrated by the equivalency of different RVEs in the stability results – 1st bifurcation load and the corresponding buckling mode. Several numerical examples including both hyperelastic and elastoplastic matrix materials are carried out that show various types of buckling modes – short wavelength buckling mode across over either multiple unit cells or one unit cell and long wavelength buckling mode of infinite length w.r.t. unit cell size. In accordance with the theoretical results [22], the numerical results also show that microscopic stability implies macroscopic stability that can be evaluated by the rank-1 convexity of the homogenized tangent moduli. Therefore, the rank-1 convexity check provides an upper bound on the critical load. In addition, the examples also show that the buckling mode can be tuned by changing the underlying matrix material properties. Moreover, in line with the previous studies [25], the dependence of bifurcation load on the load orientation is also demonstrated. It is important to note that the multiscale stability analysis presented in this study gives the 1st bifurcation load beyond which the homogenization results lose validity. To further understand the properties of the metamaterial of interests after the onset of bifurcation, e.g. energy absorption, band-gaps, etc., post bifurcation analysis may be needed which will be the future work.

## Acknowledgements

The presented work is supported in part by the US National Science Foundation through grant CMMI-1762277. Any opinions, findings, conclusions, and recommendations expressed in this paper are those of the authors and do not necessarily reflect the views of the sponsors.

## Appendix A: Equivalence of $\mathcal{V}_{min}$ and constant traction boundary condition

To prove that the minimum set of kinematical admissibility in Eq. (6) corresponds to constant traction boundary condition, consider the principle of multiscale virtual power with the constraints in Eq. (3) enforced by the Lagrange multipliers, i.e.

$$
\begin{aligned}
&-(\overline{\boldsymbol{P}}:\delta\overline{\boldsymbol{F}})+\frac{1}{V}\int_{\mathcal{B}_0}\boldsymbol{P}:\delta\boldsymbol{F}dV-\delta\boldsymbol{\lambda}.\int_{\mathcal{B}_0}\boldsymbol{u}(\boldsymbol{X},t)dV-\boldsymbol{\lambda}.\int_{\mathcal{B}_0}\delta\boldsymbol{u}(\boldsymbol{X},t)dV\\
&\qquad-\delta\boldsymbol{\mu}:\left[\boldsymbol{I}+\frac{1}{V}\int_{\partial\Omega_0^{\mu}}\boldsymbol{u}(\boldsymbol{X},t)\otimes\boldsymbol{N}(\boldsymbol{X})dS-\overline{\boldsymbol{F}}\right]\\
&\qquad-\boldsymbol{\mu}:\left[\frac{1}{V}\int_{\partial\Omega_0^{\mu}}\delta\boldsymbol{u}(\boldsymbol{X},t)\otimes\boldsymbol{N}(\boldsymbol{X})dS-\delta\overline{\boldsymbol{F}}\right]=0\\
&\qquad\qquad\forall\ \delta\overline{\boldsymbol{F}},\delta\boldsymbol{\mu}\in\text{Lin},\ \ \delta\boldsymbol{u}\in H^1(\mathcal{B}_0),\ \ \delta\boldsymbol{\lambda}
\end{aligned}
\tag{A 1}
$$

where $\boldsymbol{\lambda}$ and $\boldsymbol{\mu}$ are the Lagrange multipliers.

Assuming $\delta\overline{\boldsymbol{F}}=\delta\boldsymbol{\mu}=\boldsymbol{0}$ and $\delta\boldsymbol{\lambda}=\boldsymbol{0}$ in Eq. (A 1) gives

$$
\begin{aligned}
&\frac{1}{V}\int_{\mathcal{B}_0}\boldsymbol{P}:\delta\boldsymbol{F}dV-\boldsymbol{\lambda}.\int_{\mathcal{B}_0}\delta\boldsymbol{u}(\boldsymbol{X},t)dV-\boldsymbol{\mu}:\left[\frac{1}{V}\int_{\partial\Omega_0^{\mu}}\delta\boldsymbol{u}(\boldsymbol{X},t)\otimes\boldsymbol{N}(\boldsymbol{X})dS\right]=0\\
&\qquad\qquad\forall\ \delta\boldsymbol{u}\in H^1(\mathcal{B}_0)
\end{aligned}
\tag{A 2}
$$

It can be shown that

$$\boldsymbol{P}:\delta\boldsymbol{F}=\boldsymbol{P}:\boldsymbol{\nabla_X}\delta\boldsymbol{u}=\boldsymbol{\nabla_X}.(\boldsymbol{P}^T.\delta\boldsymbol{u})-(\boldsymbol{\nabla_X}.\boldsymbol{P}).\delta\boldsymbol{u} \tag{A 3}$$

Substituting Eq. (A 3) in Eq. (A 2) gives

$$
\begin{aligned}
&\frac{1}{V}\int_{\mathcal{B}_0}\boldsymbol{\nabla_X}.(\boldsymbol{P}^T.\delta\boldsymbol{u})dV-\frac{1}{V}\int_{\mathcal{B}_0}(\boldsymbol{\nabla_X}.\boldsymbol{P}).\delta\boldsymbol{u}dV-\boldsymbol{\lambda}.\int_{\mathcal{B}_0}\delta\boldsymbol{u}dV-\boldsymbol{\mu}:\left[\frac{1}{V}\int_{\partial\Omega_0^{\mu}}\delta\boldsymbol{u}\otimes\boldsymbol{N}dS\right]\\
&\qquad=0\quad\forall\ \delta\boldsymbol{u}\in H^1(\mathcal{B}_0)
\end{aligned}
\tag{A 4}
$$

where the dependence on $\boldsymbol{X}$ and $t$ is omitted for the sake of notational simplicity. Using divergence theorem on the first term in Eq. (A 4) gives

$$\frac{1}{V}\int_{\partial\Omega_0^\mu}(\boldsymbol{P}.\boldsymbol{N}).\delta\boldsymbol{u}dS-\boldsymbol{\lambda}.\int_{\mathcal{B}_0}\delta\boldsymbol{u}dV-\boldsymbol{\mu}:\left[\frac{1}{V}\int_{\partial\Omega_0^\mu}\delta\boldsymbol{u}\otimes\boldsymbol{N}dS\right]=0\quad\forall\,\delta\boldsymbol{u}\in H^1(\mathcal{B}_0)\qquad\text{(A 5)}$$

where the fact $\boldsymbol{\nabla}_{\boldsymbol{X}}.\boldsymbol{P}=\boldsymbol{0}$ is used together with the traction-free condition on the void boundaries, i.e. $\boldsymbol{P}.\boldsymbol{N}=\boldsymbol{0}$ on $\partial\mathcal{H}_0$. Next, combining similar terms further gives

$$\frac{1}{V}\int_{\partial\Omega_0^\mu}(\boldsymbol{P}.\boldsymbol{N}-\boldsymbol{\mu}.\boldsymbol{N}).\delta\boldsymbol{u}dS-\boldsymbol{\lambda}.\int_{\mathcal{B}_0}\delta\boldsymbol{u}dV=0\quad\forall\,\delta\boldsymbol{u}\in H^1(\mathcal{B}_0)\qquad\text{(A 6)}$$

which, by the arbitrariness of $\delta\boldsymbol{u}$, leads to the requirement that the tractions on the boundary $\partial\Omega_0^\mu$ should satisfy

$$\boldsymbol{t}_0=\boldsymbol{P}.\boldsymbol{N}=\boldsymbol{\mu}.\boldsymbol{N}\quad\text{on }\partial\Omega_0^\mu\qquad\text{(A 7)}$$

Since the Lagrange multiplier $\boldsymbol{\mu}$ is a constant 2nd order tensor throughout the entire domain of RVE, it is clear that the tractions on the opposite boundaries are equal valued with opposite directions. Moreover, it is straightforward to show that $\boldsymbol{\mu}=\overline{\boldsymbol{P}}$ by assuming $\delta\boldsymbol{u}(\boldsymbol{X})=\boldsymbol{A}_0.\boldsymbol{X}$ in $\mathcal{B}_0$ with $\boldsymbol{A}_0\in$ Lin using again divergence theorem on Eq. (A 6). Therefore, the tractions on the boundary can be expressed as

$$\boldsymbol{t}_0=\overline{\boldsymbol{P}}.\boldsymbol{N}\quad\text{on }\partial\Omega_0^\mu\qquad\text{(A 8)}$$

## Appendix B: Finite strain $J_2$ plasticity model

Table B 1 gives the finite strain $J_2$ plasticity model used for the elastoplastic metamaterial analyses. The multiplicative split of the deformation gradient is assumed, i.e. $\boldsymbol{F} = \boldsymbol{F}^e.\boldsymbol{F}^p$ where $\boldsymbol{F}^e$ and $\boldsymbol{F}^p$ represent the elastic and plastic part of the deformation gradient, respectively. Linear isotropic hardening is considered. Following is the definition of symbols that are used in Table B 1.

$\boldsymbol{\tau}$ – Kirchhoff stress tensor

$\boldsymbol{b}^e$ – Elastic Finger tensor

$\lambda_i^e$ – $i$th elastic principal stretch

$\alpha$ – Equivalent plastic strain

$\sigma_y$ – Initial yield stress

$K^p$ – Hardening modulus

$\kappa, \mu$ – Bulk and shear modulus

Table B 1. Finite strain $J_2$ plasticity model.

| | |
|---|---|
| Yield function: | $\phi(\boldsymbol{\tau},\alpha) = \lVert\boldsymbol{\tau}_{dev}\rVert - \sqrt{\frac{2}{3}}\zeta(\alpha)$ |
| where | |
| | $\boldsymbol{\tau}_{dev} = \mathbb{P}^s_{dev}:\boldsymbol{\tau}$ with $(\mathbb{P}^s_{dev})_{ijkl} = \frac{1}{2}\left(\delta_{ik}\delta_{jl} + \delta_{il}\delta_{jk}\right) - \delta_{ij}\delta_{kl}$ |
| | $\zeta(\alpha) = \sigma_y + K^p\alpha$ |
| Free energy: | $\psi^e(\boldsymbol{b}^e) = \psi^e_{vol}(J^e) + \psi^e_{iso}(\widehat{\boldsymbol{b}}^e)$ |
| | $\psi^e_{vol}(J^e) = \frac{1}{2}\kappa(\ln J^e)^2 = \frac{1}{2}\kappa(\varepsilon_1^e + \varepsilon_2^e + \varepsilon_3^e)^2$ |
| | $\psi^e_{iso}(\widehat{\boldsymbol{b}}^e) = \frac{\mu}{4}\boldsymbol{I}:\left(\ln\widehat{\boldsymbol{b}}^e\right)^2$ |
| where | |
| | $J^e = \sqrt{\det\boldsymbol{b}^e} = \lambda_1^e\lambda_2^e\lambda_3^e$ |
| | $\varepsilon_i^e = \ln\lambda_i^e, \quad i = 1,2,3$ |
| | $\widehat{\boldsymbol{b}}^e = {J^e}^{-2/3}\boldsymbol{b}^e$ |

$$\boldsymbol{b}^e = \boldsymbol{F}^e . \boldsymbol{F}^{e^T}$$

| | |
|---|---|
| Isotropic elasticity: | $\boldsymbol{\tau} = 2\dfrac{\partial \psi^e}{\partial \boldsymbol{b}^e} . \boldsymbol{b}^e$ |
| Flow rules: | $-\dfrac{1}{2}\mathcal{L}_v[\boldsymbol{b}^e] . \boldsymbol{b}^{e-1} = \gamma \dfrac{\partial \phi}{\partial \boldsymbol{\tau}} \quad \Leftrightarrow \quad \dot{\boldsymbol{C}}^i = -2\gamma \boldsymbol{F}^{-1} . \dfrac{\partial \phi}{\partial \boldsymbol{\tau}} . \boldsymbol{F} . \boldsymbol{C}^i$ <br> $\dot{\alpha} = -\gamma \dfrac{\partial \phi}{\partial \zeta} = \sqrt{\dfrac{2}{3}}\gamma$ <br> where <br> $\boldsymbol{C}^i \overset{\text{def}}{=} \boldsymbol{C}^{p-1}, \quad \boldsymbol{C}^p = \boldsymbol{F}^{p^T} . \boldsymbol{F}^p$ |
| KKT conditions: | $\gamma \geq 0, \; \phi \leq 0, \; \gamma\phi = 0$ |
| Consistency condition: | $\gamma \dot{\phi} = 0$ |